\documentclass{elsart}[10pt]
\usepackage{epsfig}
\usepackage{tabularx}
\usepackage{amsfonts}
\usepackage{latexsym}
\newtheorem{ttt}{Theorem}

\newcommand{\cal }{\mathcal}
\newcommand{\Scal}{{\cal S}}
\newcommand{\Sset}{{\bf S}}
\newcommand{\LL}{{\cal L}}
\newcommand{\MM}{{\cal M}}
\newcommand{\NN}{{\cal N}}
\newcommand{\DD}{{\cal D}}
\newcommand{\GG}{{\cal G}}

\newcommand{\Mp}{M_{\rm P}}
\newcommand{\cd}{\cdot}

\newcommand{\al}{\alpha}
\renewcommand{\b}{\beta}
\newcommand{\de}{\delta}
\newcommand{\De}{\Delta}
\newcommand{\ep}{\epsilon}
\newcommand{\ga}{\gamma}
\newcommand{\Ga}{\Gamma}
\newcommand{\ka}{\kappa}
\newcommand{\io}{\iota}
\newcommand{\La}{\Lambda}
\newcommand{\la}{\lambda}
\newcommand{\Om}{\Omega}
\newcommand{\om}{\omega}
\newcommand{\si}{\sigma}
\newcommand{\Si}{\Sigma}
\newcommand{\tha}{\theta}
\newcommand{\Th}{\Theta}
\newcommand{\vth}{\vartheta}
\newcommand{\vph}{\varphi}
\newcommand{\ra}{\rightarrow}

\newcommand{\diag}{\mbox{diag}}

\newcommand{\bm}[1]{\mbox{\boldmath $#1$}}
\newcommand{\bx}{\mbox{\boldmath $x$}}
\newcommand{\bp}{\mbox{\boldmath $p$}}
\newcommand{\by}{\mbox{\boldmath $y$}}
\newcommand{\bz}{\mbox{\boldmath $z$}}
\newcommand{\bk}{\mbox{\boldmath $k$}}
\newcommand{\bq}{\mbox{\boldmath $q$}}
\newcommand{\bn}{\mbox{\boldmath $n$}}

\newcommand{\EE}{{\cal E}}
\newcommand{\mr}[1]{\mathrm{#1}}
\newcommand{\pr}{\prime}

\newcommand{\be}{\begin{equation}}
\newcommand{\ee}{\end{equation}}
\newcommand{\bea}{\begin{eqnarray}}
\newcommand{\eea}{\end{eqnarray}}
\newcommand{\bean}{\begin{eqnarray*}}
\newcommand{\eean}{\end{eqnarray*}}
\newcommand{\dd}{\partial}

\newcommand{\gsim}{\stackrel{>}{\sim}}
\newcommand{\lsim}{\stackrel{<}{\sim}}

\newcommand{\ct}{\tau}   
\newcommand{\pt}{t}  
\newcommand{\eg}{{\em e.g. }}
\newcommand{\ie}{{\em i.e. }}

\renewcommand{\r}{\right}
\renewcommand{\l}{\left}
\newcommand{\lan}{\langle}
\newcommand{\ran}{\rangle}
\newcommand{\eqn}[1]{Eq.~(\ref{#1})}
\begin{document}

\title{{\bf Cosmic Structure Formation with Topological Defects}}
\author{R. Durrer,$^1$
 M. Kunz$^2$ and A. Melchiorri$^{2,3}$} 
\address{ $^1$D\'epartement de Physique Th\'eorique,
         Universit\'e de Gen\`eve\\
	24 quai Ernest Ansermet, CH-1211 Gen\`eve 4, Switzerland}
\address{$^2$Department of Astrophysics, Oxford University \\
         Keble Road, Oxford OX1 3RH, UK}
\address{$^3$ Dipartimento di Fisica, Universita Tor Vergata, Roma, Italy}
\maketitle
\clearpage
\tableofcontents
\clearpage
\section{Introduction} \label{Int}

Topological defects are ubiquitous in physics. Whenever a symmetry
breaking phase transition occurs, topological defects may form. The
best known examples are vortex lines in type II super conductors or in
liquid Helium, and declination lines in liquid crystals~\cite{Mermin,CDTY}.
In an adiabatically expanding universe which
cools down from a very hot initial state, it is quite natural to
postulate that topological defects may have emerged during a phase
transition in the early universe and that they may have played the role of
initial inhomogeneities seeding the formation of cosmic structure.
This basic idea goes back to Kibble (1976)~\cite{Ki}.
In this report we summarize the progress made in the investigation of
Kibble's idea during the last 25 years. Our understanding of the formation 
and evolution of topological defects is reported almost completely in the 
beautiful book by Vilenkin \& Shellard~\cite{VS} or the excellent Review by
Hindmarsh \& Kibble~\cite{KH}, and we shall hence be rather short on that 
topic. Nevertheless, in order to be self contained, we 
have included a short chapter on spontaneous symmetry breaking and defect 
formation. Our main topic is however the calculation of structure
formation with defects, results which are not included in~\cite{VS} 
and~\cite{KH}.

Besides the formation of structure in the universe, topological defects may be
relevant for the baryon number asymmetry of the
universe~\cite{baryon}. Superconducting cosmic strings~\cite{witten} or 
vortons~\cite{carter} might produce the  high energy cosmic 
rays~\cite{crays}, or even gamma ray bursts~\cite{grays}. The brane worlds 
which have have focussed a lot of attention recently, may actually just 
represent topological defects in a higher dimensional 
space~\cite{RG,brane1,brane2}. There have also been interesting results 
on chiral strings and their observational signatures~\cite{PeCa,dani}.
GUT scale cosmic strings could be detected by their very peculiar lensing
properties. For a straight cosmic string lensing is very simple~\cite{VS}. For
a more realistic network of strings, characteristic caustics and cusps in the 
lensing signal are very generically expected~\cite{LKV,UB,BU}.

The relevant energy scale for a topological defect is  $T_c$, the phase
transition temperature. Hence a good estimate for the amplitude of the 
dimensionless gravitational potential $\Psi$ induced by topological defects is
\be
  \Psi \sim 4\pi GT_c^2 = 4\pi(T_c/\Mp)^2~,
\ee
where $\Mp$ denotes the Planck mass.
The measurements of cosmic microwave background anisotropies on large 
scales by the cosmic background explorer (COBE) satellite~\cite{COBE} have
found that this potential, which is of the same order as the
temperature fluctuations on large scales, is about $10^{-5}$. Hence, for
cosmic structure formation, we are interested in phase transitions at $T_c\sim
10^{-3}\Mp \sim 10^{16}$GeV. Interestingly, this is just the scale 
of the  grand unification phase transition (GUT scale) of supersymmetric 
GUT's (grand unified theories).

Topological defects represent regions in space-time where the corresponding
field (order parameter in condensed matter physics or Higgs field in
particle physics) is frustrated. It cannot relax into the vacuum
state, the lowest energy state, by topological obstructions. They
represent positions of higher energy density and are thus inherently
inhomogeneous distributions of energy and momentum. We shall
discuss concrete examples later.

In the remainder of this introduction we give a brief overview of
the problem of structure formation and we
present the main results worked out in this report.

In Chapter~\ref{Sym} we introduce the concept of topological defect formation
during symmetry breaking phase transitions, we 
classify the defects and illustrate them with examples.\\
In Chapter~\ref{The} we present in detail the theoretical framework used
to investigate structure formation with topological defects. This
chapter together with two appendices is self contained and should enable 
a non-specialist in the
field to fully understand the often somewhat sketchy literature.  \\
In  Chapter~\ref{Num} we discuss numerical simulation of
topological defects. We distinguish global and local defects
which have to be treated in a very different way. We specify the
approximations made in different numerical implementations and discuss their
validity and drawbacks. \\
In Chapter~\ref{Res} we present the results of simulations of
structure formation with topological defects and compare them with
present observations. \\
In  Chapter~\ref{Gen} we investigate the question in whether the results 
discussed in Chapter~\ref{Res} are generic or whether they are just 
special cases. We derive properties of 
the unequal time correlators of generic causal scaling seeds. Since
these  are the sole ingredients in the calculation of the fluctuation 
power spectra, they determine the 'phase space' of defect models of 
structure formation. We discuss a model of causal scaling seeds which 
mimics the cosmic microwave background (CMB) anisotropy spectrum of 
inflation. We also consider the possibility that large scale structure
may be due to a mixture of adiabatic inflationary initial
perturbations and topological defects. We study especially the
fluctuations in the CMB. We investigate to
which extent CMB parameter estimations are degraded if we allow for an
admixture of defects. 

We end with a brief summary of the main results.

Throughout this work we use units with
$c=\hbar=k_{\rm Boltzmann}=1$. The basic unit is thus either an energy
(we usually take MeV's) or a length, for which we take cm or Mpc
depending on the situation.\\
We choose the metric signature $(-,+,+,+)$. Three-dimensional vectors are 
denoted in boldface. The variables $\bx$ and $\bk$ are comoving position and
comoving wave vector in Fourier space. Greek indices, $\mu,\nu,\cdots$ denote
spacetime components of vectors and tensors while Latin indices $i,j,\cdots$
denote three dimensional spatial components. We mostly use conformal 
time $\ct$ with $d\ct= dt/a$, where $t$ is cosmic time and $a$ is the scale 
factor. Derivatives with respect to 
conformal time are indicated by an over-dot, $\dot f = {df\over d\ct}$.

\subsection{Main results}
Before we start to discuss models of structure formation with
topological defects in any detail, let us present the main results 
discussed in this review.

We concentrate primarily on CMB  anisotropies. Since these
ani\-so\-tro\-pies are small, they can be calculated (almost fully) within 
linear cosmological perturbation theory. To compare models with other data 
of cosmic structure, like the galaxy distribution or velocities, one has to
make some assumptions concerning the not well understood relation between the 
distribution of galaxies and of the dark matter, the problem of biasing.
Furthermore, on small scales one has to study non-linear Newtonian clustering
which is usually done by $N$-body simulations. But to lay down the initial
conditions for  $N$-body simulations, one does not only need to know the 
linear power spectrum, but also 
the statistical distribution of the fluctuations which is largely unknown
for topological defects. Fluctuations induced by topological defects are 
generically non-Gaussian, but to which extent and how to characterize their
non-Gaussianity is still an open question. In this report, we therefore 
concentrate on CMB anisotropies and their polarization and shall only 
mention on the side the induced matter fluctuations and bulk velocities
on large scales.

Like inflationary perturbations, topological defects predict a
Har\-ri\-son-Zel'dovich spectrum of perturbations~\cite{Ha,Ze}. Therefore,
the fluctuation spectrum is in good agreement 
with the COBE DMR experiment~\cite{COBE}, which has measured CMB 
anisotropies on large angular scales and found that they are approximately 
constant as expected from a Har\-ri\-son-Zel'dovich spectrum of initial 
fluctuations (see \eg~\cite{Paddy}).  

Since quite some time it is known, however, that topological defect differ 
from adiabatic inflationary models in the  acoustic peaks of the CMB 
anisotropy spectrum~\cite{DGS}. Due to the isocurvature nature of defects, 
the position of the first acoustic peak is shifted from an angular harmonic 
of about $\ell \sim 
220$ to $\ell \sim 350$ (or up to $\ell\sim 500$ in the case of cosmic strings)
for a spatially flat universe. More important, the peaks are much lower in 
defect models and they are smeared out into one broad hump with small wiggles
at best. Even by changing cosmological parameters at will, this second 
characteristics cannot be brought in agreement with present CMB anisotropy 
data like~\cite{Net,DASI,Mnew}. Also the large scales bulk velocities, 
which measure fluctuation amplitudes on similar scales turn out to be too 
small~\cite{DKM}.

As the CMB anisotropy signals from cosmic strings and from global $O(N)$ 
defects are quite different, it is natural to wonder how generic these results
may be. Interestingly enough, as we shall see in Chapter~\ref{Gen}, one can 
define 'scaling causal seeds', \ie initial perturbations which obey the 
basic constraints for topological defects and which show a CMB anisotropy 
spectrum resembling adiabatic inflationary predictions very 
closely~\cite{Turok}. This 'Turok model' can nevertheless 
be distinguished from adiabatic perturbations by the CMB polarization spectrum.
Also mixed models with a relatively high degree of defect admixture, up to 
more than 50\%, are in good agreement with the data.

\subsection{Cosmic structure formation}
The geometry of our universe is to a very good approximation isotropic
and therefore (if we assume that we are not situated in a special position)
also homogeneous. The best observational evidence for this fact
is the isotropy of the cosmic microwave background which is (apart from 
the dipole anisotropy) on the level of about $10^{-5}$ --  $10^{-4}$. 

Nevertheless, on galaxy and cluster scales, the matter distribution is
very inhomogeneous. If these inhomogeneities have grown by
gravitational instability from small initial fluctuations, the
amplitude of the initial density fluctuations have to be about
$10^{-4}$ to become of order unity or larger today. 
Radiation pressure inhibits growth of perturbations as long as
the universe is radiation dominated, and even in a matter dominated
universe small perturbations grow only like the scale factor.

The discovery of large angular scale fluctuations in the CMB by the
DMR experiment aboard the COBE satellite~\cite{COBE}, which are just
about of the needed amplitude, is an important support of the
gravitational instability picture. The DMR experiment also revealed
that the spectrum of fluctuations is 'flat', which means that the
fluctuations  have a fixed amplitude $A$ when entering the Hubble
horizon. This implies that temperature fluctuations on large scales are
constant, as measured by COBE:
\be
  \l\langle\l(\frac{\De T}{ T}\r)^2(\vth)\r\rangle \simeq 10^{-10} = 
	\mbox{  constant,}
\ee
independent of the angle $\vth$, for $\vth\gg 1^o$.
These fluctuations are of the same order of magnitude as the
gravitational potential. Their smallness therefore justifies the use
of linear perturbation theory. Within linear perturbation theory, the
gravitational potential does not grow. This observation originally led
Lifshitz to abandon gravitational instability as the origin for cosmic
structure~\cite{Lif}. But density fluctuations of dust can become
large and can collapse to form galaxies and even black holes. At late
times and on scales which are small compared to the horizon, linear
perturbation theory is no longer valid and numerical N-body
simulations have to be performed to follow the formation of
structure. But also with N-body simulations one cannot compute the
details of galaxy formation which strongly depend on thermal processes
like cooling and formation of heavy elements in stars. Therefore, the
relation of the power spectrum obtained by N-body simulations to the
observed galaxy power spectrum may not be straight forward. This
problem is known under the name of 'biasing'.

Within linear perturbation analysis, structure formation is described
by an equation of the form
\be
{\cal D}X(\bm{k},\ct) = {\cal S}(\bm{k},\ct)~, \label{1gen}
\ee
where $\cal D$ is a time dependent linear differential operator, $\bm
k$ is the wave vector and $X$ is a long vector 
describing all the cosmic perturbation variables for a given $\bm
k$-mode, like the dark matter density and velocity fluctuations, the
CMB anisotropy multipole amplitudes and so on. $\Scal$ is a source term.
In Chapter~\ref{The} we will write down the system~(\ref{1gen}) explicitly.

There are two basically different classes of models for structure formation.
In the first class, the linear perturbation equations are homogeneous, 
\ie $\Scal\equiv 0$, 
and the resulting structure is determined by the initial conditions
and by the cosmological parameters of the background universe alone.
Inflationary initial perturbations are of this type. In most models of 
inflation, the initial fluctuations are even Gaussian and the entire 
statistical information is contained in the initial power spectrum. 
The evolution is given by the differential operator $\cal D$ which depends 
on the cosmological parameters.

In the second class, the linear perturbation equations are 
inhomogeneous, having a so called source term or 'seed', $\Scal$, on the right
hand side which evolves non-linearly in time. Topological defects are
of this second class. The difficulty of such
models lies in the non-linear evolution of the seeds, which in most
cases has to be determined by numerical simulations. Without
additional symmetries, like \eg~ scaling in the case of topological
defects, it is in general not possible to simulate the seed evolution
over the long time-scale and for the considerable dynamical range
needed in cosmology. We shall see in Chapter~\ref{Num} how this problem is
overcome in the case of topological defects. An additional difficulty
of models with seeds is their non-Gaussian nature. Due to non-linear
evolution, even if the initial conditions are Gaussian, the seeds are in
general not Gaussian at late times. The fluctuation power spectra
therefore do not contain the full information. All the
reduced higher moments can be relevant. Unfortunately only very little work on
these non-Gaussian aspects of seed models has been published and the
results are highly incomplete~\cite{GaMa,DJKU,GPW}.

\clearpage

\section{Symmetry Breaking Phase Transitions and the Formation of
	Topological Defects}
\label{Sym}

\subsection{Spontaneous symmetry breaking}
Spontaneous symmetry breaking is a concept which originated in condensed
matter physics. As an example consider the isotropic model of a
ferro-magnet: although the Hamiltonian is rotationally invariant, the
ground state is not. The magnetic moments point all in the same
direction.

In models of elementary particle physics, symmetry breaking is most
often described in terms of a scalar field, the Higgs field. In
condensed matter physics this field is called the order parameter. It
can also be a vector or tensor field. 

A symmetry is called spontaneously broken if the ground state is not
invariant under the full symmetry of the Lagrangian (or Hamiltonian) 
density. Since the symmetry group can be represented as a group of linear
transformations, this implies that the vacuum expectation value of the
Higgs  field is non-zero.

The essential features of a spontaneously broken symmetry can be
illustrated with a simple model which was first studied by Goldstone
(1961)~\cite{Go}. This model has the classical Lagrangian density
\be
\LL = \dd_\mu\bar{\phi}\dd^\mu\phi -V(\phi)  ~~\mbox{ with }~~~ \label{lag} 
 V = {1\over 4}\la(|\phi|^2-\eta^2)^2 ~.
\ee
Here $\phi$ is a complex scalar field and $\la$ and $\eta$ are real
positive constants. The potential in (\ref{lag}) is called the 'Mexican 
hat potential' as it looks somewhat like a Mexican sombrero. This  
Lagrangian density  is invariant under the
group $U(1)$ of global phase transformations,
\be
 \phi(x)\mapsto e^{i\al}\phi(x) ~.  \label{global}
\ee
The minima of the potential $V$ lie on the circle $|\phi|=\eta$ which
is called the 'vacuum manifold',
$ \MM= \Sset_\eta^1$~(here and in what follows $\Sset^n_R$ denotes an 
$n$-sphere of radius $R$ and $\Sset^n$ denotes an 
$n$-sphere of radius $1$).
The notion 'global' indicates that the symmetry transformation is
global, {\em i.e.}, $\al$ is independent of the spacetime position
$x$. The quantum ground states (vacuum states) $|0\rangle$ of the
model are characterized by
\be \langle 0|\phi|0\rangle  =\eta e^{i\beta} \neq 0 ~. \ee
A phase transformation changes $\beta$ into $\beta+\al$, hence a ground
state is not invariant under the symmetry transformation given in
\eqn{global}. (Clearly, the full vacuum manifold $\MM$ is invariant
under symmetry transformations and thus a mixed state which represents
a homogeneous mixture of all vacuum states is still invariant even
though no pure state is.) The only state $|u\rangle$ invariant under the 
symmetry~(\ref{global}), characterized by  $\langle u|\phi|u\rangle  = 0$,
corresponds to a local maximum of the potential. Small perturbations around
this 'false vacuum' have 'negative mass' which indicates the
instability of this state:
\be
V(\phi) = -{1\over 2}\la\eta^2|\phi|^2 + \mbox{const.}+{\cal O}(|\phi|^4)~.
\ee
The vacuum states of the broken symmetry are all equivalent and we can
thus reveal their properties by studying one of them. For convenience
we discuss the vacuum state with vanishing phase,
$\langle 0|\phi|0\rangle  =\eta$.
Expanding the field around this state yields
\be
\phi(x) = (\eta+{1\over \sqrt{2}}\vph(x))e^{i\vth(x)} ~,\label{exp}
\ee
where $\vph$ and $\vth$ are real fields. The Lagrangian density in terms 
of $\vph$ and $\vth$ becomes 
\be
\LL = {1\over 2}\left(\dd_\mu\vph\right)^2 + 
	\eta^2\left(\dd_\mu\vth\right)^2 -{1\over 2}\la\eta^2\vph^2
	+\LL_{int}(\vph,\vth) ~.
\ee
The interaction Lagrangian $\LL_{int}$ is easily determined from the
original Lagrangian, ~(\ref{lag}). This form of the Lagrangian shows that
the degree of freedom $\vph$ is massive with mass $m^2=\la\eta^2$ while
$\vth$ describes a massless particle (circular excitations), a
Goldstone boson. This simple model is very generic:
whenever a continuous global symmetry is spontaneously broken, massless
Goldstone bosons emerge. Their number equals the dimension of the
vacuum manifold, {\em i.e.}, the dimension of a group orbit (in the space of
field values). In our case the space of field values is 
$\Cset \approx \Rset^2$. 
A group orbit is a circle of dimension $1$ leading to one
massless boson, the excitations tangential to the circle which cost no
potential energy. The general result can be formulated in the
following theorem:

\begin{ttt} (Goldstone, 1961)~\cite{Go}
If a continuous global symmetry, described by a symmetry group  $G$ is 
spontaneously broken to a sub-group $H\subset G$, massless
particles emerge. Their number is equal to the dimension $n$ of the vacuum
manifold $\MM$ (the ``number of broken symmetries''). Generically, 
\[ \MM\equiv G/H ~~~\mbox{ and }~~~~n=dim G -dim H =dim\left(G/H\right)~,\]
where here $\equiv$ means topological equivalence.
\end{ttt}

In our example $G=U(1)$, $H=\{1\}$ and $n=1-0=1$.\\
Another well-known example are the three pions, $\pi^\pm$, $\pi^0$, which are
the Goldstone bosons of isospin (proton/neutron) symmetry. There
the original symmetry, $SU(2)$ is completely broken leading to
$n=dimSU(2)=3$ Goldstone bosons (see. e.g. ~\cite{IZ}).

Very often, symmetries in particle physics are gauged (local). The
simplest gauge theory is the Abelian Higgs model (sometimes also
called scalar electrodynamics). It is described by the Lagrangian
density
\be
 \LL = \overline{D}_\mu\bar{\phi}D^\mu\phi -V(\phi) -
	{1\over 4}F_{\mu\nu}F^{\mu\nu}~, \label{lagl}
\ee
where $\phi$ is again a complex scalar field and
$D_\mu =\dd_\mu -ieA_\mu$ is the covariant derivative w.r.t. the gauge field 
$A_\mu$. $F_{\mu\nu}=\dd_\mu A_\nu -\dd_\nu A_\mu$ is the gauge field-strength,
$e$ the gauge coupling constant and $V$ is the potential given
in~\eqn{lag}. 

This Lagrangian is invariant under the group of {\bf local} $U(1)$
transformations,
\[ \phi\mapsto e^{i\al(x)}\phi(x)~~,~~~~~~ A_\mu(x) \mapsto
	A_\mu(x)+{1\over e}\dd_\mu\al(x)~.\]
 The minima of the potential, $\phi=\eta e^{i\beta}$, are not
invariant, the symmetry is spontaneously broken. Expanding as before
around the vacuum expectation value
$ \langle 0|\phi|0\rangle  =\eta$, 
we find
\bea
\LL &=& [\dd_\mu\vph-(ieA_\mu+i\dd_\mu\vth)(\eta+\vph)]
	[\dd^\mu\vph+(ieA^\mu+i\dd^\mu\vth)(\eta+\vph)] \nonumber \\
	&&
   -{1\over 2}m^2\vph^2-{1\over 2}\la\vph^4-{1\over4}F_{\mu\nu}F^{\mu\nu}~,
\label{lagl2}
\eea
where, as in the global case, $m^2=\la\eta^2$. Here $\vth$ is no
longer a physical degree of freedom. It can be absorbed by a gauge
transformation. After the gauge transformation $A_\mu\mapsto
A_\mu - (1/e)\dd_\mu\vth$ the Lagrangian given in~\eqn{lagl2} becomes
\be
\LL = (\dd_\mu\vph)^2 -{1\over 2}m^2\vph^2 +{1\over 2}M^2A_\mu A^\mu
	-{1\over4}F_{\mu\nu}F^{\mu\nu} +\LL_{int}~,
\ee
with $m=\sqrt{\la}\eta$ and $M=\sqrt{2}e\eta$.
The gauge boson ``absorbs'' the massless Goldstone boson and becomes
massive. It has now three independent  polarizations (degrees of
freedom) instead of the original two. The phenomenon described here is
called the 'Higgs mechanism'. It works in the same way also
for more complicated non-Abelian gauge theories (Yang Mills
theories). 

On the classical level, what we have done here is just rewriting
the Lagrangian density in terms of different variables.
However, on a quantum level, particles are excitations
of a vacuum state, a state of lowest energy, and these are clearly not
described by the original field $\phi$ but by the fields $\vph$ and
$\vth$ in the global case and by $\vph$ and $A_i$ in the local
case. 

The two models presented here have very close analogies in condensed
matter physics:\\
a) The non-relativistic version of~\eqn{lag} is used to describe
super fluids where $\phi$ is the Bose condensate (the best known
example being super fluid He$^4$).\\
b) The Abelian Higgs model, \eqn{lagl} is the Landau Ginzburg
model of super-conductivity, where $\phi$ represents the Cooper
pair wave function.

A very physical and thorough account of the problem of spontaneous 
symmetry breaking can be found in Weinberg \cite{We2}.

It is possible that also the scalar fields in particle physics
(e.g. the Higgs of the standard model which is supposed to provide the
masses of the $W^\pm$ and $Z^0$) are not fundamental but
``condensates'' as in condensed matter physics. Maybe the fact that no
fundamental scalar particle has been discovered so far has some deeper
significance. 

\subsection{Symmetry restoration at high temperature}
 In particle physics like in condensed matter systems, symmetries
which are spontaneously broken can be restored at high
temperatures. The basic reason for this is that a system at finite
temperature is not in the vacuum state which minimizes  energy,
but in a thermal state which tends to maximize entropy. We thus have to expand
excitations of the system about a different state. More precisely, it
is not the potential energy, but the free energy
\be F = E-TS \ee
which has to be minimized. The equilibrium value of $\phi$ at
temperature $T$, $\langle\phi\rangle_T$, is in general temperature 
dependent~\cite{Li74}.
At low temperature, the entropy term is unimportant. But as the
temperature raises, low entropy becomes more and more costly and
 the state tends  to raise its entropy. The field $\phi$ becomes less
and less ordered and thus its expectation value $\langle\phi\rangle_T$ becomes smaller. 
Above a certain critical temperature, $T\ge T_c$, the expectation
value $\langle\phi\rangle_T$ vanishes. If the coupling constants are not extremely
small, the critical temperature is of order $T_c\sim \eta$.

To calculate the free energy of quantum fields at finite temperature,
one has to develop a perturbation theory similar to the $T=0$ Feynman
diagrams, where ordinary Greens functions are replaced by thermal 
Greens functions. The inverse temperature,
$\beta=1/(kT)$  plays the role of an imaginary time
component. It would lead us too far from the main topic of this
review to introduce  thermal perturbation theory, and there are
excellent reviews on the subject available, see, 
e.g.~\cite{Bellac,We2,Ka,We74,DJ,Li74}. 

Here we give a much simplified derivation of the lowest order (tree
level) thermal correction to the effective potential~\cite{VS}. In
lowest order the particles are non-interacting and their contributions to
the free energy can be summed (each degree of freedom describes one particle),
\be
 V_{\rm eff}(\phi,T) = V(\phi) +\sum_nF_n(\phi,T) ~.
\ee
Here $V(\phi)$ is the zero temperature effective potential and $F_n$
is the free energy of each degree of freedom,
\be
 F_n = \pm\int{d^3k\over (2\pi)^3}\ln(1\mp\exp(-\ep(k)/T)) ~,
\ee
as known from statistical mechanics. The upper sign is valid for
bosons and the lower one for fermions. $\ep(k)=\sqrt{k^2+m_n^2}$.

For $T\ll m_n$ the free energy is exponentially small. But it can become 
considerable at high temperature, $T\gg m_n$, where we obtain
\be
 F_n =\left\{ \begin{array}{ll}
-{\pi^2\over 90}T^4 +{m_n^2T^2\over 24} + {\cal O}(m_n^4) &
\mbox{bosons}\\
-{7\pi^2\over 720}T^4 +{m_n^2T^2\over 48} + {\cal O}(m_n^4) &
\mbox{fermions.}
\end{array}
  \right.
\ee
If symmetry restoration occurs at a temperature well above all the
mass thresholds, we can approximate $V_{\rm eff}$ by
\be V_{\rm eff}(\phi,T) = V_{\rm eff}(\phi,T=0) +{1\over 24}\MM^2T^2
-{\pi^2\over 90}\NN T^4 ~,  \label{Veff}
\ee
\be
 \NN = N_B +{7\over 8}N_F ~,~~ \MM^2 =\sum_{\rm bosons}m_n^2
+{1\over 2}\sum_{\rm fermions}m_n^2 ~.
\ee 
Here, $m_i$ is the formal mass given by $m_i^2={\dd^2V\over (\dd\phi^i)^2}$.
If the potential contains a $\phi^4$-term, the mass  includes a term 
$\propto \phi_i^2$, which leads to a positive quadratic term, 
$\propto T^2\phi_i^2$. If the temperature is sufficiently high, this term 
overcomes the negative quadratic term in the Mexican hat potential and 
$\phi=0$ becomes a global minimum of the potential. The temperature at 
which this happens is called the critical temperature.
  
In the Abelian Higgs model, the critical temperature $T_c$, 
becomes~\cite{VS}
\be
 T_c^2 = {6\la\eta^2\over \la+3e^2} \label{Tc} ~.
\ee
For non-Abelian $O(N)$ models one finds analogously~\cite{VS}
\be
 T_c^2 = {6\la\eta^2\over{N+2\over 4}\la+3(N-1)e^2} \label{TcN} ~.
\ee
The critical temperature for  global symmetry breaking, {\em i.e.}
without gauge field, is obtained in the limit $e\ra 0$. As expected, for
$e^2\lsim \la$ one finds
\be 
 T_c \sim \eta ~. \label{Tcsim}
\ee

Like in condensed matter systems, a phase transition is second order
if $\phi=0$ is a local maximum and first order if 
 $\phi=0$ is a local minimum. In the example of the Abelian Higgs
model, the order depends on the parameters $e$ and $\la$ of the model.

In $O(N)$ models, or any other model where the vacuum manifold (\ie the space 
of minima of the effective potential) of the broken symmetry phase is 
non-trivial, minimization of the effective potential fixes the 
absolute value of $\phi$ but the direction, $\phi/|\phi|$, is arbitrary. 
The field can vary in the vacuum manifold, given by the sphere 
$\Sset^{N-1} =O(N)/O(N-1)$ for $O(N)$ models. 
At low temperature, the free energy is minimized if the phase is constant 
(no gradient energy) but after the phase transition  
$\phi/|\phi|$ will vary in 
space. The size of the patches with roughly constant direction is given 
by the correlation length $\xi$ which is a function of time. In the early 
universe $\xi$ is bounded by the size of the causal horizon,
\be
 \xi(t)\le d_H(t) \sim \pt   ~~~\mbox{ for power law expansion.}
\ee
Formally $\xi$ diverges at the phase transition, but also our 
perturbative treatment is no longer valid in the vicinity of the phase 
transition since fluctuations become big. A thorough treatment of the 
physics at the phase transition is the subject of modern theory of 
critical phenomena and goes beyond the scope of this review. Very often,
the relevant correlation length is the correlation length at the Ginsburg
temperature, $T_G$, the temperature at which thermal fluctuations are 
comparable to the mass term. However, in the 
cosmological context there is also another scale, the expansion time. As
the system approaches the phase transition, it takes longer and longer to 
reach thermal equilibrium, and at some temperature, expansion is faster 
than the speed at which the system equilibrates and it falls out of thermal
equilibrium. It has been argued~\cite{Zurek} that it is the correlation length
at this moment, somewhat before the phase transition, which is relevant.

If the  phase transition is second order, the order parameter $\phi$ changes 
continuously with time. In a first order transition, the state $\phi=0$ is 
meta-stable 
(false vacuum) and the phase transition takes place spontaneously at 
different positions in space and different temperatures $T<T_c$ via bubble
nucleation (super cooling). Thermal fluctuations and/or tunneling 
take the field over the potential barrier to the true vacuum. The 
bubbles of true vacuum grow and eventually coalesce thereby completing the 
phase transition.

It is interesting to note that the order of the phase transition is not 
very important in the context of defects and structure formation. 
Even though the number of defects per horizon volume formed at the 
transition does depend on the order and, especially on the relevant 
correlation length~\cite{Zurek}, this can be compensated by a slight 
change of the phase transition temperature to obtain the  required 
density of defects.

As we have seen, a non-trivial vacuum manifold, 
$\MM\neq \{0\}$, in general implies that shortly after a phase transition
the order parameter has different values at different positions in space. 
Such non-trivial configurations are generically 
unstable and will eventually relax to the configuration $\phi=$constant, 
which has the lowest energy. Naturally, we would expect this process to 
happen with the speed of light. However, it can be slowed significantly 
for  topological reasons and intermediate long lived configurations 
with well confined energy may form, these are topological defects. 
Such defects can have important consequences in cosmology.

Several exact solutions of topological defects can be found in the 
literature, see \eg~\cite{VS}. In the case of global defects, \ie~defects 
due to {\em global} symmetry breaking, the energy 
density of the defect is mainly due to gradient energy in the scalar field 
and is therefore not well localized in space. The scalar field gradient of
local defects (defects due to the braking of a {\em local}, or gauge symmetry)
is compensated by the gauge field and the energy is well 
confined to the location of the defect. To exemplify this, we present the 
solutions for a global and a local straight cosmic string.

\subsection{Exact solutions for strings}  
\subsubsection{Global strings}
We consider  a complex scalar field, $\phi(x)\in \Cset$, with Lagrangian
\be
  {\cal L} = {1\over 2}\dd_\mu\phi\dd^\mu\overline{\phi} -V(\phi)~~~,~~~ 
   V={\la\over 4}(|\phi|^2 -\eta^2)^2  \label{Lstg}
\ee
at low temperature. The vacuum manifold is a circle of radius
 $\eta$, $\MM= \{\phi(x)\in \Cset | |\phi|=\eta\}$. At high 
temperature, $T\gg \eta$, the effective potential has a single 
minimum at $\phi=0$. As the temperature drops below the critical value 
$T_c\sim\eta$, a phase transition occurs and $\phi$ assumes a finite 
vacuum expectation value $\langle 0|\phi|0\rangle \neq 0$ which is 
uncorrelated at sufficiently distant points in physical space. If we now 
consider a closed curve in space
\[ \Ga :[0,1]\ra \Rset^3: s\mapsto \bm{x}(s)~~;~~~~ \bx(0)=\bx(1) \]
it may happen that $\phi(\bx(s))$ winds around in the circle $\MM\sim  
\Sset^1_\eta$.
We then have $\phi(\bx(s))=\eta\exp(i\al(s))$ with $\al(1)=\al(0) +n2\pi$ 
with $n\neq 0$. Since the integer $n$ (the winding number of the map
$\Ga\ra  {\Sset}^1_\eta : s\mapsto \phi(\bx(s))$) cannot change if we 
shrink the curve $\Ga([0,1])$ continuously, the function $\al(s)$ must be ill 
defined somewhere in the interior of $\Ga$, {\it i.e.} $\phi$ must assume 
the value $\phi=0$ and thus have higher potential energy somewhere in the 
interior of $\Ga([0,1])$.

If we continue this argument into the third dimension, a string of higher 
potential energy must form. The
size of the region within which $\phi$ leaves the vacuum manifold, the 
diameter of the string,  is of the order $\eta^{-1}$. For topological 
reasons, the string cannot end. It is either infinite or closed.\footnote{
The only exception may occur if other defects are present. Then a string
can end on a monopole.}

We now look for an exact solution of a static, infinite straight string 
along the $z$-direction. We make the ansatz
\be
   \phi(\bx) = \eta f_s(\rho\eta)\exp(in\varphi) ~,\label{anstring}
\ee
with $\rho=\sqrt{x^2+y^2}$ and $\tan\varphi=y/x$, $\varphi$ is the usual polar 
angle. The field equation of motion  then reduces to an 
ordinary differential equation for $f_s$,
\be
f_s'' + {1\over v}f_s' -{n^2\over v^2}f_s -{\la\over 2}f_s(f_s^2-1)=0~, 
\ee
where $v=\rho\eta$ and $'={d\over dv}$. A solution of this differential 
equation which satisfies the boundary conditions $f_s(0)=0$ and 
$f_s(v)\ra_{v\ra\infty} 1$ can be found numerically. 
It is a function of $\sqrt{\la}\rho\eta$. and behaves like 
\[
 f_s \sim 1-{\cal O}(1/v^2) \mbox{ for } \sqrt{\la}\rho\eta\gg 1 ~~~
 f_s \sim {\cal O}(v^n) \mbox{ for } \sqrt{\la}\rho\eta\ll 1 .
\]
 
 The energy momentum tensor of the string is given by
\bea
 T_0^0 =T_z^z &=& -{\la\eta^4\over 2}[f'^2 -{1\over 2}(f^2-1)^2
  + {n^2\over \la\eta^2\rho^2}f^2 ] \label{T00globst} \\
 T_\mu^\nu &=& 0 ~~~~~~~\mbox{ for all other components.}\nonumber
\eea
 The energy per unit length of a cross-section of string out to radius $R$ is
\be
    \mu(R) =2\pi\int_0^RT_0^0\rho d\rho \sim \pi\eta^2\ln(\sqrt{\la}\eta 
    R) ~.  \label{muglob}
\ee     
The log divergence for large $R$ results from the angular dependence of 
$\phi$, the gradient energy, the last term in \eqn{T00globst}, which 
decays only like $1/\rho^2$. In realistic configurations an upper cutoff 
is provided by the curvature radius of the string or by its distance to 
the next string.

Also for a single, spherically symmetric global monopoles solution the 
total energy divergies (linearly). A non-trivial results shows, however, 
that the energy needed to deform the monopole into a topologically trivial
configuration is finite~\cite{Ana}.

\subsubsection{Local strings}
We also describe a string solution of the Abelian Higgs model, the 
Nielson-Oleson or Abrikosov vortex~\cite{NiO}.

The Lagrangian density is the one of scalar electrodynamics, \eqn{lagl},
\be
\!\! \LL = (\dd_\mu +ieA_\mu)\bar{\phi}(\dd^\mu-ieA^\mu)\phi -{\la\over 
 4}(|\phi|^2\!-\!\eta^2)^2 -{1\over 4}F_{\mu\nu}F^{\mu\nu}. \label{lagl2g}
\ee
We are looking for a cylindrically symmetric, static solution of the field 
equations. For $\rho\ra\infty$ we want the solution to approach a
vacuum state, 
{\it i.e.} $\phi\ra \eta\exp(in\varphi)$ and $A_\mu\ra 
(n/e)\dd_\mu\varphi$ so that $D_\mu\phi\ra 0$ (the gauge field `screens' 
the gradient energy). 

We insert the following ansatz into the field equations
\bea
 \phi &=& \eta f_A(\rho)\exp(in\varphi) \label{fl}\\
 A_x &=& {-n\over e}(y/\rho^2)\al(\rho) ~,~~
 A_y ={n\over e}(x/\rho^2)\al(\rho) ~,~~ \label{ay}  
 A_z = 0 ~.
\eea
which leads to two coupled ordinary differential equations for $f_A$ and 
$\al$
\bea
{d^2f_A\over d\rho^2} +{1\over \rho}{df_A\over d\rho} -{n^2\over 
\rho^2}f_A(\al-1)^2 -{\la\eta^2\over 2}f_A(f_A^2-1) &=& 0 \label{fA} \\
{d^2\al\over d\rho^2} +{1\over \rho}{d\al\over d\rho} -2e^2\eta^2f_A^2(\al-1) 
 &=& 0~. 
\eea 
Solutions which describe a string along the $z$-axis satisfy the 
asymptotics above which require
\be
 f_A(0) = \al(0) = 0,~~\mbox{ and }~~~ 
 f_A(\rho)~,~\al(\rho)\ra_{\rho\ra\infty} 1 ~.
\ee
The solution to this system of two coupled ordinary differential 
equations is easily  obtained numerically. 

Asymptotically, for $\rho\ra\infty$, the $\al$-equation reduces to the 
differential equation for a modified Bessel function and we have
\be
 \al(\rho) \simeq 1-\rho\eta K_1(\sqrt{2}e\rho\eta) ,~~
|\al\!-\!1|\sim {\cal O}\left(\sqrt{\rho\eta}e^{-\sqrt{2}e\eta\rho}\right).
\ee
For large values of $\beta \equiv \la/2e^2$, the falloff of $f_A$ is 
controlled by the gauge field coupling, $\propto (\al-1)^2$. For 
$\beta\lsim 4$ the gauge field coupling can be neglected at large radii 
$\rho$ and we obtain for $\rho\ra\infty$
\be
 f_A(\rho)\sim 1-K_0(\sqrt{\la}\rho\eta) \sim 1-{\cal O}\left(
    \exp(-\sqrt{\la}\rho\eta)\right)  ~.
\ee
This field configuration leads to $F_{0i}=F_{3i}=0$, hence 
$E_i=B_1=B_2=0$ ; and 
\be
B_3 = \ep_{3ij}F_{ij} = \dd_1A_2-\dd_2A_1 = {n\over e\rho}\al' ~.
\ee
The energy per unit length of the string is
\be
\mu  = 2\pi\eta^2\int_0^{\infty}sds \left[f_A'^2+n(1-\al)^2f_A^2 +{\la\over 
4}(f_A^2-1)^2\right]~,
\ee
with $s=\eta\rho$. All the terms in the integral are regular 
 and decay exponentially for large $s$. The energy per unit 
length of gauge string is finite. 
The gradient energy which leads to the divergence for the global string 
is `screened' by the gauge field. The integral is of the order $\mu\sim 
2\pi\eta^2$. This value is exact in the case $\beta=|n|=1$, where it can 
be computed analytically. In the general case with $|n|=1$, we have
$
  \mu = 2\pi\eta^2 g(\beta) ~,
$
where $g(\beta)$ is a slowly varying function of order unity.

The thickness of a Nielson Oleson string is about $\eta^{-1}$ and 
on length scales much larger than $\eta^{-1}$ we can approximate its 
energy momentum tensor by
\be
(T_\mu^\nu) =\mu\de(x)\de(y)\diag(1,0,0,1) ~. \label{Tmunustloc}
\ee

\subsection{General  remarks on topological defects}
In three spatial dimensions four different types of defects can form.
The question whether and what kind of topological 
defects form during a symmetry breaking phase transition is determined
by the topology of the vacuum manifold $\MM$:
\begin{itemize}
\item If $\MM$~ is {\bf disconnected}, {\bf domain walls} from. Example: if 
the symmetry $\phi \ra -\phi$ for a real scalar field is spontaneously 
broken, $\MM = \{-\eta,\eta\}$. Domain walls form when discrete symmetries 
are broken. Discrete symmetries are not continuous and therefore cannot be 
gauged. Hence domain walls are always global defects.
\item If there exist loops in $\MM$~ which cannot be continuously shrunk 
into a point, $\MM$~ is {\bf not simply connected}, {\bf strings} form.
Example: if $U(1)$ is completely broken by a complex scalar field, 
$\MM={\Sset}^1$, see previous subsection.
\item If $\MM$~ contains {\bf non-contractible spheres}, {\bf monopoles} form.
Example: if $O(3)$ is broken to $O(2)$ by a three component scalar 
field, $\MM = {\Sset}^2=O(3)/O(2)$.
\item If $\MM$~ contains {\bf non-contractible 3-spheres}, {\bf textures} form.
Example: if $O(4)$ is broken to $O(3)$ by a four component scalar field, $\MM =
 {\Sset}^3=O(4)/O(3)$.
\end{itemize}
These topological properties of $\MM$~ are best described by the homotopy 
groups, $\pi_n(\MM)$. The group $\pi_3(\MM)$ is relevant for the existence 
of textures, $\pi_2(\MM)$ decides about monopoles, $\pi_1(\MM)$ is relevant 
for strings and $\pi_0(\MM)$ for domain walls~\cite{kibb1}. If a symmetry 
group $G$ is spontaneously broken to a subgroup $H\in G$, the vacuum manifold 
is in general equivalent to the quotient space, $\MM \simeq G/H$. In the 
monopole example above, we have $G=O(3)$ and $H=O(2)$. The vacuum 
manifold is $\MM\simeq {\Sset}^2 \simeq O(3)/O(2)$.

\subsection{Defect formation and evolution in the expanding
	universe}
Our universe which is to a good approximation an expanding Friedman 
universe was much denser and hotter in the past. During expansion
the universe may cool through a certain critical temperature $T_c$ at 
which a symmetry $G$ is spontaneously broken down to $H\subset G$. 
If $\MM=G/H$ is topologically non-trivial, 
topological defects can form during the phase transition.
This scenario is called the Kibble mechanism~\cite{kibb1}. We apply the 
Kibble mechanism to estimate the energy density in defects from
phase transitions with different vacuum manifolds at a given
temperature $T_c$. 
Consider a field theory with symmetry group $G$ and Higgs-field $\phi$ with
a self-interaction potential $V(\phi)$. For illustration we use 
$\phi\in\Cset$, $G=U(1)$ and 
	\be\label{pot3}
	V(\phi)=\frac{1}{4}\la (\overline\phi\phi-\eta^{2})^{2}.
	\ee
At finite temperature, the free energy is of the form
	\be\label{pot4}
	V_{T}(\phi)={A\over 2}T^{2}\overline\phi\phi+V_{0}(\phi),
	\ee
where $A$ is a real constant given by combinations of $\la$ and other 
coupling constants (\eg~gauge 
couplings, Yukawa couplings). The sign of $A$ depends on the number of 
fermions. We assume $A>0$, \ie~that there are only few fermions and 
sufficiently small Yukawa-couplings. Then, from Eqs.~(\ref{pot3}) 
and~(\ref{pot4}), we see that the effective masses of the field $\phi$ at 
temperature $T$ and $T_{c}$ are
	\[m^{2}(T)\equiv V_{T}''(\phi=0)=AT^{2}-\la\eta^{2},\]
	\[m(T_{c})=0, ~~ T_{c}=\eta\left(\frac{\la}{A}\right)^{1/2}~ ,
	\qquad \mbox{for }\; \la\sim A\sim 1, ~~T_{c}\sim\eta.\]
At $T=T_{c}$, this field theory undergoes a second order phase 
transition: the equilibrium 
point $\phi=0$ becomes unstable for $T<T_{c}$ ($m^{2}$ becomes negative)
and $\phi$ assumes a non-vanishing vacuum expectation value.

For another form of $V_{T}$, the equilibrium 
$\phi=0$ at $T=T_{c}$ can be meta-stable so that the phase transition is
of first order. Hence, to decide whether the transition is of first or 
second order, it is important that we can rely on the form of the 
effective potential 
$V_{T}(\phi)$ which is obtained by perturbation theory or by numerical lattice
calculations. This is in general a difficult problem. For the electro-weak 
theory, \eg,~ it was discovered only recently that the electroweak 
transition is probably not a real phase transition but just a 
continuous cross-over~\cite{EWtrafo}.

The correlations of the field $\phi$ are described by the thermal Greens
functions:
	\[G(|\bx-\bx'|)=\langle\phi^{*}(\bx,t)\phi(\bx',t)\rangle.\]
For massive particles at $T>T_{c}$ where $\omega_{k}^{2}=k^{2}+m^{2}(T)$
one can write
	\[G(|\bx-\bx'|)=2\int_{0}^{\infty}\frac{d^{3}k}{(2\pi)^{3}}
	\;\frac{1}{\omega_{k}}\frac{e^{i{\bf k}(\bx-\bx')}}{e^{\omega_{k}/T}-1}
	+G^{0},\]
where $G^{0}$ are the zero temperature contributions.
For $T\rightarrow T_{c}$ such that $m(T)\ll T$ we have, with $r\equiv 
|\bx-\bx'|$:
	\[G(r)\simeq \left\{\begin{array}{ll}
		T^{2}/6, &\mbox{for}\; r\ll 1/T \\
		\exp[-m(T)r]T/(2\pi r) ,\;\;&\mbox{for}\; r\gg 1/T ~.
		\end{array}\right.\]
For $T\rightarrow T_{c}$, $m(T)\rightarrow 0$ and therefore, at $T_C$
 $G\sim r^{-1}$ for large $r$. The correlation length $\xi$ 
for the phase transition (of $2^{nd}$ order) is defined as
	$\xi:=\frac{1}{T_{c}}$.
This definition is different from the definition used in solid state physics.
There one defines the correlation length as the length above which the 
correlation decreases exponentially. In this sense, the correlation length
would be infinite at $T_c$ ($r_{c}=1/m(T)$). In cosmology the correlation 
length cannot diverge because of causality. It is bounded from above 
by the distance a photon can travel during the age of the universe until 
$t_c$.
This distance is (for non-inflationary expansion) given by
	\[l_{H}=a(\ct)\ct\cong t_{c},\qquad t_{c}:=\int a(\ct)\,d\ct.\]
Hence another meaningful definition of the correlation length would be
	$\xi:=l_{H}\cong t_{c}$. 
Often also the correlation length at the Ginzburg temperature or at the 
temperature (before the phase transition) at which the system drops out 
of thermal equilibrium~\cite{Zurek} is chosen. 
For the following it is not important which of the above definitions we use.
We only require that $ \xi\le l_{H}$. We now suppose 
that directly after the phase transition, the vacuum expectation 
value $\langle\phi\rangle$ takes arbitrary uncorrelated values in points
with distance $r>\xi$, but stays continuous (finite gradient energy!). If
$\pi_{n}(\MM)$ is non-trivial for $n\leq 3$,
the map
	\[\langle\phi\rangle:\:{\Sset}^{n}\rightarrow \MM,\qquad
	\bx\mapsto \langle \phi(\bx) \rangle\]
 for a large enough $n$-sphere ${\Sset}^{n}$ in physical space, 
may represent a non-trivial element of $\pi_{n}(\MM)$. Then 
$\phi({\Sset}^{n})$ cannot be contracted continuously to a point on 
${\cal M}$ and,
somewhere inside ${\Sset}^{n}$, $\langle \phi\rangle$ has to leave the vacuum
manifold, $\langle\phi\rangle(p)\not\in \MM$. These positions of higher 
potential energy 
are topological defects. The type of defect formed depends on the order 
$n$ of the non-trivial homotopy group:
	\begin{itemize}
	\item $n=0$: 2-dimensional defects, domain walls, $d=3$
	\item $n=1$: line-like defects, cosmic strings, $d=2$
	\item $n=2$: point-like defects, monopoles,  $d=1$
   \item $n=3$: event-like defects, texture, $d=0$. 
	\end{itemize}
Here $d$ is the spacetime dimension of the defect, $d=4-1-n$. If a vacuum 
manifold has several nontrivial homotopy groups with $n\le 3$, generically 
only the lowest $n$ defects survive and the higher order defects are instable.
As an example, in the isotropic to nematic phase transition of liquid 
crystals~\cite{CDTY}  $O(3)$ is broken  to $O(2)\times \Zset_2$ 
leading to $\MM =O(3)/(O(2)\times \Zset_2 \equiv \Sset^2/\Zset_2$. This allows 
for texture, monopoles and strings, but textures decay into  monopole 
anti-monopole pairs and monopole/anti-monopole pairs are connected by strings
and attract each other until they annihilate. Only strings scale~\cite{CDTY}.

The case  of texture, $n=3$  can be described in this context only if 
either the universe is closed and physical space is a three sphere of if 
$\phi$ is asymptotically 
parallel, i.e. $\phi(\bx,t)\stackrel{|\bx|\to\infty}
{\longrightarrow}\phi_{0}$. Then the points $|\bx|\rightarrow 
\infty$ can be identified in all directions and we can regard $\phi$ as a 
map from $\Rset^{3}\cup\{\infty\}\equiv {\bf S}^{3}$ to $\MM$ and ask 
whether this map is topologically trivial or not. In the cosmological 
context  this concept violates causality.
However, the texture case  allows for a texture winding number 
density whose integral 
over all of space only takes integer values if $\phi$ is asymptotically 
constant (or space is a three sphere). The integral of the winding number 
density  over a region of space tells us whether the 
field configuration inside contains textures.

According to the above description of the process of defect formation after
a cosmological phase transitions, called the Kibble mechanism\cite{kibb1},
we typically expect on the order of one defect per horizon volume.
Simulations and analytical arguments show that the actual 
number is somewhat larger for cosmic strings and significantly smaller 
for texture.  

If defects are local, the scalar field gradiants are compensated by the 
gauge field and they do not interact at large distance other than 
gravitationally in the simplest model, where no massless charged 
particles 'live' inside the defect. An exeption to this are superconducting 
cosmic string~\cite{witten}. For
example local monopoles do not annihilate once they are
formed and their density just scales with the expansion of the Universe, 
like $1/a^3$. Since they are non-relativistic, $m>T$, their energy density
scales the same way and they soon dominate the total energy density of
the universe. Every simple GUT group produces monopoles when it
breaks down to the standard model symmetry group, $SU(3)\times
SU(2)\times U(1)$. The observed absence of monopoles therefore represents a
serious problem for the unification of standard cosmology with grand unified 
theories~\cite{KT}. Local texture, on the contrary, soon thin
out and do not induce sufficiently strong perturbations to generate
structure in the universe. Only local strings scale, {\em i.e.}
contribute a constant small fraction to the energy density of
the universe, and are therefore possible candidates of topological
defects for structure formation.
If the group $\pi_1(\MM)$ is non-Abelian ($\pi_1$ is the only homotopy 
group which can be non-Abelian), the cosmic string network becomes 
'frustrated' and does not scale. Such a low energy frustrated string 
network has been proposed as candidate for the cosmic dark energy~\cite{SpBu}. 

The situation is different for global defects. There, the main
contribution to the energy density comes from the Higgs field and scales
as $1/t^2$, like the background energy density in the universe (up to
logarithmic corrections for global strings). The only exception are
domain walls which are forbidden, since they soon come to dominate,
leading to a very inhomogeneous universe. Recently, however  'soft domain 
walls'~\cite{soft}, \ie domain walls forming at a late time phase transition, 
have been studied. 
\clearpage
\section{Theoretical Framework}
\label{The}
\subsection{Linear cosmological perturbations with seeds}
A basic tool for cosmic structure formation is linear cosmological
 perturbation theory. The fact that CMB anisotropies are small shows
that at least initially also perturbations in the matter density have been
much smaller than unity and therefore they may be treated within linear
perturbation theory. It is generally assumed (an assumption which is
supported by several observational facts, see \eg~\cite{vrr}) that
perturbations are still linear on scales above about $10h^{-1}$Mpc.
On smaller scales non-linear N-body simulations are needed to compute 
the evolution of density fluctuations.

The principal difference in perturbation theory 
in models with topological defects as compared to the more familiar 
inflationary models, is the fact that here cosmic perturbation equations are 
not homogeneous. The perturbations are induced by 'seeds' which are not
present in the background energy momentum tensor.. 
The defect energy momentum tensor enters in the
perturbation equation as 'source' or 'seed' term on the right hand
side, but the defects themselves evolve according to the background
space-time. Perturbations in the defect evolution are of second order. 
(This procedure has sometimes also been termed the 'stiff
approximation'~\cite{as}, but it is actually nothing else than
consistent linear perturbation theory.).

 Gauge-invariant perturbation equations for cosmological models
with seeds have been derived in Refs.~\cite{D90,Review}. 
Here we follow the notation and use the results presented in 
Ref.~\cite{Review}. Definitions of all the gauge-invariant perturbation
variables used in this Review in terms of perturbations of the metric, the
energy momentum tensor and the photon and neutrino brightness are given 
in Appendix~A for completeness. 

We consider a background universe with density parameter 
$\Om_0=\Om_m+\Om_\La=1$,
consisting of photons, cold dark matter (CDM),  baryons and
neutrinos. At very early times $z\gg z_{dec}\sim 1100$, photons and 
baryons form a perfectly coupled ideal fluid.
 As time evolves, and as the electron density drops due to recombination
of primordial helium and hydrogen, Compton
scattering becomes less frequent and higher moments in the photon
distribution develop. This process has to be described by a Boltzmann
equation. Long after recombination, free electrons are so sparse that
the collision term can be neglected, and photons evolve according
to the collisionless Boltzmann or Liouville equation. During the epoch of
interest here, neutrinos are always collisionless and thus obey the
Liouville equation. 

In the next section, we parameterize in a completely general
way the degrees of freedom of the seed energy momentum
tensor. Section~\ref{einst} is devoted to the perturbation of Einstein's
equations and the fluid equations of motion. Next we treat the evolution
of CMB photons by the Boltzmann perturbation equation, including polarization.
The detailed derivations as well as the expressions for the CMB anisotropy
and polarization power spectra are given in Appendix~\ref{AppBoltz}. We 
finally explain how to compute the power spectra of density fluctuations,
CMB anisotropies and peculiar velocities by means of the derived perturbation
equations and the unequal time correlators of the seed energy momentum
tensor which are obtained by numerical simulations.

\subsection{The seed energy momentum tensor}

Since the energy momentum tensor of the seeds, $\Th_{\mu\nu}$, does not
contribute to the background Friedman universe, it is gauge invariant by itself
according to the Stewart-Walker Lemma~\cite{StW}.

$\Th_{\mu\nu}$  can be calculated by solving the matter equations for
the seeds in the Friedman {\em background} geometry. Since $\Th_{\mu\nu}$
has no background component it satisfies the unperturbed
``conservation'' equations. We decompose
$\Th_{\mu\nu}$ into scalar, vector and tensor contributions.
They decouple within linear perturbation theory and it is thus  possible
to write  the equations for each of these contributions separately.
As always (unless noted otherwise), we work in Fourier space, $\bk$ is 
the comoving wave number and $k=|\bk|$.
We parameterize the scalar $(S)$ vector $(V)$ and tensor
$(T)$ contributions to $\Th_{\mu\nu}$ in the form
\bea  \Th_{00}^{(S)} &=& M^2f_{\rho}
     \label{3seed00} \\
     \Th_{j0}^{(S)}  &=& iM^2k_jf_v 
     \label{3seed0j} \\
   \Th_{jl}^{(S)}   
    &=& M^2\left[(f_p + \frac{1}{3}k^2 f_{\pi})\de_{jl} - k_jk_lf_\pi\right]
  \label{3seedjl}\\
     \Th_{j0}^{(V)}  &=& M^2w^{(v)}_j       \\
   \Th_{jl}^{(V)}    &=&iM^2\frac{1}{2}\l(k_jw^{(\pi)}_l+k_lw^{(\pi)}_j\r) \\
   \Th_{jl}^{(T)}  &=& M^2\tau^{(\pi)}_{ij}  ~. \label{Tseedjl}
\eea
Here $M$ denotes a
typical mass scale of the seeds. In the case of topological defects we
set $M=\eta$, where $\eta$ is the symmetry breaking scale~\cite{Review}.
The vectors $\bm w^{(v)}$ and $\bm w^{(\pi)}$ are transverse and
$\tau^{(\pi)}_{ij}$ is a transverse traceless tensor,
\[ \bk\cd\bm w^{(v)} = \bk\cd\bm w^{(\pi)}=
	k^i\tau^{(\pi)}_{ij}=\tau^{(\pi)\;j}_j = 0 ~.\]

From the full energy momentum tensor $\Th_{\mu\nu}$ which 
contains scalar, vector and tensor contributions, the scalar parts
$f_v$ and $f_{\pi}$ of a Fourier mode are given by 
\be ik^j\Th_{0j} = -k^2M^2f_v  ~, \label{3fv}\ee
\be-k^ik^j(\Th_{ij} - \frac{1}{3}\de_{ij}\de^{kl}\Th_{kl}) =
   \frac{2}{3}k^4M^2f_{\pi}   \; . \label{3fpi} \ee
On the other hand $f_v$ and $f_{\pi}$ are also
determined in terms of $f_{\rho}=\Th_{00}/M^2$ and $f_p=\Th_{ii}/(3M^2)$ 
by energy and momentum conservation,
\be \dot{f}_{\rho} +k^2f_v + \frac{\dot{a}}{ a}(f_{\rho} +3f_p) = 0 
\label{ec}~, \ee
\be \dot{f}_v + 2 \frac{\dot{a}}{ a}f_v - f_p +\frac{2}{3}k^2f_{\pi} = 0 ~.
 \label{3f}  \ee
Once $f_v$ is known it is  easy to extract
\be M^2w^{(v)}_j =  \Th_{0j} - ik_jM^2f_v~. \label{2wv}\ee 
For $w^{(\pi)}_i$ we use
\be ik^j(\Th_{lj}- \Th_{lj}^{(S)}) = -{k^2M^2\over 2}w^{(\pi)}_l~.
   \label{3wpi}
\ee
Again,  $w^{(\pi)}_l$ can also be obtained in terms of  $w^{(v)}_l$ by
means of momentum conservation,
\be \dot{w}^{(v)}_l +2\frac{\dot{a}}{ a}w^{(v)}_l +
	\frac{1}{2}k^2w^{(\pi)}_l = 0 ~.\ee
Finally,
\bea 
  M^2\tau^{(\pi)}_{ij} &=&\Th_{ij} -M^2\left[(f_p+{k^2\over 3}f_\pi)\de_{ij}
 -k_ik_jf_\pi \right. \nonumber \\
 && \left. -{i\over 2}(k_iw^{(\pi)}_j + k_jw^{(\pi)}_i) \right]~. \label{3tau}
\eea

The geometry perturbations  induced by the seeds are
characterized by the Bardeen potentials, $\Phi_s$ and $\Psi_s$, for scalar 
perturbations, by the potential for the shear of the extrinsic curvature, 
$\bm\Si^{(s)}$, for vector perturbations and by the gravitational wave
amplitude, $H_{ij}^{(s)}$, for tensor perturbations. Detailed
definitions of these variables and their geometrical interpretation
are given in Ref.~\cite{Review} (see also Appendix~A). Einstein's
equations for an unperturbed cosmic background fluid with seeds relate 
the seed perturbations of the geometry to the energy momentum tensor of the 
seeds. Defining the dimensionless small parameter 
\be
\ep\equiv 4\pi GM^2~, 
\ee
we obtain to first order in $\ep$
\bea
k^2\Phi_s &=&\ep(f_\rho+3\frac{\dot{a}}{ a}f_v)  \label{Phis}\\
 \Phi_s +\Psi_s &=& -2\ep f_{\pi}  \label{Psis}  \\
 -k^2\Si^{(s)}_i &=& 4\ep w^{(v)}_i  \label{Sis}\\
\ddot{H}^{(s)}_{ij} +2 \frac{\dot{a}}{ a}\dot{H}^{(s)}_{ij} +
  k^2H^{(s)}_{ij} &=&
  2\ep\tau^{(\pi)}_{ij} ~. \label{Hs}
\eea
Eqs.~(\ref{Phis}) to (\ref{Hs}) would determine the geometric
perturbations  if the cosmic fluid were perfectly unperturbed. In a 
realistic situation, however, we have to add the fluid perturbations 
which are defined in the next subsection.
Only the total geometrical perturbations are
determined via Einstein's equations. In this sense, Eqs.~ (\ref{Phis}) 
to (\ref{Hs}) should be regarded as definitions for
$\Phi_s~,\Psi_s~,{\bf\Si}^{(s)}$ and $H^{(s)}_{ij}$.

A description of the numerical calculation of the energy momentum tensor of the
seeds for global defects and cosmic strings is given in Chapter~\ref{Num}.

\subsection{Einstein's equations and the fluid equations \label{einst}}
\subsubsection{Scalar perturbations}
Scalar perturbations of the geometry have two degrees of freedom which
can be cast in terms of the gauge-invariant Bardeen
potentials,  $\Psi$ and $\Phi$~ \cite{Ba,KS}. For Newtonian forms of matter,
$\Psi=-\Phi$ is nothing else than the Newtonian gravitational
potential. For matter with significant anisotropic stresses, $\Psi$ and $-\Phi$
differ. In geometrical terms, the former represents the lapse function
of the zero-shear hyper-surfaces while the latter is a measure of
their 3-curvature~\cite{Review}. In the presence of seeds, the
Bardeen potentials are given by
\bea 
\Psi &=&\Psi_s +\Psi_m~,  \label{dec1}\\
\Phi &=&\Phi_s +\Phi_m ~, \label{dec2}
\eea
where the indices $_{s,m}$ refer to contributions from
 a source (the seed)  and the cosmic fluid respectively.
The seed Bardeen potentials are given in Eqs.~(\ref{Phis}) and (\ref{Psis}).

To describe the scalar perturbations of the energy momentum tensor of a
given matter component, we use the gauge invariant variables  $D_g$
for density fluctuations, corresponding to the 
usual density fluctuation in the 'flat gauge', $V$, for the
potential of peculiar velocity fluctuations, corresponding to the usual 
velocity potential in the longitudinal gauge and $\Pi$, a potential
for anisotropic stresses (which vanishes for CDM and baryons). A
definition of these variables in terms of the components of the 
energy momentum tensor of the fluids and the metric perturbations can
be  found in Refs.~\cite{KS} or~\cite{Review} and in Appendix~A. 

Subscripts and superscripts $_\ga$, 
$_c$, $_b$ or $_\nu$ denote the photons, CDM, baryons and neutrinos 
 respectively.

Einstein's equations yield the following relation for the
matter part of the Bardeen potentials~\cite{DS}
\bea
\Phi_m &=& \frac{4\pi Ga^2}{ k^2}\big[\rho_\ga D _g^{(\ga)}+
\rho_c D _g^{(c)} + \rho_b D _g^{(b)}+
 \rho_\nu D _g^{(\nu)} -\{4\rho_\ga+ 3\rho_c \nonumber\\
 +\!3\rho_c &+&\! 3\rho_b +\! 4\rho_\nu\}\Phi   +\!3\frac{\dot a}{ a}k^{-1}
 \{\frac{4}{3}\rho_\ga V_\ga+\!\rho_c V_c  +\!\rho_bV_b
	+\!\frac{4}{ 3}\rho_\nu V_\nu\}\big] \label{Phm}\\
\Psi_{m} &=&-\Phi_{m} - \frac{8\pi Ga^2}{ k^2}\left(p_\ga\Pi_\ga
+ p_\nu\Pi_\nu\right)~. \label{Psm}
\eea
Note the appearance of $\Phi=\Phi_s+ \Phi_m$ on the r.h.s. of Eq.~(\ref{Phm}).
Using the decompositions (\ref{dec1},\ref{dec2}) we can solve for $\Phi$
and $\Psi$ in terms of the fluid variables and the seeds. With the
help of Friedman's equation, Eqs.~(\ref{Phm}) and~(\ref{Psm}) can
then be written in the form
\bea
\Phi &=& \frac{1}{\frac {2}{ 3}\left(\frac{\dot{a}}{ a}\right)^{-2}k^2
		{\scriptstyle +4x_\ga +3x_c
  +3x_b+4x_\nu}}\big[x_\ga D_g^{(\ga)} + x_cD_g^{(c)}
+xD_g^{(b)}     \nonumber\\ 
 &&  + x_\nu D_g^{(\nu)} +\frac{\dot{a}}{a}k^{-1}\left( 4x_\ga V_\ga
	+3x_cV_c+3x_bV_b   \right.  \nonumber\\
 &&  \left. + 4x_\nu V_\nu\right)
+\frac{2}{ 3}k^2\left(\frac{\dot{a}}{ a}\right)^{-2}\Phi_s \big] \label{Phi}\\
\Psi &=&-\Phi-2\ep f_\pi-\left(\frac{\dot{a}}{ a}\right)^2k^{-2}(x_\ga\Pi_\ga
	+x_\nu\Pi_\nu)~. \label{Psi} 
\eea

Here we have normalized the
scale factor such that $a=1$ today. The density parameters
$\Om_{\scriptstyle\bullet}$ always represent the values of the corresponding
density parameter today (Here $_{\scriptstyle\bullet}$ stands for
$_c~,~_\ga~,~_b$ or $_\nu$.). To avoid
any confusion, we have introduced the variables $x_{\scriptstyle\bullet}$
for the time dependent density parameters,
\bea
x_{\ga,\nu} &=& \frac{\Om_{\ga,\nu}}{ \Om_\ga + \Om_c a
+ \Om_b a+ \Om_\nu  +\Om_\La a^4} \\
x_{c,b} &=& \frac{\Om_{c,b}a}{ \Om_\ga + \Om_c a
+ \Om_b a + \Om_\nu +\Om_\La a^4}~.
\eea 

The fluid variables for photons and neutrinos are obtained by
integration over directions of the scalar brightness perturbations, which 
we denote by $\MM_S(\tau,\bk,\bn)$ and $\NN_S(\tau,\bk,\bn)$ respectively.
They are given in Appendix~\ref{AppBoltz}.

The evolution of  CDM perturbations is determined by energy and momentum
conservation,
\be
\dot{D}_g^{(c)} +kV_c = 0 ~,~~~~
\dot{V}_c +\left(\frac{\dot{a}}{ a}\right)V_c = k\Psi ~. \label{DVc}
\ee
During the very tight coupling regime, $z\gg z_{dec}$, we may neglect
the baryon contribution in the energy momentum conservation of the
baryon-photon plasma. We then have
\bea
\dot{D}_g^{(\ga)} +\frac{4}{ 3}kV_\ga = 0~,~~~~
\dot{V}_\ga -k\frac{1}{ 4}D_g^{(\ga)} = k(\Psi-\Phi)~,\label{DVga}\\
\mbox{and }~~ D_g^{(b)} = \frac{3}{4}D_g^{(\ga)}~,~~~ 
V_b = V_\ga ~. \label{DVbt}
\eea
The conservation equations for neutrinos are not very useful,
since they involve anisotropic stresses and thus do not close. At the
temperatures of interest to us, $T\ll1$MeV, 
neutrinos have to be evolved by means of the Liouville equation which
we discuss in the next section.

Once the baryon contribution to the baryon-photon fluid becomes
non-negligible, and the imperfect coupling of photons and baryons has
to be taken into account (for a 1\% accuracy of the results, the
redshift corresponding to this epoch is around $z\sim 10^7$), we
evolve also the photons with a Boltzmann equation. The equation of
motion for the baryons is then
\bea
\dot{D}_g^{(b)} +kV_b &=& 0~, \label{Db}\\ 
\dot{V}_b +\left( \frac{\dot{a}}{ a}\right)V_b &=& k\Psi
-\frac{4a\si_Tn_e\Om_\ga }{ 3\Om_b}[V_\ga-V_b]~. \label{Vb}
\eea
The last term in Eq.~(\ref{Vb}) represents the photon drag force
induced by non-relativistic Compton scattering, $\si_T$ is the
Thomson cross section, and $n_e$ denotes the number density of free electrons.
The scale factor $a$ enters since our derivative is taken w.r.t conformal time.
At very early times, when $\si_Tn_ea\gg 1/\tau$, the 'Thomson drag' just
forces $V_b=V_\ga$, which together with Eqs.~(\ref{DVga}) and
(\ref{Db}) implies the first eqn. of (\ref{DVbt}).

An interesting phenomenon often called 'compensation' can be important
on super horizon scales, $k\tau\ll 1$. If we neglect ani\-so\-tro\-pic 
stresses of photons and neutrinos and take into account that 
${\cal O}(V) = {\cal O}(k\tau\Psi)$ and  ${\cal O}(D_g) = {\cal O}(\Psi)$
for $k\tau\ll 1$, Eqs.~(\ref{Phi}) and (\ref{Psi}) lead to
\be
 {\cal O}(\Phi) =  {\cal O}\left((k\tau)^2\Phi_s -2\ep f_\pi\right) 
	~. \label{comp}
\ee
Hence, if anisotropic stresses are relatively small, $\ep f_\pi\ll \Phi_s$,
the resulting gravitational potential on super horizon scales is much
smaller than the
one induced by the seeds alone. One must be very careful not to over
interpret this 'compensation' which is not strictly related to
causality, but is due to the initial condition $D_g~,V\ra_{\ct\ra 0}0$. 
A thorough discussion of this issue is found in
Refs.~\cite{DS,MC,UDT}. As we shall see, for textures 
$\Phi_s$ and $\ep f_\pi$ are actually of the same order. Therefore
Eq.~(\ref{comp}) does not lead to compensation, but it indicates
that CMB anisotropies on very large scales (Sachs-Wolfe effect) are
dominated by the amplitude of seed anisotropic stresses. Nevertheless, for 
purely scalar or coherent perturbations, as we shall see 
$f_\pi \propto (k\ct)^2\Phi_s$ and hence 'compensation' is important.

The quantities which we want to calculate and compare with
observations are the CDM density power spectrum and the peculiar
velocity power spectrum today
\be
 P(k) = \langle|D_g^{(c)}(k,\tau_0)|^2\rangle ~\mbox{ and }~~
P_v(k) = \langle|V_c(k,\tau_0)|^2\rangle ~.
\ee
Here $\langle\cdots\rangle$ denotes an ensemble average over models.
Note that even though $D_g$ and $V$ are gauge invariant quantities
which do not agree with, {\em e.g.}, the corresponding quantities in
synchronous gauge, this difference is very small on sub-horizon scales
(of order $1/k\tau$) and can thus be ignored.

On sub-horizon scales the seeds decay, and CDM perturbations evolve
freely. We then have, like in inflationary models~\cite{Pee},
\be P_v(k) = H_0^2\Om_m^{1.2}P(k)k^{-2} ~.\ee

\subsubsection{Vector perturbations}
 Vector perturbations of the geometry have two degrees of freedom
which can be cast in a divergence free vector field. A gauge-invariant
quantity describing vector perturbations of the geometry is $\bm\Si$, a
vector potential for the shear tensor of  the $\{\ct=$const.$\}$
hypersurfaces. Like for scalar perturbations, we split 
 $\bm\Si$ into a source term  coming from the seeds
given in the previous section, and a
part due to the vector perturbations in the fluid,
\be \bm\Si = \bm\Si_s + \bm\Si_m  \label{Si} ~.\ee
The perturbation of Einstein's equation for
${\bm\Si}_m$ is~\cite{Review}
\be
k^2\bm\Si_m = 6\left(\frac{\dot{a}}{ a}\right)^2[\frac{4}{
	3}x_\ga\bm\om_\ga + x_c\bm\om_c 
 	+ x_b\bm\om_b + \frac{4}{3}x_\nu\bm\om_\nu] \label{Sim}~.
\ee
Here $\bm\om_{\scriptstyle\bullet}$ is the fluid vorticity which
generates the vector type shear of the equal time hyper-surfaces
(see Appendix~A). By definition, vector perturbations are transverse,
\be \bm\Si\cd\bk= \bm\Si_m\cd\bk= \bm\Si_s\cd\bk=
\bm\om_{\scriptstyle\bullet}\cd\bk = 0~. \label{vector}
\ee

It is interesting
to note that vector perturbations in the geometry do not induce any
vector perturbations in the CDM (up to unphysical gauge modes), since
no geometric terms enter the momentum conservation for CDM vorticity,
\[ \dot{\bm\om}_c + \frac{\dot{a}}{ a}\bm\om_c = 0 ~,\]
hence we may simply set $\bm\om_c = 0$.
This is also the case for the tightly coupled baryon radiation
plasma. But as soon as higher moments in the photon distribution build
up, they feel the vector perturbations in the geometry (see next
section) and transfer it onto the baryons
via the photon drag force, 
\be
\dot{\bm\om}_b +\left(\frac{\dot{a}}{ a}\right)\bm\om_b =
\frac{4a\si_Tn_e\Om_\ga }{3\Om_b}[\bm\om_\ga-\bm\om_b]~. \label{wb}
\ee
The photon vorticity is derived via the integral over the vector
type photon brightness perturbation, $\MM_V$, using
\be
\bm\om^\ga={\bf V}^\ga -\bm\Sigma^\ga
\ee
and
\be
{\bf V}^\ga = \frac{1}{4\pi}\int \bn \MM^{(V)} d\Om ~,
          \label{omga}  
\ee
where the integral is over  photon directions, $\bn$ (see 
Appendix~\ref{AppBoltz}).
The vector equations of motion for photons and neutrinos are discussed
in the next section.

\subsubsection{Tensor perturbations}
Metric perturbations also have  two tensorial degrees of freedom,
gravity waves, which
are represented by the two helicity states of a transverse traceless
tensor (see Appendix~A). As before, we split the geometry perturbation
into a part induced by the seeds and a part due to the matter fluids,
\be
 H_{ij} = H^{(s)}_{ij} +H^{(m)}_{ij} \label{Hij} ~.
\ee
The only matter perturbations which generate gravity waves are tensor
type anisotropic stresses which are present in the photon and neutrino
fluids. Numerically one finds that the effect of anisotropic stresses of 
photons and neutrinos contributes less than 1\% to the final 
result~\cite{DKM}, and hence may be neglected  by setting 
$H^{(m)}_{ij}=0$.
 
\subsection{Boltzmann equation, polarization and CMB power spectra}
When particle interactions are less frequent, the fluid
approximation is not  sufficient, and we have to describe the
given particle species by a Boltzmann equation, in order to take into account
phenomena like collisional and directional dispersion. In the case of
massless particles like massless neutrini or photons, the Boltzmann
equation can be integrated over energy, and we obtain an
equation for the brightness perturbation which depends only on
momentum directions~\cite{Review}.  As before, we
split the brightness perturbation into a scalar, vector and
tensor component, and we discuss the perturbation equation of each
of them separately,\footnote{We could in principle add higher 
spin components to the distribution functions. But they are not 
seeded by gravity and
since photons (and neutrinos) interact at high enough temperatures, they
are also absent in the initial conditions.} 
\be
\MM  =\MM_S +\MM_V +\MM_T~.
\ee
The function $\MM$ depends on the wave vector $\bk$, the
photon direction $\bn$ and conformal time $\ct$.
Linear polarization of photons induced by Compton scattering
is described by the Stokes parameter $Q$ and $U$,
depending on the same variables. An explicit derivation of the Boltzmann 
equation 
including polarization is presented in Appendix~\ref{AppBoltz}. Here we just 
repeat the necessary definitions and results.

The brightness anisotropy $\MM$ and the non-vanishing Stokes parameters $Q$ 
and $U$ can be expanded as
\bea
 \MM(\ct,\bk,\bn) &=& \sum_{\ell}\sum_{m=-2}^2\MM_{\ell}^{(m)}(\ct,k)
	{ _0 G^m_{\ell}}(\bn),\\
 Q(\ct,\bk,\bn)\pm iU(\ct,\bk,\bn) &=& \sum_{\ell}\sum_{m=-2}^2
	(E_{\ell}^{(m)} \pm iB_{\ell}^{(m)}){ _2G^m_{\ell}}(\bn).
\eea
The spin weighted spherical harmonics $ _sG^m_{\ell}$ are described in 
Appendix~\ref{AppBoltz}. Up to a normalization constant, the $ _0G^m_{\ell}$ 
coincide with the usual spherical harmonics. The coefficients $m=0,m=\pm1$ 
and $m=\pm2$ describe 
the scalar $(S)$, vector $(V)$ and tensor $(T)$ components respectively. 
The Boltzmann equation for the coefficients $X^{(m)}_{\ell}$ is given by
\bea
\lefteqn{ \dot\MM_{\ell}^{(m)} -k\l[{_0\ka^m_{\ell}\over 2\ell-1}
	\MM_{\ell-1}^{(m)}
-{_0\ka^m_{\ell+1}\over 2\ell+3}\MM_{\ell+1}^{(m)}\r] = }\nonumber \\
  &&	-n_e\si_Ta\MM_{\ell}^{(m)}  +S_{\ell}^{(m)} ~~~ (\ell\ge m)\\
\lefteqn{ \dot E_{\ell}^{(m)} -k\l[{_2\ka^m_{\ell}\over 2\ell-1}
E_{\ell-1}^{(m)}  -{2m\over \ell(\ell+1)}B_{\ell}^{(m)}
-{_2\ka^m_{\ell+1}\over 2\ell+3}E_{\ell+1}^{(m)}\r] = } \nonumber \\ && 
  -n_e\si_Ta[E_{\ell}^{(m)} + \sqrt{6}C^{(m)}\de_{\ell,2} \\
\lefteqn{ \dot B_{\ell}^{(m)} - k\l[{_2\ka^m_{\ell}\over 2\ell-1}
  B_{\ell-1}^{(m)}   +{2m\over \ell(\ell+1)}E_{\ell}^{(m)}
-{_2\ka^m_{\ell+1}\over 2\ell+3}B_{\ell+1}^{(m)}\r] = } \nonumber \\ &&  
  -n_e\si_TaB_{\ell}^{(m)}~.
\eea
where we  set
\be \begin{array}{cc}
 S_0^{(0)} =n_e\si_Ta\MM^{(0)}_0, & S^{(0)}_1 =n_e\si_Ta4V_b +4k(\Psi-\Phi), \\
 S^{(0)}_2 =n_e\si_TaC^{(0)}, & S^{(1)}_1 =n_e\si_Ta4\om_b,  \\
  S^{(1)}_2 =n_e\si_TaC^{(1)} +4\Si, &   S^{(2)}_2 =n_e\si_TaC^{(2)} +4\dot H 
\end{array}
\ee
and $C^{(m)}= {1\over 10}[\MM^{(m)}_2 -\sqrt{6}E^{(m)}_2] $.
The coupling coefficients are
\[ 
	_s\ka^m_{\ell} =\sqrt{{(\ell^2-m^2)(\ell^2-s^2)\over \ell^2}}.
\]
In Appendix~\ref{AppBoltz}) we express the fluid variables in terms of 
integrals of the photon brightness over directions.

The CMB temperature and polarization power spectra
are given in terms of the expansion coefficients $\MM_{\ell}^{(m)}$,
 $E_{\ell}^{(m)}$ and $B_{\ell}^{(m)}$ as
\be
 (2\ell+1)^2C_\ell^{XY(m)} = {n_m\over 8\pi}\int k^2dkX_{\ell}^{(m)}
	Y_{\ell}^{(m)*}~,  \label{PSpecC2}
\ee
where $n_m=1$ for $m=0$ and $n_m=2$ for $m=1,2$, accounting for the number 
of modes. Since $B$ is parity odd, the only non-vanishing cross correlation 
spectrum is $C^{TE}$.

\subsection{Neutrinos}

Analogously to the photon brightness perturbation it is useful
to introduce the neutrino one as well, which we will call
$\NN$,
\be
\NN  =\NN^{(S)} +\NN^{(V)} +\NN^{(T)}.
\ee
Since the neutrinos are collisionless during the entire epoch under 
consideration, they satisfy the collisionless equations which are 
obtained from the Boltzmann equation for the intensity by setting $\si_T=0$.
For simplicity, we expand $\NN$ not in terms of the functions
$_0G^{(m)}_{\ell}$, but use the more basic approach with Legendre polynomials.
\bea
\NN^{(S)} &=& \sum_\ell (-i)^\ell (2 \ell+1) \nu_{\ell}^{(S)} P_\ell(\mu)  \\
\NN^{(V)} &=& \sqrt{1\!-\!\mu^2}\l[ \NN_1^{(V)}(\mu) \cos\phi 
	\!+\! \NN_2^{(V)}(\mu) \sin\phi\r] \label{vnu1}\\
 \NN_{1,2}^{(V)} &=& \sum_\ell (-i)^\ell (2 \ell+1) \nu_{\ell(1,2)}^{(V)} 
  P_\ell(\mu)\\
\NN^{(T)} &=& (1-\mu^2) \l[ \NN_+^{(T)} \cos(2\phi) + 
		\NN_\times^{(T)} \sin(2\phi)\r] \label{tnu1}\\
 \NN_{+,\times}^{(T)} &=& \sum_\ell (-i)^\ell (2 \ell+1) 
	\nu_{\ell(+,\times)}^{(T)} P_\ell(\mu)~.
\eea
The Liouville equation for the coefficients $\nu_{\ell\bullet}^{(S,V,T)}$
then becomes
\bea
\dot{\nu}^{(S)}_\ell -  {k \over 2 \ell + 1}
	\left[\ell\nu^{(S)}_{\ell- 1} -
   (\ell+ 1)\nu^{(S)}_{\ell + 1}\right]
&=& {4 \over 3} k (\Psi-\Phi)\de_{\ell,1}\\
\dot{\bm\nu}^{(V)}_\ell  -\frac{k }{2 \ell +1}\l[  \ell\bm \nu_{\ell-1}^{(V)}
 -(\ell+1) \bm\nu_{\ell+1}^{(V)} \r]
  &=& 4 k \de_{\ell,1} \bm\Si\\
\dot{\nu}^{(T)}_{\ell,\epsilon} -{k \over 2\ell +1}
	\left[\ell\nu^{(T)}_{\ell-1,\epsilon} -
   (\ell+1)\nu^{(T)}_{\ell+1,\epsilon}\right]
&=& 4{\dot H}_{\epsilon}~.
\eea

\subsection{Computing power spectra in seed models}

The generation of the seeds,\eg~ topological defects during a symmetry 
breaking phase transition,
is an inherently random process. The exact  seed distribution in our 
universe is just one realization and cannot be predicted. 
Only statistical properties, expectation values, can be calculated. 
Yet the source functions to the Boltzmann equation are elements of the seed 
energy momentum tensor, not their expectation values. 

In principle one could
calculate the induced random variables $D_g^{(c)}(\bk,\ct_0)$,
$V_c(\bk,\ct_0)$, $\MM^{(m)}_{\ell}(k,\ct_0)$ etc  for 100 to 
1000 realizations of a given
model and determine the expectation values  $P(k)$, $P_v(k)$ and $C_\ell$ 
by averaging. This procedure has been adapted in Ref.~\cite{ABR} for a
seed energy momentum tensor modeled by a few random parameters and in 
Ref.~\cite{ZD} where the CMB anisotropies on large scales have been 
determined in $\bx$-space by direct line of sight integration and averaging
over several observer positions.

In a more realistic calculation of topological defects, where the seed 
energy momentum tensor comes entirely from
numerical simulations, this procedure is  not feasible. The first and most
important bottleneck is the dynamical range of the simulation which
is about $40$ in the largest $(512)^3$ simulation which have been
performed~\cite{PST,DKM}. They need about 1 to 2 Gbyte of RAM and run
in about one hour CPU time on a modern work station or PC.
To determine the $C_\ell$'s for $2\le\ell\le 1000$ we need a dynamical
range of about 10,000 in $k$-space. This means $k_{\max}/k_{\min} \sim
10'000$, where $k_{\max}$ and $k_{\min}$ are the maximum and minimum wave
numbers which contribute to the $C_\ell$'s to achieve and accuracy of
about 10\%. A dynamical range of 10,000 requires at least a $(100,000)^3$
simulations which needs about 10,000 Terabytes  RAM! Correspondingly the 
CPU time required for such a simulation is about 1000 years.

With brute force, this problem is thus not tractable with present or
near future computing capabilities. But there are a series of theoretical
observations which reduce the problem to a feasible one:

As we have seen, for each wave vector $\bk$ given, we have to solve a system of 
linear perturbation equations with random sources,
\be
\DD X = \Scal~. \label{lineq}
\ee
Here $\DD$ is a time dependent linear differential operator, 
$X$ is the vector of the
 matter perturbation variables specified in the previous
subsections (photons, CDM, baryons and neutrini;
total length up to $4000$), and $\Scal$ is the random source term, 
consisting of linear combinations of the seed energy momentum tensor.

For given initial conditions, this equation can be solved by
means of a Green's function (kernel), $\GG(\ct,\ct')$, in the form
\be
X_j(\ct_0,\bk) =\int_{\ct_{in}}^{\ct_0}\! d\ct\GG_{jm}(\ct_0,\ct,\bk)
	\Scal_m(\ct,\bk)~.   \label{Gsol}
\ee
We want to compute power spectra or, more generally, quadratic 
expectation
values of the form
\[  \langle X_j(\ct_0,\bk)X_m^*(\ct_0,\bk')\rangle ~,\]
which, according to Eq.~(\ref{Gsol}) are given by
\bea
\lefteqn{\langle X_j(\ct_0,\bk)X_l^*(\ct_0,\bk')\rangle =} \nonumber \\
&& \int_{\ct_{in}}^{\ct_0}\! d\ct\GG_{jm}(\ct,\bk)
  \int_{\ct_{in}}^{\ct_0} \! d\ct'\GG^*_{ln}(\ct',\bk')\times
\langle\Scal_m(\ct,\bk)\Scal_n^*(\ct',\bk')\rangle~. \label{power}
\eea
The only information about the source random variable which we really
need in order to compute power spectra are therefore the unequal time
 two point correlators
\be
\langle\Scal_m(\ct,\bk)\Scal_n^*(\ct',\bk')\rangle~. \label{2point}
\ee
This nearly trivial fact has been introduced by Hindmarsh~\cite{Hind} and
exploited by many workers in the field. For example in Ref.~\cite{ACFM},
 where decoherence of models with seeds has been discovered, and later in
Refs.~\cite{PST,Aetal,KD,DS,DKM} and others. The eigenvector method 
discussed here has been introduced in~\cite{Tu}. (The CMB anisotropy spectrum
from cosmic texture shown in this paper is, however, incorrect.)

To solve the enormous problem of dynamical range, one then uses
 causality, statistical isotropy and 'scaling'. 

Seeds are called 'scaling' if their correlation functions 
$C_{\mu\nu\rho\la}$ defined by
\bea 
\Th_{\mu\nu}(\bk,\ct) &=& M^2\tha_{\mu\nu}(\bk,\ct) ~, \\
\langle\tha_{\mu\nu}(\bk,\ct) \tha_{\rho\la}^*(\bk',\ct')\rangle  
	&=& C_{\mu\nu\rho\la}(\bk,\ct,\ct')\de(\bk-\bk')
\label{cor}
\eea
are scale free; {\em i.e.} the only  dimensional parameters in
$C_{\mu\nu\rho\la}$ are the variables  $\ct,\ct'$ and $\bk$ themselves. 
The $\de$-function in $\bk$-space is a simple consequence of statistical 
homogeneity.
Up to a certain number of dimensionless functions $F_n$ of 
$z=k\sqrt{\ct\ct'}$ and $r=\ct/\ct'$, the correlation functions are 
then determined by the requirement of statistical isotropy, symmetries 
and by their dimension. Causality requires the 
functions $F_n$ to be analytic in $z^2$. A more detailed investigation
of these arguments and their consequences is given in Chapter~\ref{Gen}.
There we also show that statistical isotropy and energy
momentum conservation reduce the correlators~(\ref{cor}) to five
such functions $F_1$ to $F_5$. 

In cosmic string simulations, energy and momentum are not
conserved. Cosmic string loops oscillate and emit gravitational waves (see 
Refs. \cite{VV,D89}). They lose their energy by radiation of gravitational
waves and, in 'cusps' or tiny wiggles into massive particles~\cite{VS}. 
In this case 14 functions of $z^2$ and
$r$ are needed to describe the unequal time correlators~\cite{CHM}.

Since analytic functions generically
are  constant for small arguments $z^2\ll 1$, $F_n(0,r)$ actually
determines $F_n$ for all values of $k$ with $z=k\sqrt{\ct\ct'}\lsim
0.5$. Furthermore, the correlation functions decay inside the horizon
and we can safely set them to zero for $z\gsim 40$ where they have
decayed by about two orders of magnitude (see Figs.~\ref{fig1} 
to~\ref{figtensor} in Chapter~\ref{Res}). 
Making use of these generic
properties of the correlators, we have reduced the dynamical range
needed for our computation to about 40, which can be attained with
$(512)^3$ simulations feasible on present computers.

Clearly, all correlations between scalar and vector, scalar and tensor
as well as vector and tensor perturbations have to vanish.

The source correlation matrix $C_{\mu\nu\rho\si}(\bk,\ct,\ct')$ can be
considered as  kernel of a positive hermitian operator
in the variables $x=k\ct=zr^{1/2}$ and $x'=k\ct'=z/r^{1/2}$, which 
can be diagonalized:
\be
 C(x,x') = \sum_n \la_n v_{n}(x)v_{n}^{*}(x')~
	\label{EVS}
\ee  
(the variable $\bk$ and the space time indices are supressed in this 
and the following expressions).
 The series $\left(v_{n}\right)$ is an orthonormal series of 
eigenvectors of the operator $C$ (ordered according to the amplitude
of the corresponding eigenvalue)  for a given weight function $w$.
We then have\footnote{Here the assumption that the operator $C$
is trace-class enters. This hypothesis is verified
numerically by the fast convergence of the sum~(\ref{EVS}).}
\be
 \int  C(x,x')v^{(n)}(x') w(x')dx' = \la_n
	v^{(n)}(x)~.
\ee 
The  eigenvectors  and  eigenvalues depend on the weight
function $w$ which can be chosen to optimize the convergence speed
of the sum~(\ref{EVS}). For $O(N)$ models,
scalar perturbations typically need 20 eigenvectors whereas vector
and tensor perturbations need five to ten eigenvectors for an
accuracy of a few percent (see Fig.~\ref{fig13} in Chapter~\ref{Res}).

Inserting the expansion (\ref{EVS}) in Eq.~(\ref{power}), leads to
\be
\langle X_i(\bk,\ct_0)X_j^*(\bk,\ct_0)\rangle = \sum_n\la_n X_i^{(n)}(k\ct_0)
	X_j^{(n)*}(k\ct_0) \label{EVpower} ~,
\ee
where $ X_i^{(n)}(\ct_0)$ is the solution of Eq.~(\ref{lineq}) with
deterministic source term $v_i^{(n)}$,
\be
X_j^{(n)}(\ct_0,\bk) =\int_{\ct_{in}}^{\ct_0}d\ct\GG(\ct_0,\ct,\bk)_{jl}
	v_l^{(n)}(x,\bk)~.
\label{EVsol}
\ee

For the CMB anisotropy spectrum this gives
\be
 C_\ell = \sum_n^{n_S}\la_n^{(S)}C_\ell^{(Sn)} +  
	\sum_n^{n_V}\la_n^{(V)}C_\ell^{(Vn)}
  +  	\sum_n^{n_T}\la_n^{(T)}C_\ell^{(Tn)} ~.
\ee
$C_\ell^{(\bullet n)}$ is the CMB anisotropy
induced by the deterministic source $v^{(\bullet n)}$, and $n_{\bullet}$ is the
number of eigenvalues which have to be considered to achieve good accuracy.
Here we have also used that the unequal time correlation matrix contains 
uncorrelated scalar, vector and tensor blocks.

Instead of averaging over random solutions of Eq.~(\ref{Gsol}), we can
thus integrate Eq.~(\ref{Gsol}) with the deterministic source term
$v^{(n)}$ and sum up the resulting power spectra. The computational
requirement for the determination of the power spectra of one seed 
model with given source term is thus on the order of
$n_S+n_V+n_T$ inflationary models.
This eigenvector method has first been applied in Ref.~\cite{PST}.

This completes the formal developments needed to compute structure 
formation with defects. In the next chapter we discuss the numerical 
simulations which have been performed to obtain the unequal time correlators 
of the seed energy momentum tensor.
These are then diagonalized and the eigenfunctions are entered as sources
in the system of linear equations derived in this chapter. The results from
this procedure are described in~Chapter~\ref{Res}.
\clearpage
\section{Numerical Implementation}
\label{Num}
In the previous chapter we have learned that the only input needed for
the computation of power spectra and other two point correlation
functions are the unequal time correlators of the defect energy
momentum tensor. In this chapter we discuss how they are obtained
in practice. Their general structure will be analysed in Chapter~\ref{Gen}.

\subsection{Global defects}

\subsubsection{The $\si$ model approximation}

We consider a spontaneously broken scalar field with O($N$) symmetry. 
If we are not interested in the microscopical structure of the
field in the vicinity of the core but only in its behaviour
on large scales, we can force the field to stay on the vacuum
manifold with a Lagrange multiplier $\lambda$ and drop the potential. 
The bulk part of the energy of global strings, global monopoles and
global texture is contained in the field gradient 
at large distances of the core and is not affected by this approximation.

\be
   \LL = \dd_\mu\phi\cd\dd^\mu\phi +\lambda(\phi^2-\eta^2)~.
\ee
Varying the action with respect to $\phi$ and $\lambda$ leads to 
\be
  \Box\phi +\lambda\phi = 0~,~~~~~ \phi^2-\eta^2 =0~.
\ee
Multiplying the first equation with $\phi$ and using $\phi^2=\eta^2 $ 
yields
\be
\lambda = (\phi\cd\Box\phi)/\eta^2 ~\mbox{ hence }~~~
\Box\phi -(\phi\cd\Box\phi)\phi/\eta^2 = 0~.
\ee
Applying $\dd_\mu\dd^\mu$ on $\phi^2-\eta^2$ we find 
$(\phi\cd\Box\phi)=-(\dd_\mu\phi\cd\dd^\mu\phi)$
so that
\[\Box\phi + (\dd_\mu\phi\cd\dd^\mu\phi)\phi/\eta^2 =0 ~. \]
Setting $\beta=\phi/\eta$, we finally obtain the equation of motion
\be
\Box\beta + (\dd_\mu\beta\cd\dd^\mu\beta)\beta =0 \label{simo}
\ee  
for the field $\beta\in {\bf S}^n$, where $n=1,2$ and $3$ for global strings, 
monopoles and texture respectively. Eq.~(\ref{simo}) is
the equation of motion for the non-linear $\sigma$-model for a scalar field
on $\Sset^n$.

\subsubsection{The energy momentum tensor of the seeds}

The energy momentum tensor is given by the variation of the
action with respect to the metric. It is
\be
T_{\mu\nu} = \dd_\mu\b\,\dd_\nu\b
	-\frac{1}{2}g_{\mu\nu}\l(\dd_\la\b\cdot\dd^\la\b\r).
\ee

The required source functions can be directly derived from
this expression using (\ref{3seed00}) to (\ref{Tseedjl}). 
For the {\bf scalar}
sources we use 
\bea
\Phi_s &=& \frac{1}{k^2}\l(f_\rho+3\frac{\dot{a}}{a}f_v\r)~~~~ 
\Psi_s = -\Phi_s-2 f_\pi ~~\mbox{ with}\\
f_\rho &=& \frac{1}{2}\l(\widehat{\dot{\b}^2}+\widehat{(\nabla\b)^2}\r)\\
f_v    &=& - \frac{i k^j}{k^2}\widehat{\l(\dot{\b}\cdot\b_{,j}\r)} \\
f_\pi  &=& -\frac{3}{2}\frac{k^i k^j}{k^4}\l(\widehat{\b_{,i}\cdot\b_{,j}}
		-\frac{1}{3}\de_{ij}\widehat{(\nabla\b)^2}\r),
\eea
where~ $\widehat{ }$ ~denotes the Fourier transform (note that 
$\hat{\b}\hat{\b}\neq \widehat{\b^2}$ !!).

{\bf Vector} sources are determined by
\be
 w_j^{(v)} = \widehat{\dot{\b}\cdot\b_{,j}}-\frac{k_j}{k^2} k^l 
	\widehat{\dot{\b}\cdot\b_{,l}}\quad,
\ee
and it is sufficient to calculate the correlation function
of one of them, \eg~$w_1$, as the transversal character of $\bm w$  imposes
\[ \lan w_i^{(v)}(\bk,\ct)w_j^{(v)}(\bk,\ct')\ran 
= (k_i k_j-k^2 \de_{ij})\sqrt{\ct\ct'}W(k\ct,k\ct').\]
For the same symmetry reasons the {\bf tensor} type
correlators are also determined by one function, $F_5$, alone, see the
equation (\ref{Ctau}) in Chapter~\ref{Gen}.
Therefore we can again pick one special index selection to determine 
$F_5$. Since $F_5$ does not depend on the direction of $\bk$, we can choose
the special coordinate system $k_1=k_2=0$, leading to
$F_5=\sqrt{\ct\ct'}\lan T_{12}(\ct) T_{12}(\ct')\ran$. 

\subsubsection{The large $N$ limit}

Before discussing numerical simulations of global defect, we study the 
limit where the number of components of the scalar field becomes very 
large~\cite{Bray}..
As we shall see, in this limit, the single non-linear term in the $\si$-model
can be replaced by its average, and the equation of motion becomes linear 
and can be solved. This solution has been found in Ref.~\cite{TSlN}.
Using the equation of motion~(\ref{simo}) in a FLRW metric,
$\Box\b = 1/a^2(\ddot{\b}-2 \dot{a}/a \dot{\b}-\De\b)$, we find 
\be
\ddot{\b}-2 \frac{\dot{a}}{a} \dot{\b}-\De\b
= T^\mu_\mu \b . \label{eqmo2}
\ee
In a first step we impose scaling on the trace of the energy
momentum tensor by setting $T^\mu_\mu = T(\bx)/(a\ct)^2$ where
$T(\bx)$ is now dimensionless. In the large-$N$ approximation,
the fluctuations of quadratic quantities, like the energy momentum
tensor, are of order $1/N$, so we neglect them for the field
evolution, and we replace $T(\bx)$ with its average $\bar{T}$.
To solve the resulting equation, we change into Fourier space and
replace $\dot{a}/a$ by $\al/\ct$, which is exact in perfectly matter
($\al=2$) and radiation ($\al=1$) dominated universes, and an acceptable 
approximation otherwise. This leads to
\be
\ddot{\b}+2 \frac{\al}{\ct}\dot{\b}+\l(k^2 -\frac{\bar{T}}{\ct^2}\r)\b=0 .
\ee

This equation is solved by
\be
\b(\bk,\ct) = \ct^{1/2-\al}\Big(f_1(\bk)\,J_\nu(k\ct)+
	f_2(\bk)\,J_{-\nu}(k\ct)\Big),
\ee
where $J_\nu$ denotes the Bessel function of order $\nu$ and 
$\nu=\bar{T}+(1/2-\al)^2$.

The functions $f_i(\bk)$ are random variables, and we can take them
to be Gaussian distributed and uncorrelated at all points,
\be
\l\langle f_i^l(\bk) f_j^{m*}(\bq) \r\rangle = 
C |\bk|^n \de(\bk-\bq) \de_{ij} \frac{\de^{lm}}{N} ~. \label{distr}
\ee
We discard the solution with negative $\nu$, since it diverges for
$\ct\rightarrow 0$. Furthermore, we choose the solution starting
as white noise as function of $\bk$, leading to $n=-2\nu$. Enforcing
the $\si$ model condition that the field cannot leave the vacuum
manifold ($\l\langle \b(\bx,\ct)^2 \r\rangle = 1$) fixes
$C$ and requires $\nu=\al+1$. The absolute normalisation is 
actually not important for the
calculation of CMB anisotropies, since the fluctuations will be
normalised to the COBE data points. In this case $C$ merely
determines the required energy scale of symmetry breaking.

Using $\chi(x)\equiv J_\nu(x)/x^\nu$ as well as
$\vph(x) \equiv {3\over 2} \chi(x)-J_{\nu+1}(x)/x^{\nu-1}$ we
can write the solution as
\be
\b(\bk,\ct) = \sqrt{A} \ct^{3/2} \chi(k\ct)
	\b_{\rm in}(\bk) \label{lNsol}
\ee
and its time derivative as
\be
\dot{\b}(\bk,\ct) = \sqrt{A} \ct^{1/2} \vph(k\ct) \b_{\rm in}(\bk)~.
\ee
It is now easy to derive the energy momentum tensor of the seeds
using the expressions of the last section. As a worked out example,
we take a closer look at $f_\rho$. The 
equal time correlator (ETC) is
\bea
\lefteqn{\l\lan f_\rho(\bk,\ct) f_\rho^*(\bk',\ct) \r\ran} \nonumber \\
 &=& \frac{A^2\ct^2}{4}\int d^3\!q d^3\!p\, \Big\{\vph(q\ct)\vph(|\bk-\bq|\ct)
	\vph(p\ct)\vph(|-\bk'-\bp|\ct) \nonumber \\
&& -\ct^4 \bq (\bk-\bq)  \bp (\bk'+\bp) \chi(q\ct)\chi(|\bk-\bq|\ct)
	\chi(p\ct)\chi(|\bk'+\bp|\ct) \nonumber \\
&& +\ct^2 \bp (\bk'+\bp) \vph(q\ct)\vph(|\bk-\bq|\ct) 
	\chi(p\ct)\chi(|\bk'+\bp|\ct) \nonumber \\
&& - \ct^2 \bq (\bk-\bq) \vph(p\ct)\vph(|\bk'+\bp|\ct)
	\chi(q\ct)\chi(|\bk-\bq|\ct)\Big\} \nonumber \\
&&	\l\lan\b_{\rm in}(\bq) \b_{\rm in}(\bk-\bq)\b_{\rm in}(\bp) 
	\b_{\rm in}(-\bk'-\bp)\r\ran .
\eea

The expectation value of the initial fields is given by
Eq.~(\ref{distr}) and the requirement that $\b_{\rm in}$ be a
Gaussian random variable implies 
\bea
\lefteqn{ \l\lan\b_{\rm in}(\bq) \b_{\rm in}(\bk-\bq)\b_{\rm in}(\bp) 
	\b_{\rm in}(-\bk'-\bp)\r\ran} \nonumber \\
&=&\frac{C^2 \de(\bk-\bk')}{A^2 N} \l[\de(\bp+\bq)+\de(\bp-(\bq-\bk))\r] .
\eea
This allows us to perform the integral over $d^3p$.
We introduce the dimensionless variables $\bx\equiv \bq\ct$ and 
$\by \equiv \bk\ct$. To simplify the notation,
we replace all occurrences of an expression like $a(x) b(|\by-\bx|)$
by $(ab)$.
\bea
\l\lan|f_\rho^2|\r\ran (\by,\ct) 
 &=& \frac{C^2}{2N \ct} \int d^3\!x\, \l\{(\vph\vph)^2
+ [\bx(\by\!-\!\bx)]^2(\chi\chi)^2 \nonumber \r. \\
&& \l. -2 [\bx(\by\!-\!\bx)] (\vph\vph)(\chi\chi)\r\}, \nonumber\\
&=& \frac{\pi C^2}{N \ct} \int dx\,d\mu\,x^2 \l\{(\vph\vph)^2
+ [xy\mu\!-\!x^2]^2 (\chi\chi)^2- \r.\nonumber \\
&&\l. 2 [xy\mu - x^2] 
	(\vph\vph)(\chi\chi)\r\}.
\eea
In the last equation we performed the integration over one
angular variable and introduced $\mu=(\hat{\bx}\cdot\hat{\by})$.

In this way all required unequal time correlators can be
derived. They are shown in Figs.~\ref{fig1} to \ref{figtensor}. A more 
explicit
treatment can be found e.g. in \cite{martin}.

\subsubsection{Numerical simulation of global texture}

Fields with a finite number of components $N$ cannot be treated
analytically. The unequal time correlators have to be calculated numerically
on a grid. A useful approach for global field simulation is to 
minimize the discretized action~\cite{PST2}. There one does
not solve the equation of motion directly, but use a discretized
version of the action
\be
S=\int d^4\!x \,a^2(\ct) \l[\frac{1}{2} \dd_\mu \b \cd \dd^\mu \b
	+\frac{\la}{2} \l(\b^2-1\r)\r]\quad,
\ee
where $\la$ is a Lagrange multiplier which fixes the field to the
vacuum manifold (this corresponds to an infinite Higgs
mass). Tests have shown that this formalism agrees
well with the complementary approach of using the equation of
motion of a scalar field with Mexican hat potential and setting
the inverse mass of the particle to the smallest scale that can
be resolved in the simulation (typically of the order of $10^{-35}$ GeV),
but tends to give better energy momentum conservation.

As we cannot trace the field evolution from the unbroken phase
through the phase transition due to the limited dynamical range,
we choose initially a random field at a comoving time $\ct=2\De x$.
Different grid points are uncorrelated at all earlier
times~\cite{aberna}.

\begin{figure}
\centerline{\psfig{figure=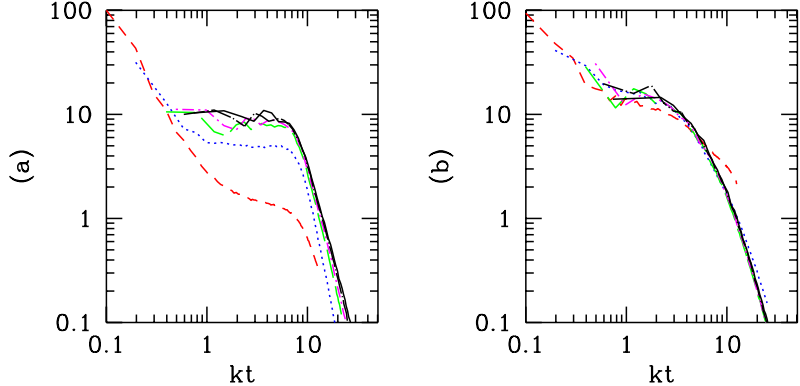,width=11.5cm}}
\caption{\label{fig20}
 The ETCs $C_{11}(z,1)=\langle|\Phi|^2\rangle(k\ct)$ (panel a) and 
 $C_{22}(z,1)=\langle|\Psi|^2\rangle(k\ct)$ (panel b) are shown for different
 times. In grid units the times are
 $\ct=4$ (dashed), $\ct=8$ (dotted), $\ct=12$ (long dashed),  $\ct=16,~20$
 (dash dotted, long dash dotted) and $\ct=24$ (solid). Clearly $C_{22}$ 
 scales much sooner than $C_{11}$.  To safely arrive in the scaling
 regime one has to wait until $\ct\sim 16$ and $C_{ij}(k\ct=0)$ is best
 determined at $\ct\ge 20$ but $k\ct< 1$.}
\end{figure}

The use of finite differences in the discretized action as well as in
the calculation of the energy momentum tensor introduce immediately
strong correlations between neighboring grid points. This problem
manifests itself in an initial phase of non-scaling
behaviour, the length of which varies between $10\De x$ and $20\De x$,
depending  on the variable considered.
 It is very important to use
results from the scaling regime only (cf. Fig.~\ref{fig20}).

In order to reduce the time necessary to reach scaling and to
improve the  overall accuracy, one has to choose the finite differences 
in an optimal way. One possibility is to calculate
all values in the center of each cubic cell defined by the lattice.
The additional smoothing introduced by this  improves energy-momentum
conservation by several percent. 

To calculate unequal time correlators (UTC), the values of the  
observables under consideration are saved once scaling is reached at 
time $\ct_c$ and
then correlated at all following time steps. While there is
some danger of contaminating the equal time correlator (ETC), which contributes
most strongly to the $C_\ell$'s, with non-scaling sources, this method
ensures that the constant for $k\ct \ra 0$ is determined with maximal
precision for the ETCs.  This is very
important as the constants $C_{ij}(0,1)$  fix
the relative size of scalar, vector and tensor contributions of the
Sachs-Wolfe part and severely influence the resulting $C_\ell$'s.
In contrast, the CMB spectrum seems quite stable under small
variations of the shape of the UTCs.

The resulting UTCs are obtained numerically as functions of the 
variables $k$, $\ct$ and $\ct_c$ with $\ct\geq \ct_c$ and $\ct_c$ fixed. They 
are linearly interpolated to the required range. One then
constructs a hermitian  matrix in $k\ct$ and $k\ct'$, 
with the values of $k\ct$ chosen on a linear scale to maximize the 
information content, $0\leq k\ct\leq x_{\mathrm max}$. The choice of 
a linear scale ensures good convergence
of the sum of the eigenvectors after diagonalization 
(see Fig.~\ref{fig13}), but still retains
enough data points in the critical region, ${\cal O}(x) =1$,
 where the correlators
start to decay. In practice one chooses as the endpoint $x_{\mathrm max}$
of the range sampled by the simulation the value at which the correlator
decays by about two orders of magnitude, typically $x_{\mathrm max}\approx 40$.
The eigenvectors that are fed into the Boltzmann code
are then interpolated using cubic splines with the condition
$v^{(n)}(k\ct)\ra 0$ for $k\ct \gg x_{\mathrm max}$.

\begin{figure}
\centerline{\psfig{figure=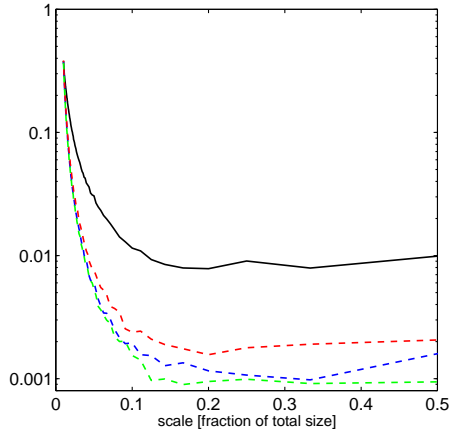,width=6.5cm}}
\caption{\label{fig21}
Energy momentum conservation of numerical simulations
is shown. The lines represent the sum of the terms which has to vanish
if energy (solid) respectively momentum (dashed) is conserved,
divided by the sum of the absolute values of these terms. The abscissa
indicates the wavelength of the perturbation as fraction of the size
of the entire grid. (from~\cite{DKM}}
\end{figure}

There are several methods to test the accuracy of simulations: 
One of them is energy momentum conservation. In Ref~\cite{DKM} it is found 
to be better than 10\% on all scales larger than about 4 grid units, as is 
shown in Fig.~\ref{fig21}. Another possibility is a comparison with the exact
spherically symmetric solution in non-expanding space~\cite{PST2}.

\begin{figure}
\centerline{\psfig{figure=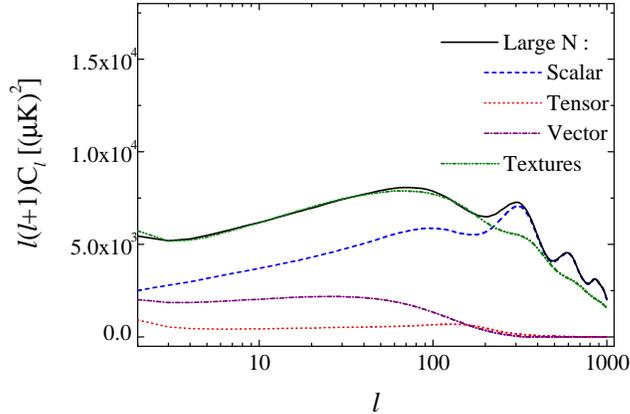,width=8.5cm}}
\caption{\label{fig19}
The $C_\ell$ power spectrum is shown for the large-$N$ limit
(bold line) and for the texture model. The main difference is
clearly that the large-$N$ curve shows some acoustic oscillations
which are nearly entirely washed out in the texture case.}
\end{figure}

The overall shape and amplitude of the unequal time correlators
are quite similar to those found in the analytic large-$N$ approximation
\cite{TS,KD,DK}  (see Figs.~\ref{fig1} to \ref{figtensor}). 
The main difference of the large-$N$ approximation is
that there the field evolution, Eq.~(\ref{eqmo2}), is approximated by a
linear equation. The non-linearities in the large-$N$ seeds, which are
due solely to the energy momentum tensor being quadratic in the
fields, are much weaker than in the texture model where the field
evolution itself is non-linear. Therefore, decoherence which is a
purely non-linear effect, is much weaker in the
large-$N$ limit. This is actually the main difference between the two
models as can be seen in Fig.~\ref{fig19}. Otherwise the similarity
of the results obtained by simulating the full non-linear problem
and by considering the simplified linear limit is quite remarkable.
The entire class of global $O(N)$ models behaves in this
way, and potentially many other global defect models as well.

\subsection{Cosmic strings}

So far we discussed only theories with global defects. Yet in modern particle
physics, local (gauge) symmetries play a much more important role than global
symmetries. As we have seen in Chapter~\ref{Sym}, strings are the only local 
defects which  scale  and which are therefore potential candidates to 
seed structure formation. 
Most research concentrates on local $U(1)$ theories, the best known 
cosmic strings. But also other models, even with non-abelian strings, have been
considered (see \eg~\cite{SpPe,SpBu}).

The main difference for the numerical treatment of local strings as 
compared to global defects is due to the existance of a gauge 
field which compensates the gradient energy of
the field. All the energy is therefore concentrated in the tiny region 
of order the symmetry breaking scale $l_M \sim 1/M\sim 1/T_c$ where the 
scalar field leaves the vacuum manifold. Let us estimate how thin cosmic 
strings really are: COBE normalisation requires the phase
transition to take place at the GUT scale, $T_c\sim 10^{16}$GeV. 
This corresponds to $l_m\sim 1/T_c \sim 10^{-30}$ cm. Clearly, a numerical
simulation with a grid of not much more than $(512)^3$ cells which should 
simulate the entire Hubble volume, $\sim (10^{28}{\rm cm})^3$, cannot   
resolve this scale by more than $55$ orders of magnitude. Therefore, 
strings are approximated as infinitely thin, and
it can be shown that they obey to a very good approximation the Nambu-Goto 
action of fundamental string theory~\cite{VS}. Corrections are of the order of 
the string thickness devided by the string curvature scale, and therefore
irrelevant for cosmology.

Furthermore, the string network needs to loose energy by gravitational 
radiation in order to scale. Hence,
 energy momentum conservation cannot be enforced for the defect
field alone, leading to 14 UTCs insted of only five.
As the interactions of cosmic strings are not known, 
one must also make ad hoc assumptions concerning the decay products. This
choice  has considerable impact on the results
\cite{CHM,Nathalie}.

We do not describe the numerical simulations to evolve 
cosmic strings. Detailed accounts of this problem can be found in the 
literature~\cite{VS,KH,MB,VHS}. Let us, nevertheless, point out the main 
problem. It is very difficult to simulate cosmic strings in expanding space, 
due to the large difference between the Hubble scale and the scale of 
small scale structure. Hence, it remains unclear up to date, whether string
simulation in an expanding universe can capture enough of the small scale 
structure, the tiny wiggles and loops which develop due to the string 
self-interaction, to produce meaningful results. On the other hand, string 
simulations in flat space, do not satisfy energy momentum conservation of 
expanding space.
 To adress the small scale structure problem, most of the recent results in 
the literature, actually all except~\cite{ACSSV,Aetal,Avel}, use flat space 
simulations or semi-analytical methods to calculate the string UTC's. It is 
not clear to us which procedure gives the best results, but since all the 
obtained CMB spectra disagree significantly with observations, this question 
has somehow lost its urgency.

As initial configuration of a string simulation, one usually lays down 
string segments according to the so called Vachaspati-Vilenkin algorithm. 
These are then evolved with the Nambu-Goto equation of motion. The physical 
problem of the 'decay product' of cosmic strings is related to the 
numerical problem of small 
scale structure: Most string codes find that the network develops structure
(wiggles, tiny loops) on the smallest scales which the simulation can resolve.
The phyical scale of these small wiggles and loops is still unknown. It may
even be, that the loops become smaller and smaller due to self-intersection,
until their size is  of the order of their thickness and they decay into
elementary particles. This picture, which is in contrast to the decay into 
gravity waves, is avocated in Ref.~\cite{VAH}. There have also been several 
attempts to take into account these wiggles 
in semi-analytic models~\cite{PoVa}.

\clearpage

\section{Result}
\label{Res}

\subsection{The unequal time correlators}
As explained in Chapter~\ref{The} to compute the observable CMB and matter,
 power spectra we need the  unequal time correlators of the seed energy 
momentum tensor. 

More precisely, for the {\em scalar} part we need the correlators
\bea
  \langle\Phi_s(\bk,\ct)\Phi_s^*(\bk,\ct')\rangle &=&
	\frac{1}{k^4\sqrt{\ct\ct'}}C_{11}(z,r)~, \label{Cs11}\\
  \langle\Phi_s(\bk,\ct)\Psi_s^*(\bk,\ct')\rangle &=& 
 	\frac{1}{ k^4\sqrt{\ct\ct'}}C_{12}(z,r)~, \label{Cs12}\\
  \langle\Psi_s(\bk,\ct)\Psi_s^*(\bk,\ct')\rangle &=& 
	\frac{1}{ k^4\sqrt{\ct\ct'}}C_{22}(z,r)~, \label{Cs22}
\eea
as well as $C_{21}(z,r) =C^*_{12}(z,1/r)$. The functions $C_{ij}$ 
are analytic in $z^2$. The pre-factor $1/(k^4\sqrt{\ct\ct'})$ comes from the
fact that the correlation functions $\langle f_\rho f^*_\rho\rangle$,  
$ k^4\langle f_\pi f^*_\pi\rangle$ and  $k^2\langle f_v f^*_v\rangle$ have to
be analytic and from dimensional considerations (see Ref.~\cite{DK}). 

\begin{figure}[ht]
\centerline{\epsfig{figure=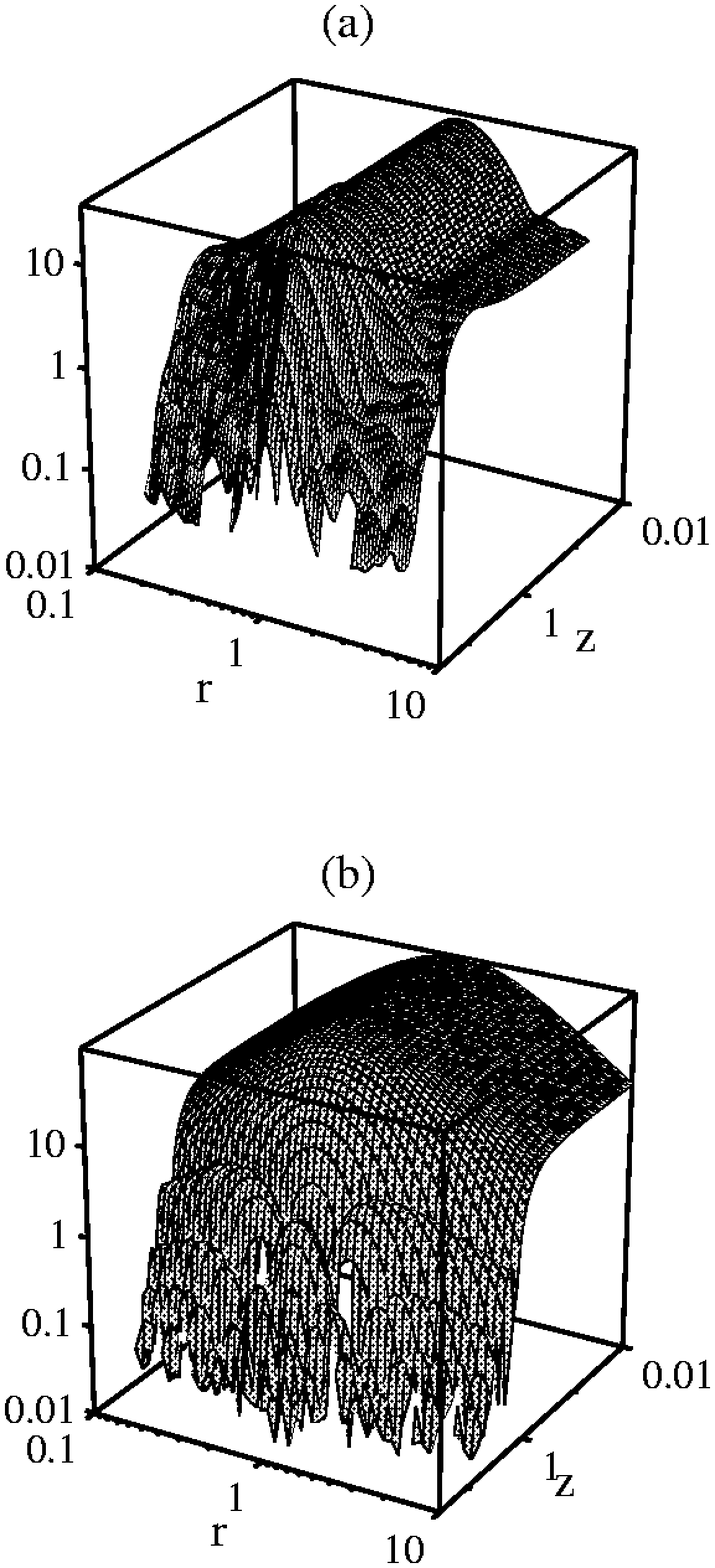,width=4.5cm}
\epsfig{figure=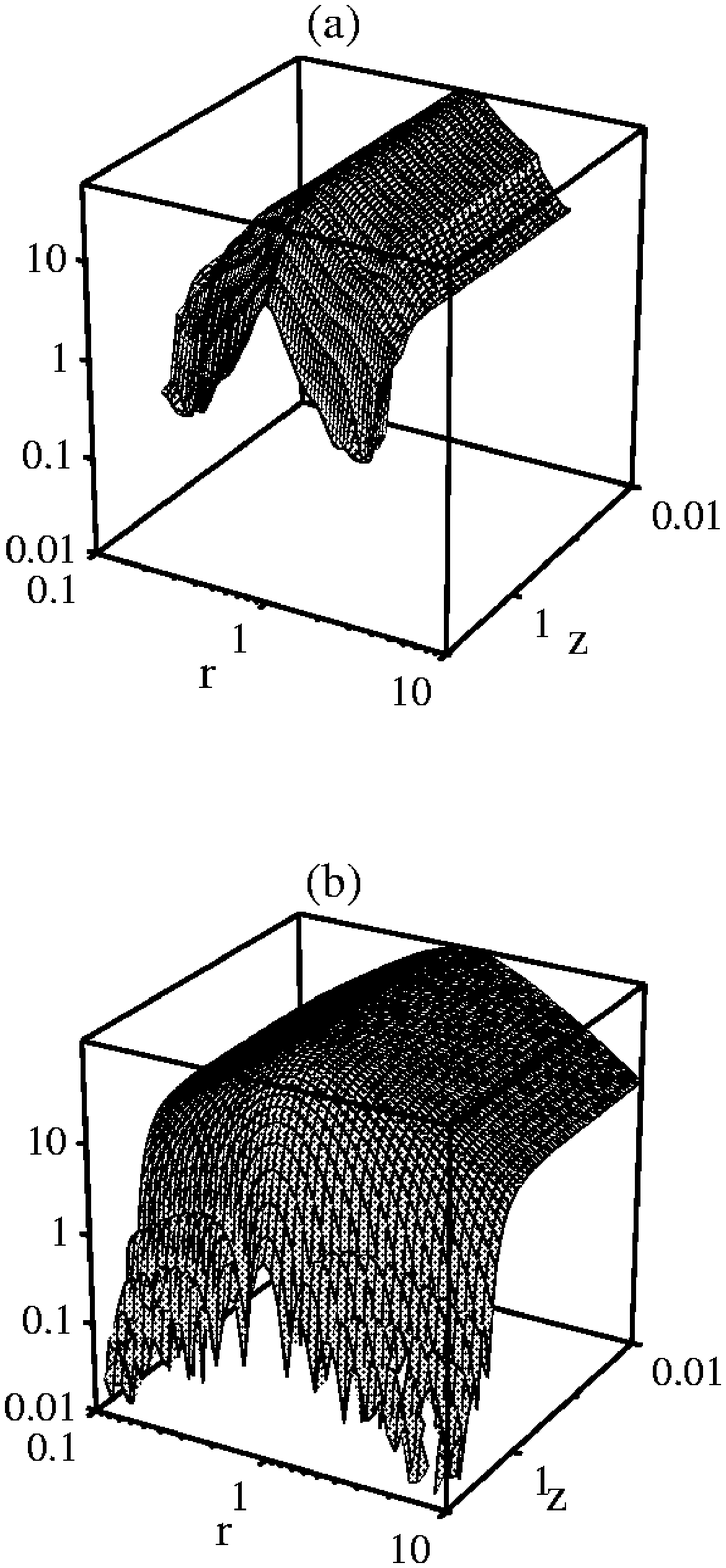,width=4.5cm}
\epsfig{figure=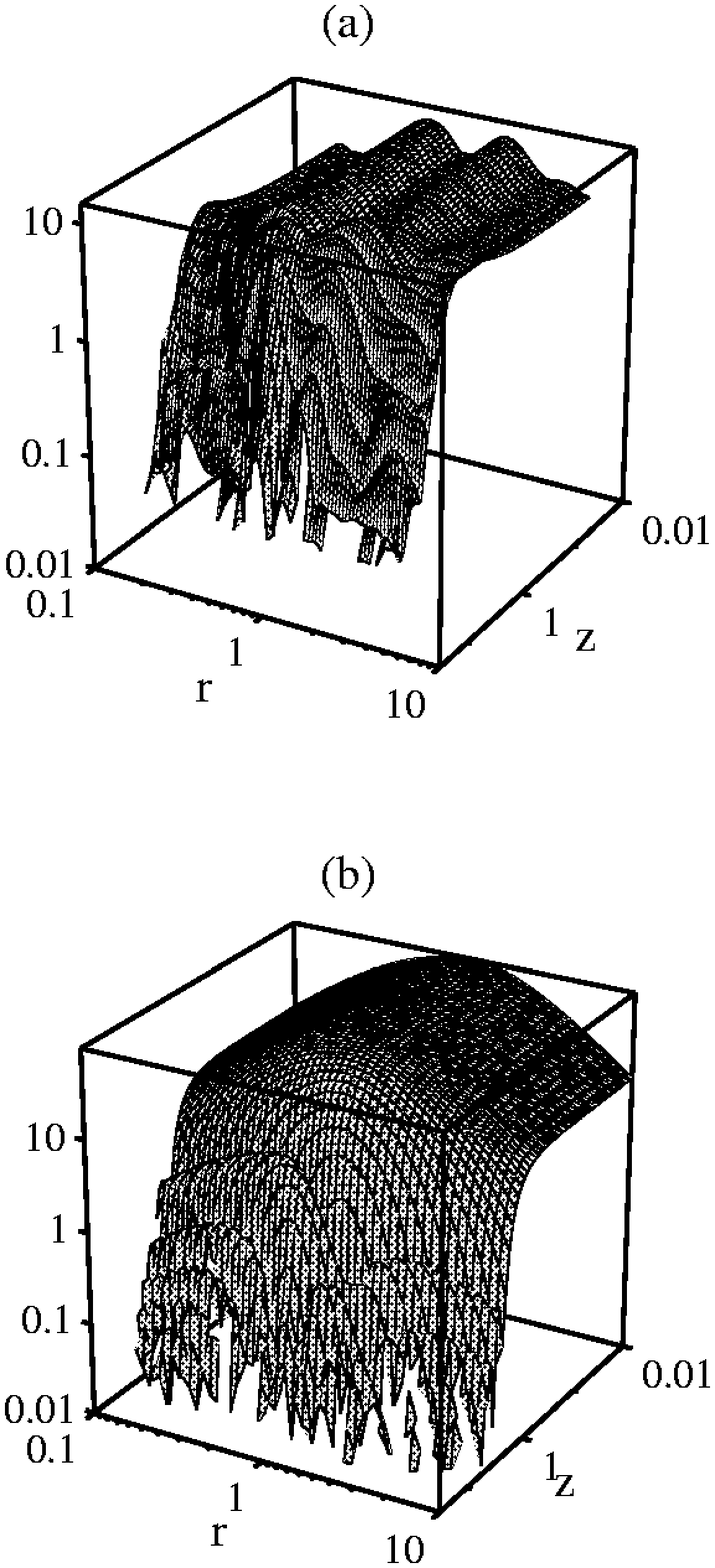,width=4.5cm}}
\caption{\label{fig1}The two point correlation functions 
 $C_{11}(z,r) = 
 k^4\sqrt{\ct\ct'}\langle\Phi_s(\bk,t)\Phi_s^*(\bk,\ct')\rangle$ (left),
 $C_{22}(z,r) = 
 k^4\sqrt{\ct\ct'}\langle\Psi_s(\bk,\ct)\Psi_s^*(\bk,\ct')\rangle$ (center) and
 $|C_{12}(z,r)| = 
 k^4\sqrt{\ct\ct'}|\langle\Phi_s(\bk,\ct)\Psi_s^*(\bk,\ct')\rangle|$ (right).
 Panels (a) represent the result from numerical simulations of
 the texture model; panels (b) show the large-$N$ limit.  For
 fixed $r$ the correlator is constant for $z<1$ and then decays. Note
 also the symmetry under $r\ra 1/r$ for $C_{11}$ and $C_{22}$ which
 is lost for $C_{12}$ (from~\cite{DKM}).}
\end{figure}

The functions $C_{ij}$ are shown
in Fig.~\ref{fig1}. Panels~(a) are obtained from numerical
simulations.  Panels~(b) represent the same correlators for the
large-$N$ limit of global $O(N)$-models (see~\cite{TSlN,KD}).

\begin{figure}[ht]
\centerline{\epsfig{figure=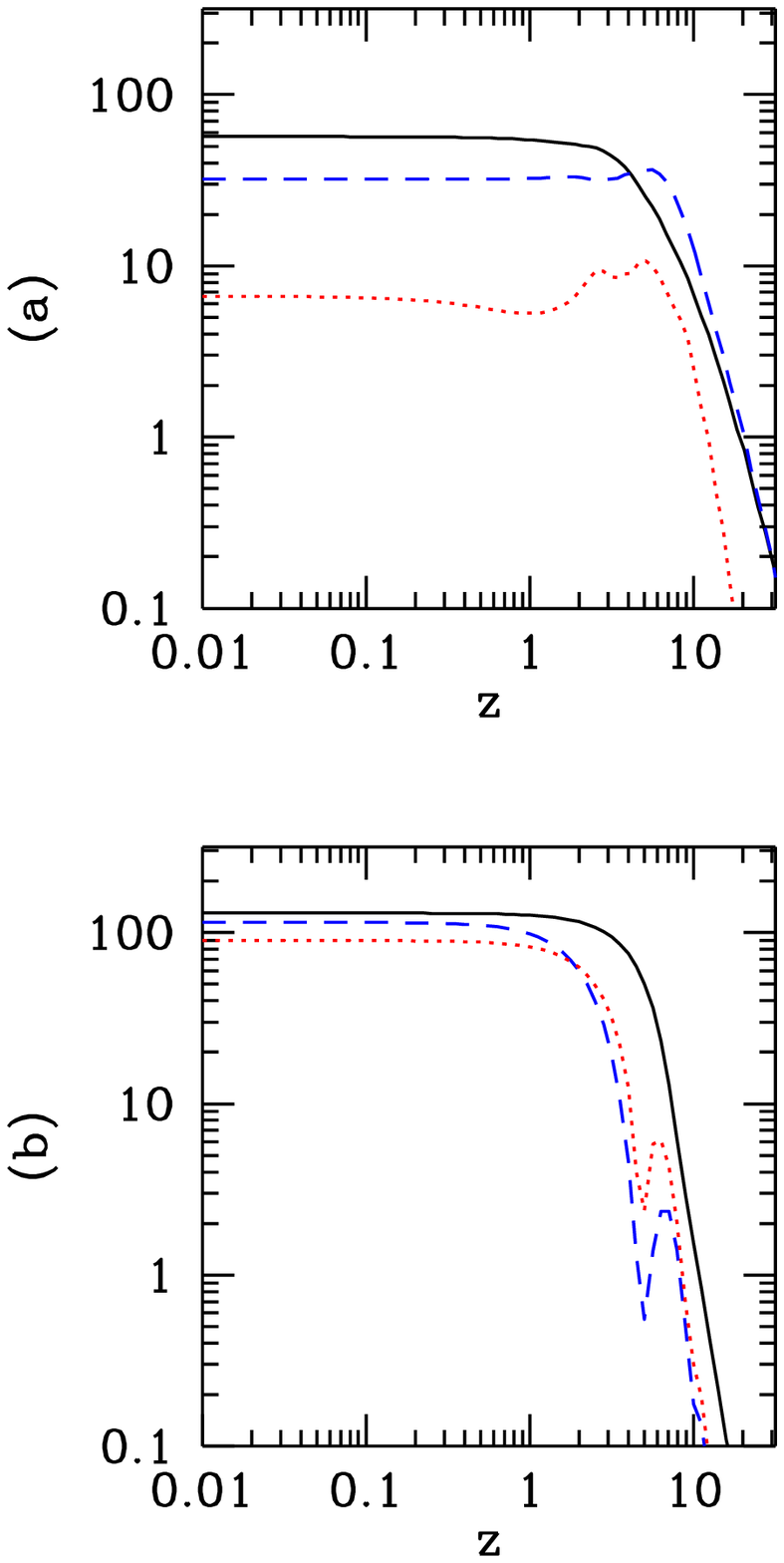,width=5cm}
            \epsfig{figure=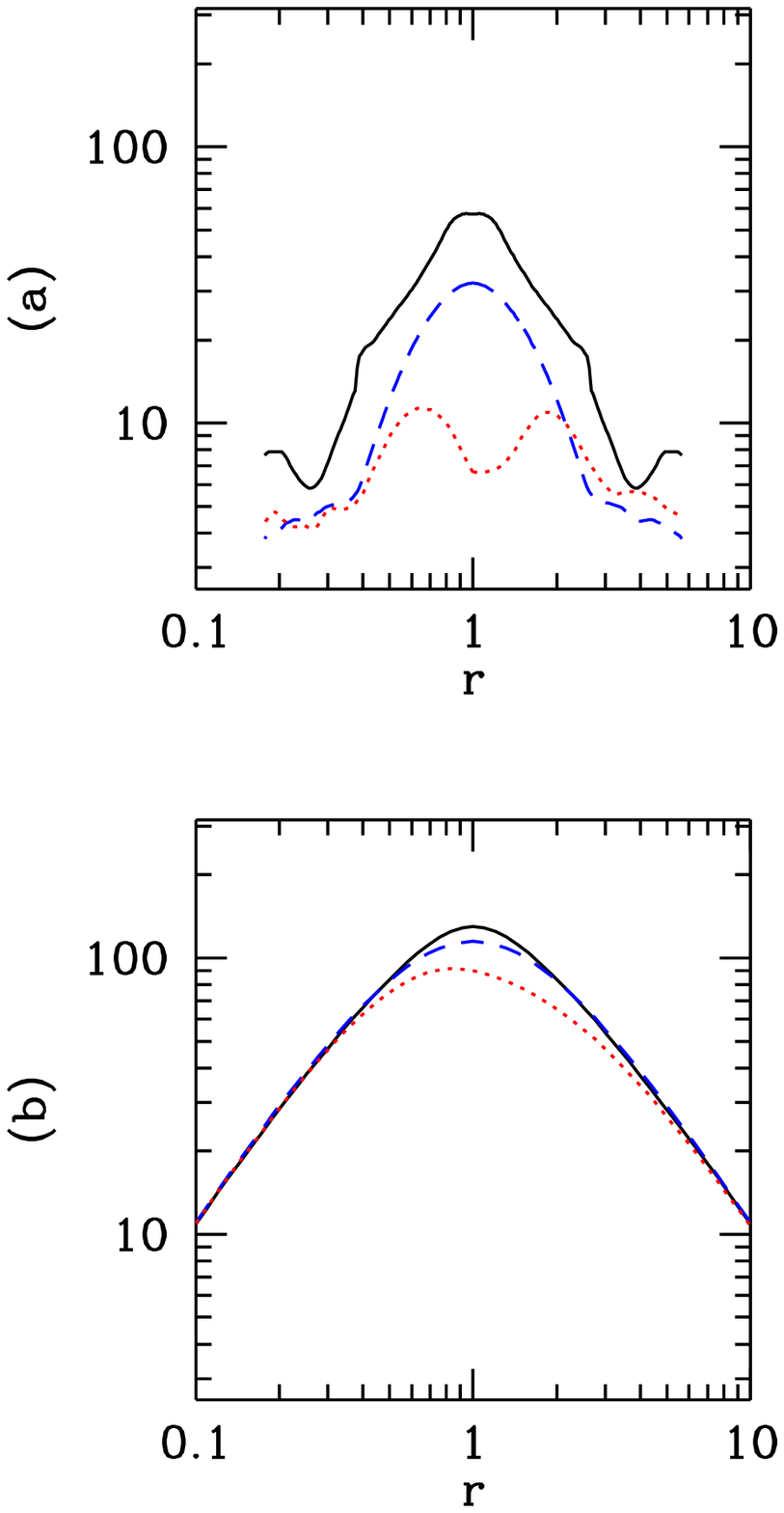,width=5cm}}
\caption{\label{fig4}
 On the left the correlators  $C_{ij}(z,1)$ are
 shown, while the right figure depicts $C_{ij}(0,r)$ with $r=\ct'/\ct$.
 The solid, dashed and dotted lines represent $C_{22}~,~C_{11}$
 and $|C_{12}|$ respectively. Panels (a) are obtained from numerical
 simulations of the texture model and panels (b) show the large-$N$
 limit. A striking difference is that the large-$N$ value for
 $|C_{12}|$ is relatively well approximated by the perfectly coherent
 result $\sqrt{|C_{11}C_{22}|}$ while the texture curve for $|C_{12}|$
 lies nearly a factor 10 lower (from~\cite{DKM}).}
\end{figure}

In Fig.~\ref{fig4} we show $C_{ij}(z,r=1)$, and the
'constant' of the Taylor expansion for $C_{ij}$ is given as a function
of $r$, {\em i.e.},  $C_{ij}(0,r)$.

\begin{figure}[htb]
\centerline{\psfig{figure=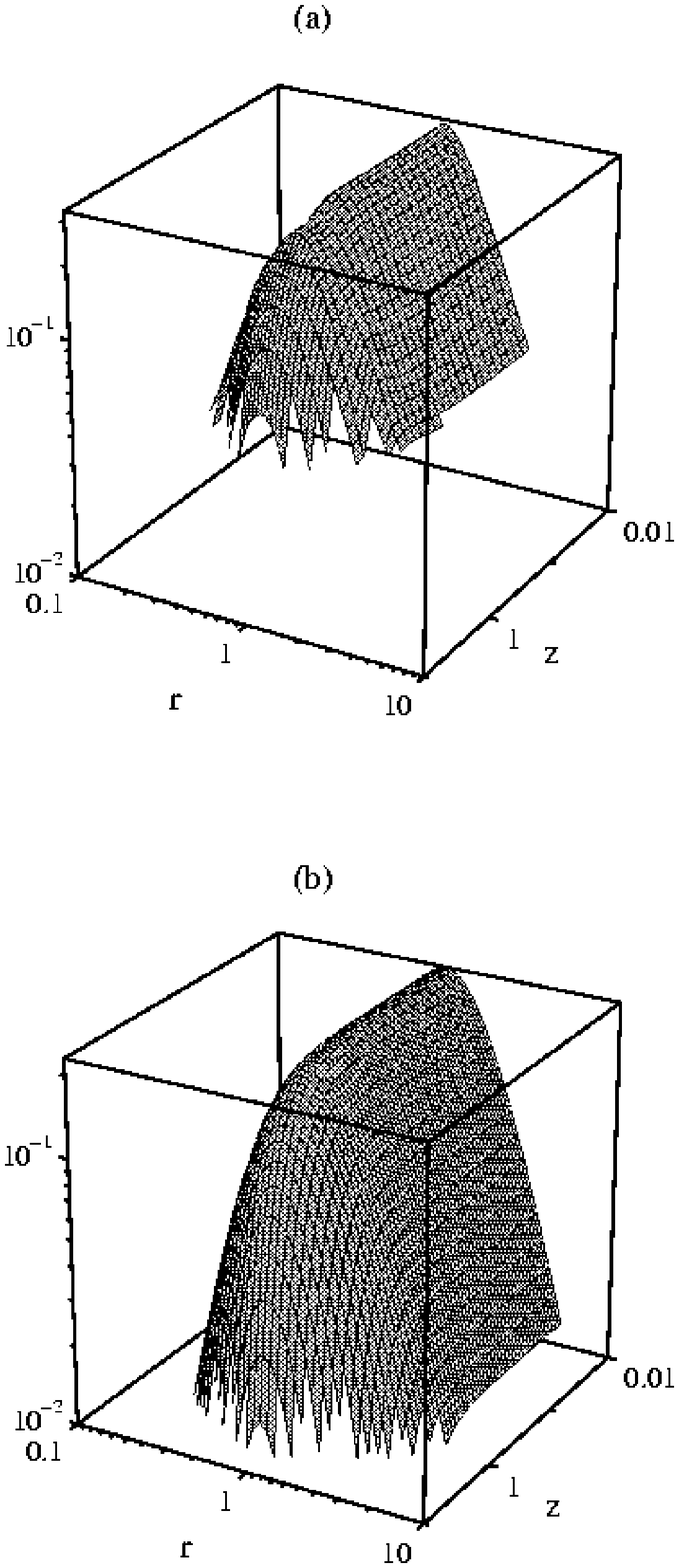,width=5.5cm}
\quad\begin{minipage}[b]{6cm}\psfig{figure=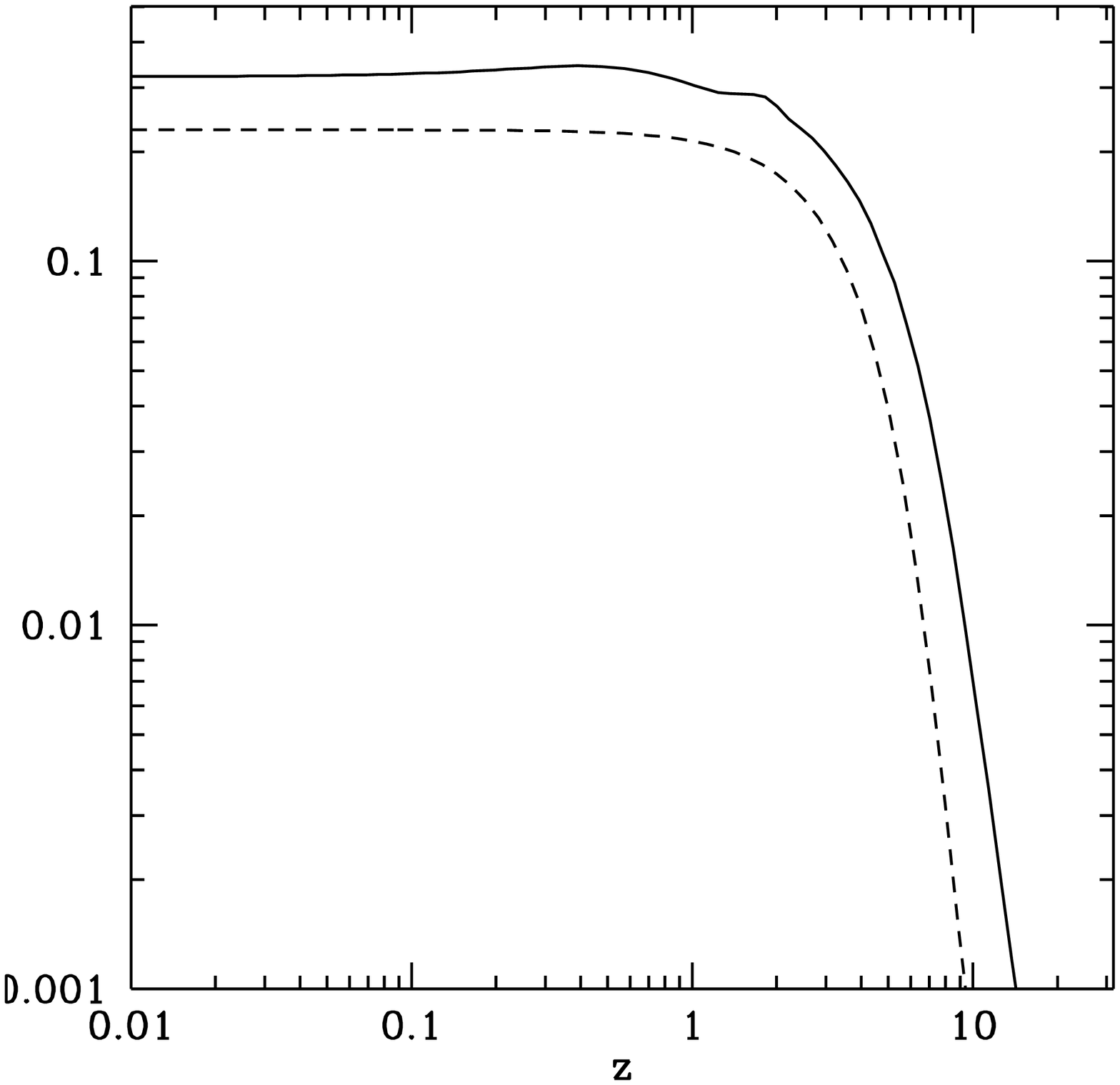,width=5cm}\\
\psfig{figure=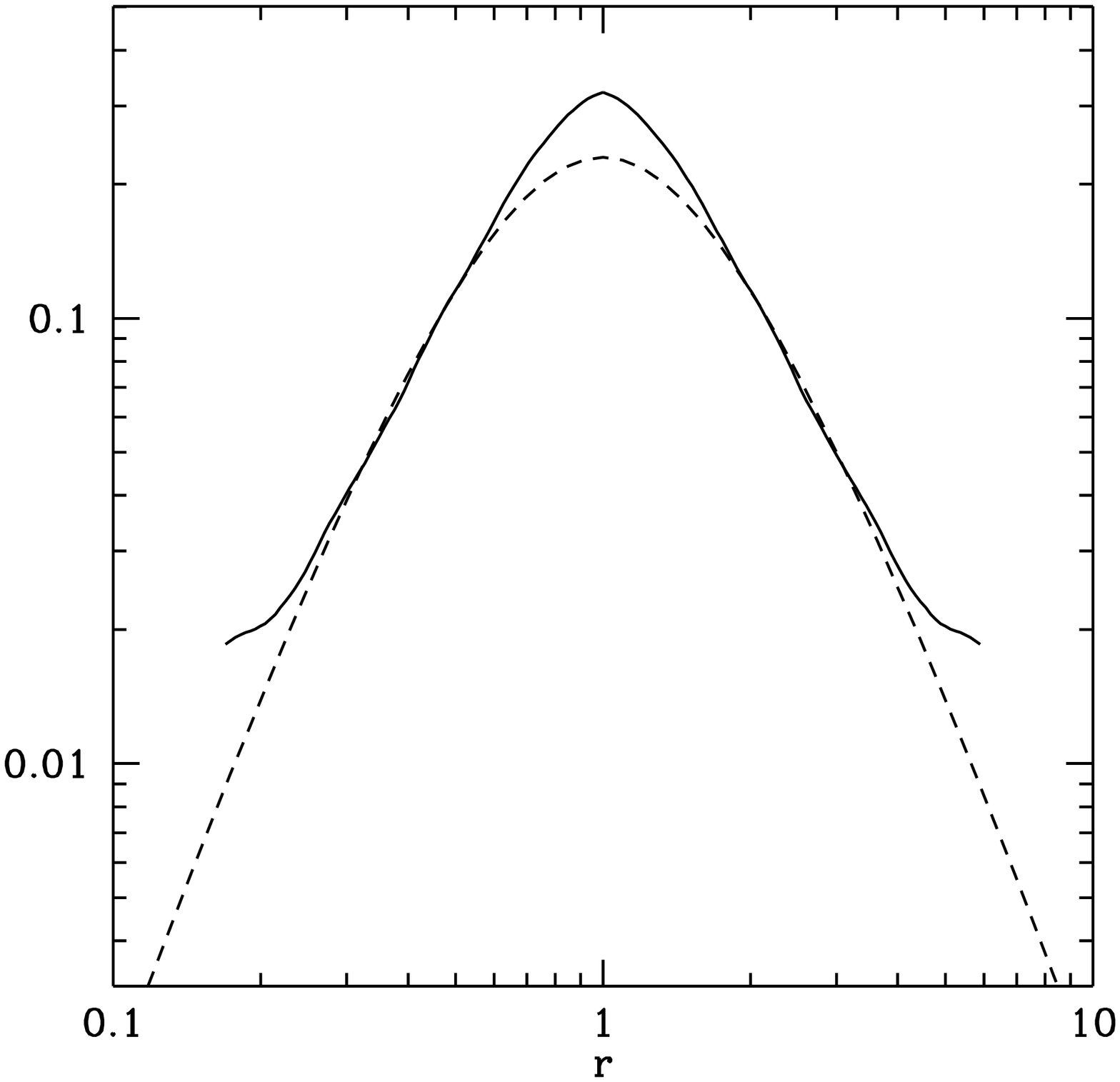,width=5cm}\end{minipage}}
\caption{\label{figveccor}
The vector correlator  $W(z,r)$ is shown on the left. The texture
simulations, panel (a), and the large-$N$ limit, panel (b), give
very similar results, up to a slight difference in amplitude. 
On the right we see $W(z,1)$ (top)
 and $W(0,r)$ (bottom). The solid line
represents the texture simulations and the dashed line is the
large-$N$ result.  The
`wings' visible in the lower texture curve are probably not due to a
resolution problem but the beginning of oscillations (from~\cite{DKM}).}
\end{figure}

{\em Vector} perturbations are induced by $\bm\Si^{(s)}$ which is
seeded by $\bm{w}^{(v)}$. Transversality and dimensional arguments
require the correlation function to be of the form
\be
 \langle w_i^{(v)}(\bk,t)w_j^{(v)*}(\bk,\ct')\rangle =
\sqrt{\ct\ct'}(k^2\de_{ij}-k_ik_j)W(z,r)~. \label{Cv} 
\ee
Again, as a consequence of causality, the function $W$ is analytic in
$z^2$ (see~\cite{DK}). The functions $W(z,r)$, $W(z,1)$ and  $W(0,r)$
are plotted in Fig.~\ref{figveccor}.

\begin{figure}[htb]
\centerline{\psfig{figure=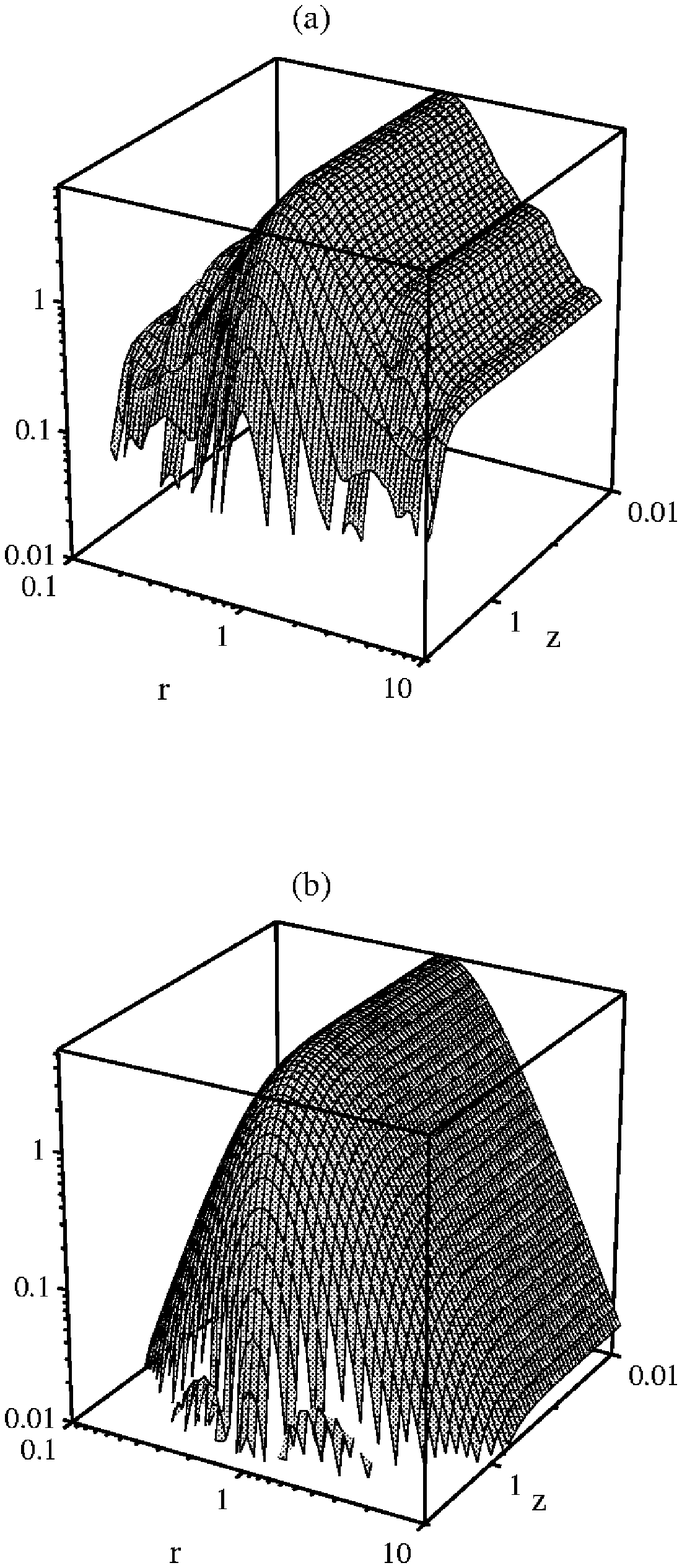,width=5cm}
\quad\begin{minipage}[b]{6cm}\psfig{figure=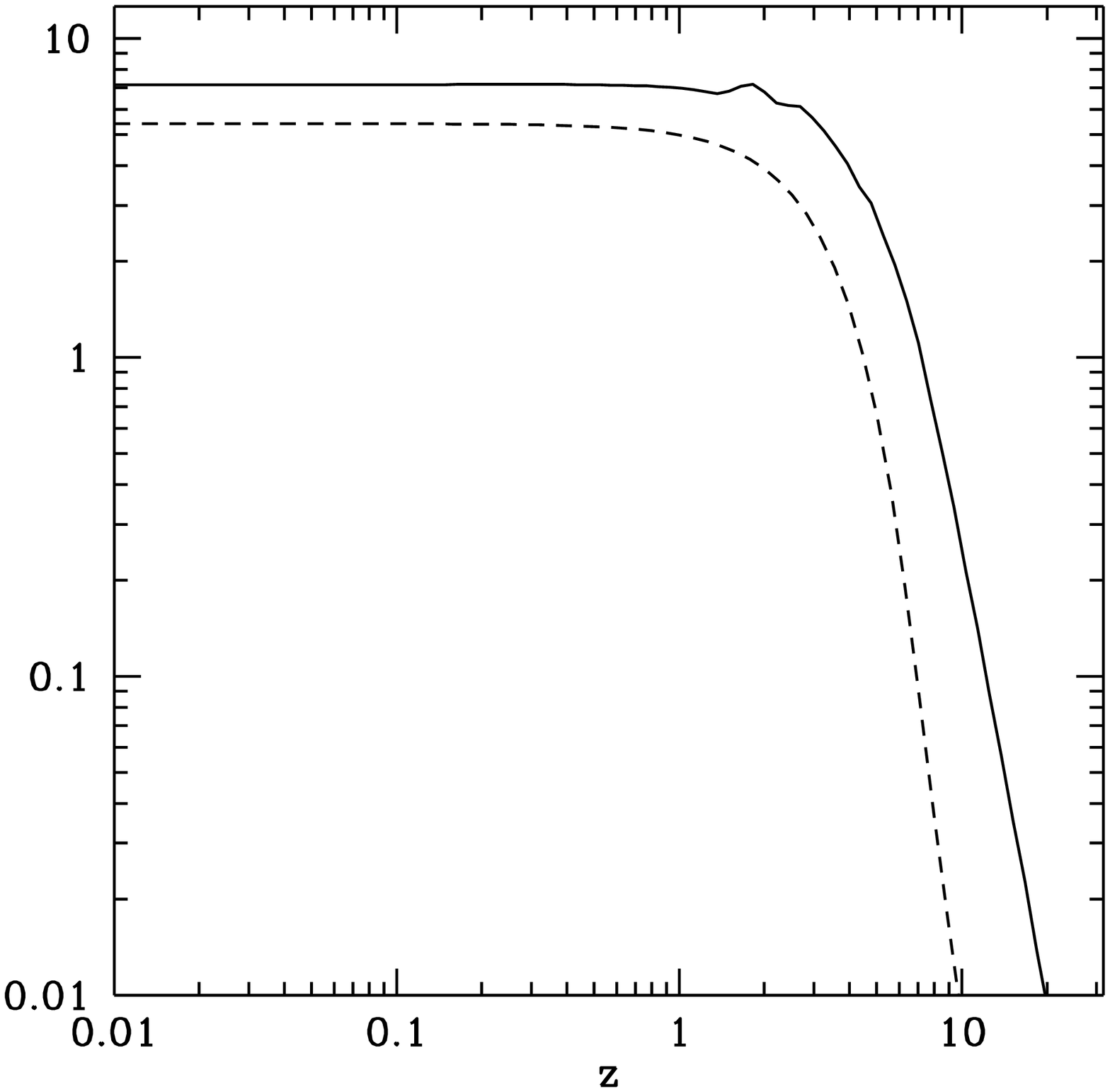,width=5cm}\\
\psfig{figure=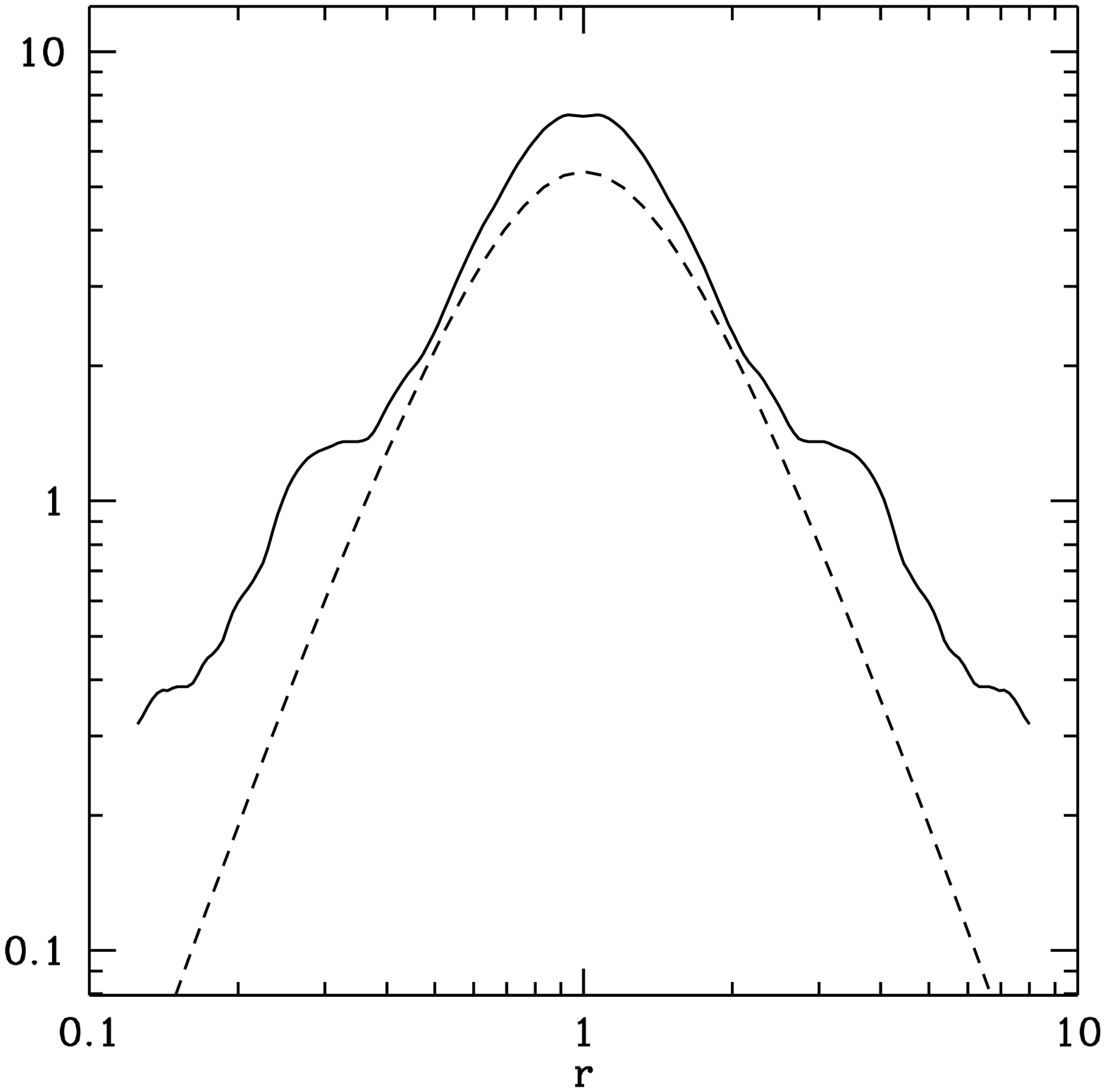,width=5cm}\end{minipage}}
\caption{\label{figtensor}
The tensor correlator  $T(z,r)$ is shown on the left, the texture
simulations in panel (a), and the large-$N$ limit in panel (b).
On the right we see $T(z,1)$ (top) and $T(0,r)$ (bottom). The solid lines
represent texture simulations and the dashed lines are the
large-$N$ result. (from~\cite{DKM})}
\end{figure}

Symmetry, transversality and tracelessness,  together with 
statistical isotropy require the {\em tensor} correlator to be 
of the form (see~\cite{DK})
\bea
\lefteqn{\langle \tau^{(\pi)}_{ij}(\ct)\tau^{(\pi)*}_{lm}(\ct')\rangle =
	\frac{1}{\sqrt{\ct\ct'}}T(z,r)[\de_{il}\de_{jm}
+\de_{im}\de_{jl} }  \nonumber \\
&&   -\de_{ij}\de_{lm} + k^{-2}(\de_{ij}k_lk_m +
	\de_{lm}k_ik_j -\de_{il}k_jk_m - \de_{im}k_lk_j  \nonumber \\
&&	-\de_{jl}k_ik_m -\de_{jm}k_lk_i) + k^{-4}k_ik_jk_lk_m] \label{Ct}
    ~.\eea
The functions $T(z,r)$ as well as $T(z,1)$ and $T(0,r)$ are shown in
Fig.~\ref{figtensor}. 

Expanding the unequal time correlators in terms of eigenfunctions and
eigenvectors as explained in Chapter~\ref{The}, one finds that ae expansion 
with a linear weight actually converges faster than one with a logarithmic 
weight. This is illustrated in  Fig.~\ref{fig13} below.

\begin{figure}[htb]
\centerline{\psfig{figure=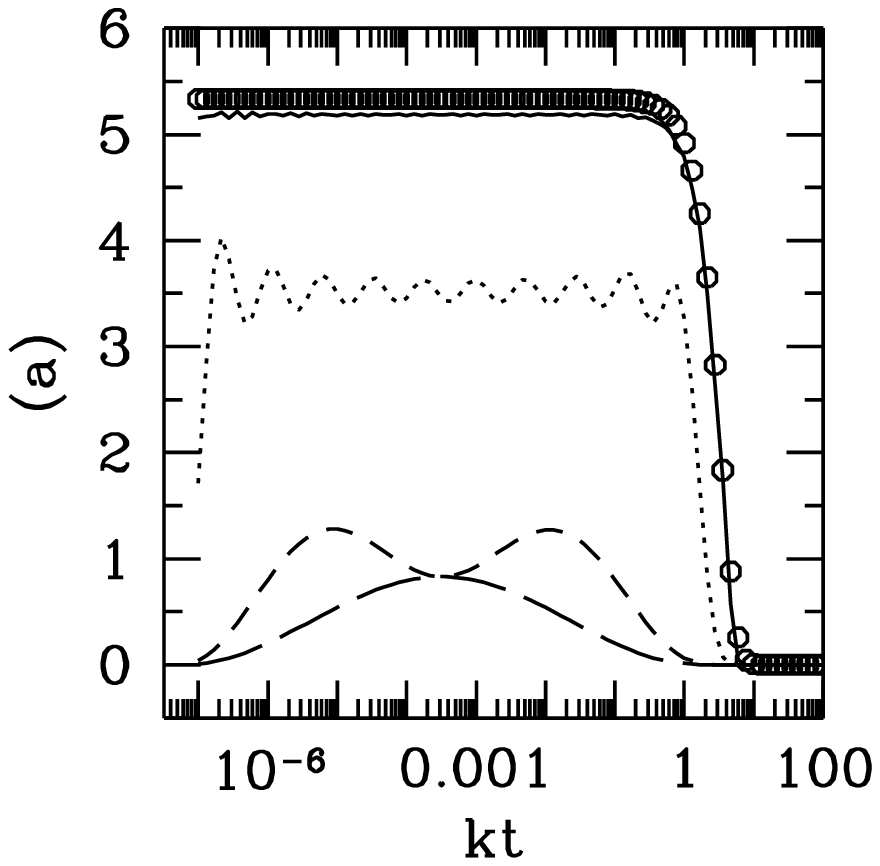,width=6cm}
\quad \psfig{figure=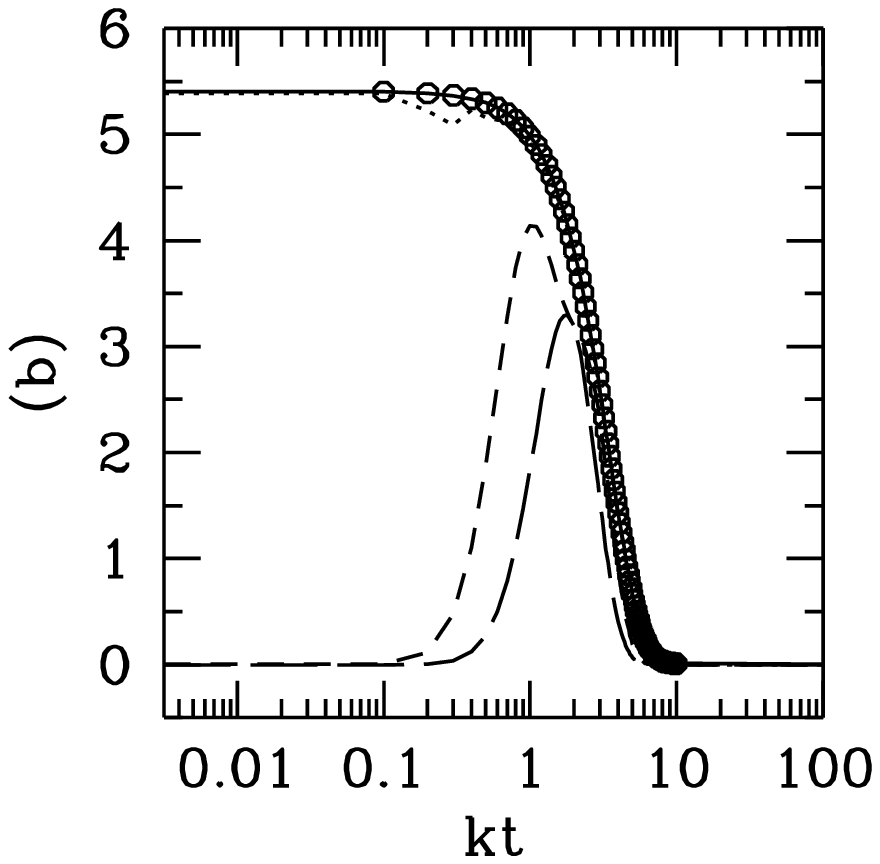,width=6cm}}
\caption{\label{fig13}
The sum of the first few eigenfunctions of $T(x,x)$ is shown
for two different weight functions, (a) logarithmic, $w=1/x$ and (b)
linear, $w=1$. The first (long dashed), first and second (short
dashed), first ten (dotted) and first thirty (solid)
eigenfunctions are summed up. The open circles represent the full
correlation function. Clearly, the eigenfunctions obtained by linear  
weighting converge
much faster. Here we only show the equal time diagonal of the
correlation matrix, but the same behavior is also found in the
$C_\ell$ power spectrum which is sensitive to the full correlation matrix 
(from Re.~\cite{DKM}).}
\end{figure}

\begin{figure}
\centerline{\psfig{figure= 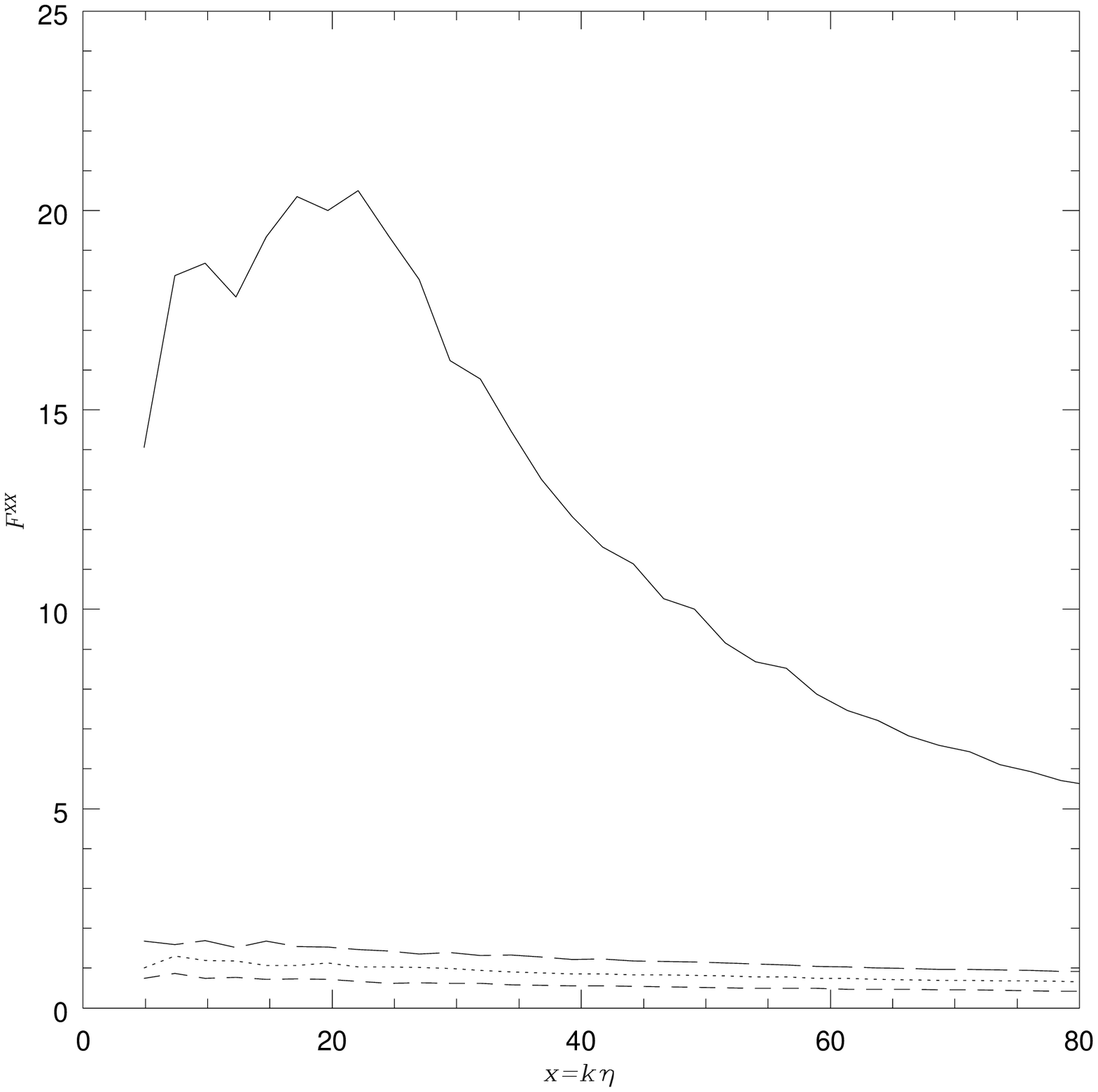
,width=7cm}\psfig{figure= 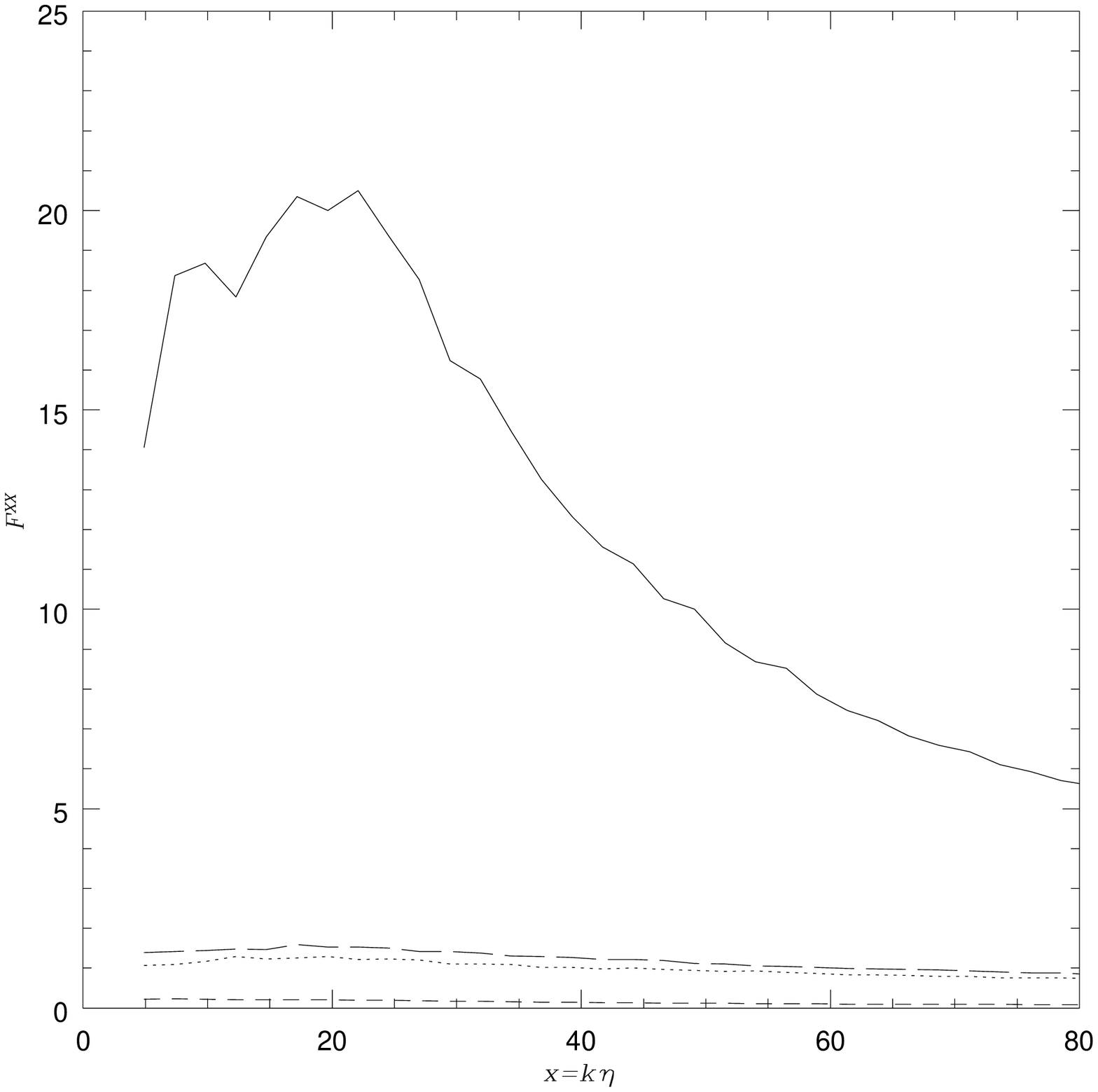,width=7.cm }}
\caption{\label{em_string}
The size of the equal time contribution from the energy
density (solid line) and the anisotropic stresses (left). Pressure,
velocity and vorticity are of about the same size as the
anisotropic stresses (right) (from Magueijo \& Brandenberger~\cite{MB}).
}\end{figure}
In Fig.~\ref{em_string} we also show the equal time correlators of the
energy density and pressure, velocity and anisotropic stresses of local 
cosmic strings. It is interesting to note to which extent the spatial 
components of the energy momentum tensor are smaller than  the energy 
density. In contrary to global defects, local strings 
provide a nearly non-relativistic source.

A source is called totally coherent~\cite{Ma,DS} if the unequal time
correlation functions can be factorized. This means that only one 
eigenvector is relevant. A simple totally coherent approximation, 
which however misses some important characteristics of defect
models, can be obtained by replacing the
correlation matrix by the square root
of the product of equal time correlators,
\be \langle \Scal_i(\ct)\Scal_j^*(\ct')\rangle \ra \pm \sqrt{\langle
|\Scal_i(\ct)|^2\rangle  \langle |\Scal_j(\ct')|^2\rangle}  ~. \label{coapp}
\ee 
This approximation is exact if the source evolution is linear. Then the
different $\bk$ modes do not mix and the value of the source term at fixed
$\bk$ at time $\ct$ is given by its value at initial time multiplied
by some transfer function, $\Scal(\bk,\ct) =T(\bk,\ct,\ct_{in})
\Scal(\bk,\ct_{in})$. In this situation, (\ref{coapp})
becomes an equality and the model is perfectly coherent. Decoherence
is due to the non-linearity of the source evolution which induces a
'sweeping' of power from one scale into another. Different wave
numbers $\bk$ do not evolve independently.

It is interesting to note that the perfectly coherent approximation
(\ref{coapp}) leaves open a choice of sign which has to be positive
if $i=j$, but which is undetermined otherwise. According to Schwarz
inequality the correlator $ \langle \Scal_i(\ct)\Scal_j^*(\ct')\rangle $ is
bounded by
\[  \!\!\!\!
 -\sqrt{\langle |\Scal_i(\ct)|^2\rangle\langle |\Scal_j(\ct')|^2\rangle}\le
	\langle \Scal_i(\ct)\Scal_j^*(\ct')\rangle \le
\sqrt{\langle |\Scal_i(\ct)|^2\rangle  \langle |\Scal_j(\ct')|^2\rangle}.
\]
Hence, for scales/variables for which the Greens function is not
oscillating (e.g. Sachs Wolfe scales) the full result always lies
between the 'anti-coherent' (minus sign) and the coherent result.
This behavior has been verified numerically~\cite{DKM}.

\begin{figure}[htb]
\centerline{\psfig{figure=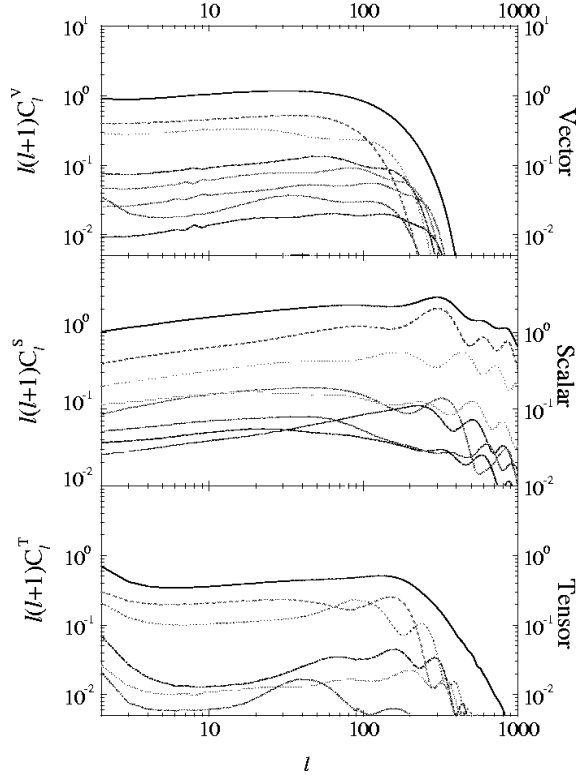,width=7.5cm}}
\caption{\label{fig1416}
The scalar, vector and tensor contributions for the
texture model of structure formation are shown. The dashed lines show
the contributions from single eigenfunctions while the solid line
represents the sum. Note that the single contributions to the scalar and
tensor  spectrum do show oscillations which are however washed out in
the sum. (Vector perturbations do not obey a wave equation and  
thus do not show oscillations.)}
\end{figure}

The first evidence that acoustic peaks are suppressed in defect models
has been obtained in the perfectly coherent approximation in Ref.~\cite{DGS}.
In Fig.~\ref{fig1416} we show the contributions to the $C_\ell$'s from more
and more eigenvectors. A perfectly coherent model has only one
non-zero eigenvalue.

\begin{figure}[htb]
\centerline{\psfig{figure=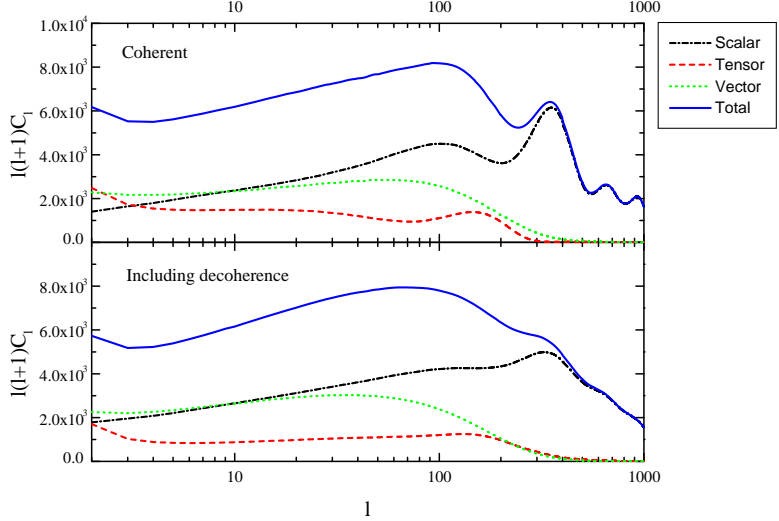,width=7cm,angle=-90}}
\caption{\label{fig17}
The $C_\ell$ power spectrum for the texture scenario is shown
in the perfectly coherent approximation (top panel) and in the full
eigenfunction expansion. Even in the coherent approximation, the
acoustic peaks are not higher than the Sachs Wolfe
plateau. Decoherence just washes out the structure but does not
significantly damp the peaks.}
\end{figure}

A comparison of the full result with the
totally coherent approximation is presented in Fig.~\ref{fig17}. 
There one sees
that decoherence does smear out the oscillations present in the fully
coherent approximation, and does somewhat damp the
amplitude. Decoherence thus prevents the appearance of a series of acoustic 
peaks. The absence of power on the angular scale of the peaks, however,
is not a consequence of decoherence. Is mainly due to the
anisotropic stresses of the source, which lead to perturbations in the
geometry inducing large scale $C_\ell$'s (Sachs Wolfe), but not to density
fluctuations, which are responsible for the acoustic peaks. Large 
anisotropic stresses are also at the origin of vector and tensor 
fluctuations which contribute more than 50\% on large scales. These 
results were first computed fully in Ref.~\cite{PST} and followed by a 
parameter study in Ref~\cite{DKM}. Thy are in agreement with Ref.~\cite{DGS}
but disagree strongly with Ref.~\cite{TC}.

In the real universe, perfect scaling of the seed correlation functions 
is broken by the radiation--matter transition, which
takes place at the time of  matter and radiation equality, 
$\ct_{eq}\simeq$   $20h^{-2}\Om_m^{-1/2}$Mpc. The time $\ct_{eq}$ is an 
additional
scale which enters the problem and influences the seed
correlators. Only in a purely radiation or matter dominated universe are
the correlators strictly scale invariant. This means actually that
the $k$ dependence of the correlators $C$, $W$ and $T$ cannot really
be cast into a dependence on $x$ and $x'$, but that these functions depend
on $\ct,\ct'$ and $k$ in a more complicated way. In principle, one thus
has to calculate and  diagonalize the seed correlators  
for each wave number $k$ separately and the huge gain of dynamical
range is lost as soon as scaling is lost.

In the actual case at hand, however, the deviation from scaling is
weak, and most of the scales of interest for structure formation 
(especially those which are still in the linear regime) enter the horizon only
in the matter dominated era. The behavior of the correlators in the
radiation dominated era is of minor importance.
To solve the problem, one can calculate the  eigenvalues
and eigenfunctions twice, in a pure radiation and in a pure matter
universe and  interpolate the source term from the radiation to the
matter epoch. Denoting by $\la_m, v_m$ and $\la_r, v_r$ a given pair
of eigenvalue and eigenvector in a matter and radiation universe
respectively, we choose as our deterministic source function
\bea
 v(\ct) &=& y(\ct)\sqrt{\la_r}v_r(k\tau) +(1-y(\ct))\sqrt{\la_m}v_m(k\tau) \\
\mbox{ with, {\em e.g.},}&& \nonumber\\
 y(\ct) &=& \frac{\ct_{eq}}{ \ct+\ct_{eq}} ~~\mbox{ or }~~~ 
	y(\ct) =\exp(-\ct/\ct_{eq})~,
\eea 
or some other suitable interpolation function.
 The effect of the radiation dominated early state of the
universe is relatively unimportant for the scales of interest for CMB 
anisotropies and linear gravitational clustering
The difference between the pure matter era result and the
interpolation is very small~\cite{DKM}.
This seems to be quite
different for cosmic strings where the fluctuations in the radiation
era are about twice as large as those in the matter era~\cite{Paul}.
The radiation dominated era has, however, little effect on the key feature
of CMB anisotropies from topological defects; namely the absence of 
acoustic peaks.

In models with cosmological constant, there is actually a second break
of scale invariance at the matter--$\La$ transition. There one can
proceed in the same way as outlined above. Since defects cease to
scale and disappear rapidly in an exponentially expanding universe,
the eigenvalues for the $\La$ dominated universe all vanish. 

\begin{figure}[ht]
\centerline{\psfig{figure=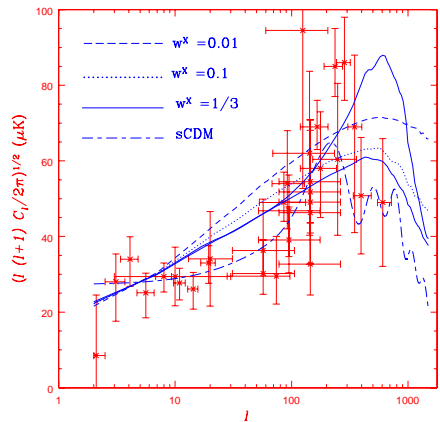 ,width=6cm}}
\caption{\label{stringcl}
 The angular CMB power spectrum of cosmic strings, from a simulation
by Contaldi,  Hindmarsh and  Magueijo~\cite{CHM}.
The figure shows the result for strings decaying into a fluid obeying the
equation of state $p=w\rho$. with $w=1/3,~0.1,~0.01$ respectively. 
The old data overlaid to the graph is to be ignored.}
\end{figure}

In Fig.~\ref{stringcl} we show the CMB power spectrum from cosmic strings.
Predictions from different research groups working on the 
subject~\cite{Aetal,Avel,CHM,Ma,Nathalie,PoVa} 
agree only partially. Contrary to global defects models, there seems to be a 
broad peak at rather high $\ell\sim 400$ to $500$. The height of the peak
is very much a matter of debate. In some work it is completely 
absent~\cite{Aetal}, while in other it is quite high~\cite{CHM,PoVa}. We can
maybe understand this difference between global and local defects
as being due to the larger difference between the source functions
in the radiation and matter dominated era: For the global $O(N)$
defects, the ratio between the sources in the two epochs is about $1.2$,
while for strings it is rather in the vicinity of $4$~\cite{WASA}.
This difference can  explain the  more prominent peak
at high $\ell$ for cosmic strings. An additional intriguing difference
to global defect is the strong domination of the energy density
over the rest of the energy momentum tensor (see Fig.~\ref{em_string}).
The dependence of the spectrum on the equation of state of the decay product,
$p_X=w_X\rho_X$ is remarkable.

A feature which
the two classes of defects have in common is decoherence which
smears out all secondary peaks. Furthermore, cosmic strings seem to suffer
as well from insufficient power on large scales in the dark matter.
The clear sequence of acoustic peaks recently observed~\cite{Net,DASI,Mnew}
rules out any seeds with strongly non-linear evolution, such as $O(N)$
(with low $N$), both for global and local gauge theories, as well as
 any other non-linear mechanism for seeding cosmic perturbations!

\subsection{non-Gaussianity}

An interesting difference between structure formation with topological 
defects and inflationary models is also that the latter have generically
Gaussian perturbations while the former don't. Even if the defect energy 
momentum tensor would be Gaussian initially, non-linear evolution induces
non-Gaussianities. (Recently, however, a model leading to Gaussian 
fluctuations from cosmic defects has been investigated~\cite{AM}.)

Even if the fields themselves would be Gaussian (which they are not), 
the energy momentum tensor, which is quadratic in the fields, would obey a
$\chi^2$-distribution.
But the $\si$-model condition $\sum_{i=1}^N(\b_i)^2=1$
cannot be satisfied if the fields $\b_i$ are Gaussian random variables.
In the large $N$ model, the $O(N)$-model in the limit
$N\ra \infty$, the energy momentum tensor is a infinite sum of variables
$\dd_\mu\beta_i\dd_\nu\beta_i$ which all obey the same distribution. This 
large $N$ model is nearly Gaussian as a consequence of the central limit 
theorem. The fact that the variables $\b_i$ have to satisfay the 
normalization condition however implies that the variables 
$\dd_\mu\beta_i\dd_\nu\beta_i$ are not statistically independent.

Clearly, $O(N)$-models with $N>4$ components do not lead to topological 
defects in $3$-dimensional space, but in the large $N$-limit the equations 
of motion can be solved analytically and, as we have already seen, the 
model is very useful to study certain features of $O(N)$ defects. The fact
that non-Gaussianity becomes weaker as $N$ becomes larger, indicates that 
cosmic strings are probably the most non-Gaussian defects and textures are
the least non-Gaussian.

For defects to be non-Gaussian,
means that their energy momentum tensor does not obey 
Gaussian statistics. Therefore certain reduced higher oder moments 
do not vanish. It is not evident how to find the best observable to 
locate the non-Gaussianity. Some suggestions for variables which might 
be useful in the case of defects have been proposed in Refs.~\cite{FM,GPW}.
Furthermore, a given observational 
variable like,\eg, the integrated Sachs-Wolfe effect, may be
the sum of many non-Gaussian but equally distributed contributions
and hence  be very closely Gaussian due to the central limit theorem.

Unfortunately, it is not known how strong non Gaussian features are
in the CMB or in the dark matter distribution of topological defect models. It 
is also not clear on which scales they are strongest. Due to the arguments 
indicated above, we expect cosmic strings to be most non-Gaussian. A well 
distinguished non-Gaussian feature in CMB anisotropies from cosmic strings 
is the Kaiser-Stebbins effect~\cite{KS}. This is a disconinuity in the 
CMB temperature due to a moving string between the observer and the CMB.
Even though this effect is easily obtained analytically for a straight 
cosmic string, it is difficult to get a handle on it in a string network 
of many bent and twisted strings which also contain small scale 
structure. One expects the Kaiser-Stebbins effect to be relevant for the 
anisotropies on about arc-minute scale.

Old partial results on non-Gaussianity from numerical simulations 
(see \eg.~\cite{PST2,ZD}) are probably strongly affected by finite size 
effects and not very reliable. On the other hand,
with the more successful method to compute power spectra just by 
determining the unequal-time two point distribution of the defects, one loses
all information about higher order correlations and thus about non-Gaussianity.
Since the known defects are such a bad fit to the present CMB data, 
nobody has been sufficiently motivated to study and solve the difficult 
problem of the non-Gaussianity which defects may induce in the CMB 
anisotropies or in the matter distribution. A semi-analytic study of the 
bi-spectrum, the third moment of CMB anisotropies for cosmic strings can be 
found in Ref.~\cite{GPW}.

 We know that also non-linear Newtonian clustering induces 
non-Gaussianities. The reduced $n$-point functions in the matter distribution
due to non-linear clustering scale like $ D^{2n-2}$, where $D$ is the matter 
density perturbation amplitude (see \eg~\cite{Fry}). In contrary, a 
non-Gaussianity already present at the linear level simply scales like $D^n$ 
in the $n$-point function~\cite{DJKU}. Therefore, initial non-Gaussianities, 
like those of defects are best detected on large scales, where 
non-linearities are negligible since $D\ll 1$. 
On small scales, where non-linearities are important, the non-Gaussianity 
from Newtonian clustering is stronger than the one form the initial conditions.
Therefore, the fact that non-linearities in the observed galaxy 
distribution closely follow the behavior expected from Newtonian clustering, 
does not constrain topological defects very strongly.

\subsection{CMB anisotropies and polarization}
The $C_\ell$'s for the global texture model are shown in 
Fig.~\ref{fig17}, bottom panel. 

\begin{figure}
\centerline{\psfig{figure=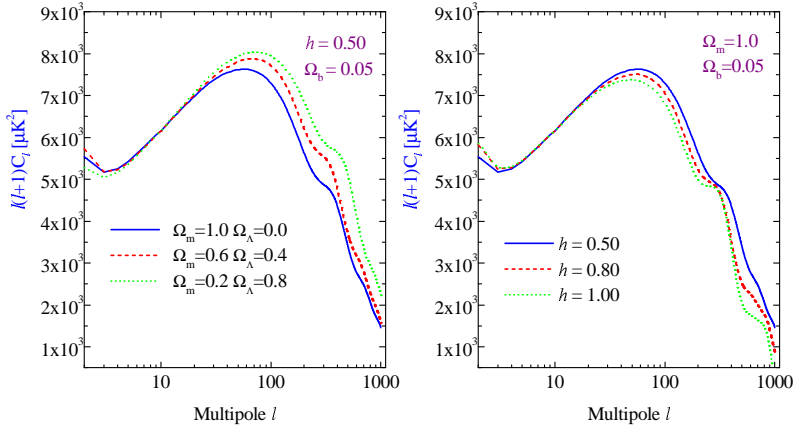,width=12cm}}
\caption{\label{fig23}
The $C_\ell$ power spectrum is shown for different values of
cosmological parameters. In the right panel we choose $\Om_\La=0$,
$\Om_{CDM}=0.95$, $\Om_b=0.05$ and vary $h$. In the left panel we
fix $h=0.5$, $\Om_b=0.05$ and vary  $\Om_\La$. We only consider
spatially flat universes, $\Om_0=1$.}
\end{figure}

Vector and tensor modes are of the same order as 
the scalar component at COBE-scales. For the 'standard' texture model 
one has 
$C_{10}^{(S)} : C_{10}^{(V)} : C_{10}^{(T)} \sim  0.9 : 1.0 : 0.3$.
These results were found numerically and by  analytical 
arguments in Refs.~\cite{PST,Aetal,ABR,DKM} and \cite{DK}.
Due to tensor and vector contributions, even assuming perfect coherence 
(see Fig.~\ref{fig17}, top panel), the total power spectrum does not  
increase from large to small scales.
Decoherence leads to smoothing of  oscillations in the power
spectrum at small scales and the final spectrum has a smooth
shape with a broad, low isocurvature 'hump' at $\ell\sim 100$ and a
small residual of the first acoustic peak at $\ell \sim 350$.
There is no structure of peaks at small scales. The power
spectrum is well fitted by the following fourth-order 
polynomial in $x= \log \ell$:

\be {\ell ( \ell +1 ) C_{\ell}\over 110C_{10}} = 1.5 -2.6 x +
3.3 x^2 -1.4 x^3 +0.17x^4 ~.\ee

The effect of decoherence is less important
for the large-$N$ model, where oscillations and peaks are still visible
 (see Fig~\ref{fig19}, bottom panel). As argued before, this is due to 
the fact that 
the non-linearity of the large-$N$ limit is  only in the quadratic 
energy momentum tensor.  Since decoherence is inherently due to
non-linearities, we expect it to be stronger for lower values of $N$. 

In Fig.~\ref{fig23} we plot the global texture $C_{\ell}$ power spectrum 
for different choices of cosmological parameters. The  variation of 
parameters leads to  similar effects like in the inflationary case,
but with  smaller amplitude.
At small scales ($\ell \ge 200$), the $C_\ell$'s tend to decrease with
increasing $H_0$ and they increase when a cosmological
constant $\Omega_{\Lambda} = 1 - \Omega_m$ is introduced.
Nonetheless, the amplitude of the anisotropy power spectrum at
high $\ell$s remains in all cases on the same level like the one at
low  $\ell$s, without showing the substantial peak found in 
inflationary models and in the data.
The absence of acoustic peaks is a stable prediction
of global $O(N)$ models.
 The models are normalized to the full CMB data set,
 which leads to slightly larger values of the normalization 
parameter than pure COBE normalization.
An avarage value for different choices of cosmological parameters is
$\ep=4\pi G\eta^2\sim 1.6$

We compare the texture results with the three best current experimental
data sets from Boomerang-98~\cite{Net}, Maxima~\cite{Mnew} and 
DASI~\cite{DASI}. We take in to account the calibration
uncertainty of $10 \%$ for Boomerang and of $5 \%$ for Maxima
and DASI. We also include the COBE dataset using Lloyd Knox's 
RADPack packages~\cite{Knox}.

\begin{figure}
\centerline{\psfig{figure=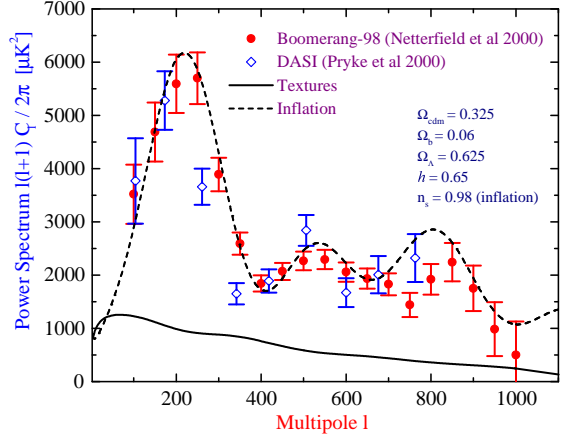,width=10cm, angle=-90}}
\vspace*{-2.5cm}
\caption{\label{fig24}
The $C_\ell$ spectrum obtained in the standard texture model
is compared with the recent BOOMERANG and DASI data.
A standard inflationary model, which gives a much better fit to
the data, is also plotted.}
\end{figure}

No matter how precisely an experiment will measure the CMB sky,
the result will be always affected by the {\it intrinsic} statistics of
the perturbations, {\it i.e.} the cosmic variance.
In the recovery of the power spectra of the above experiments,
the temperature distribution is often assumed to be Gaussian. 
Deviation from Gaussianity could lead to an enhancement of the
cosmic variance, which can be as large as a factor of $7$ (see \cite{xiao}).
Inevitably therefore, the quoted error-bars in the CMB power spectrum 
could be in principle underestimated by the assumption of the
Gaussian statistic.
This effect is, however only relevant for relatively low $\ell$s.
Furthermore, preliminarly statistical analysis of the Maxima 
map~\cite{wu} have found no-evidence for non-Gaussianities. Keeping this 
caveat in mind, but missing a more precise alternative, we indicate the 
minimal, Gaussian error calculated according to the published data.
The numerical seeds taken from~\cite{DKM} are assumed to be about 10\% 
accurate. In Fig.~\ref{fig24} we plot the Boomerang and DASI 
data  together with the   theoretical predictions for a texture model 
with $h=0.65$,$\Omega_{\Lambda} = 0.625$, $\Omega_{cdm} = 0.315$ and 
$\Omega_{b} = 0.06$, and the corresponding
inflationary model with $n_s=0.98$ which gives a best fit to the data.
All of the experiments detect a very clear first peak at 
$\ell \approx 200$ which is incompatible
with all $O(N)$ models~\cite{DGS,PST,SPT,DKM} as well as with the 
cosmic string results~\cite{CHM,PoVa,Nathalie}
In Table~\ref{tabeCl}, we report the $\chi^2$ values from a comparison 
of the CDM-texture model with each  experiment, separately.
As we can see, while in reasonable agreement with the large angular
scale data from the COBE experiment, the new data at intermediate
and small angular scales from BOOMERanG, DASI and MAXIMA rule
out the model with high significance.

In addition, the data shows indications of a series of peaks
pointing to coherent acoustic oscillations. If the evidence
for this peak sequence strenghtens once new data becomes available,
it will provide strong evidence against non-linear mechanisms
for the generation of perturbations. Furthermore, it will
enable us to place stringent limits on the relative amounts of
isocurvature and adiabatic contributions to the perturbations.
We present a more general discussion in the next chapter.

\begin{table}
\begin{center}
\begin{tabular}{||c|c|c||}
\hline
Experiment & Data Points & $\chi^2$\\
\hline
COBE        & 24 &  28 \\
COBE+B98    & 43 & 317 \\
B98         & 19 & 221 \\
COBE+MAXIMA & 36 & 119 \\
MAXIMA      & 12 &  32 \\
COBE+DASI   & 33 & 246 \\
DASI        &  9 &  75 \\
\hline
\end{tabular}
\end{center}
\vspace{0.3cm}
\caption{\label{tabeCl}
Topological defects versus CMB data. While in reasonable agreement with
the COBE data, the power spectrum inferred from textures is in strong 
disagreement with the present intermediate and small angular scale 
observations. The BOOMEranG data alone, in particular, rule out 
the model at extremely high significance.}
\end{table}

\begin{figure}
\centerline{\psfig{figure=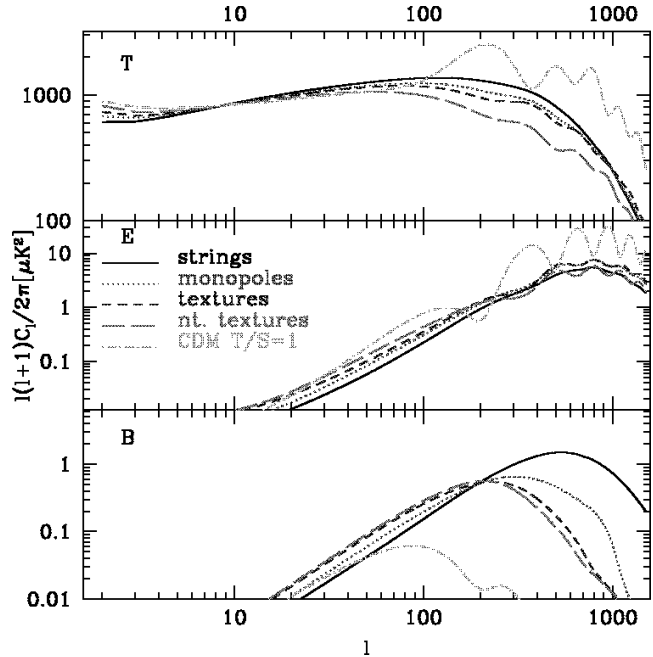,width=8.5cm}}
\caption{\label{polfig}Power spectra of temperature (T), electric type 
polarization (E)
and magnetic type polarization (B) for global strings, monopoles, textures
and  an $O(N)$ scalar field with $N>4$ are shown. For comparison, 
also  the corresponding spectra in a standard CDM model with $T/S=1$ (which 
maximizes the B component) is plotted. All defect models predict 
a much larger component of B polarization on small angular scales 
(from Seljak, Pen and Turok~\cite{SPT}).}
\end{figure}

Finally, we compare $E$- and $B$-type polarisation induced by defects 
with the result of an inflationary model with equal amplitude of scalar 
and tensor temperature fluctuations in Fig.~\ref{polfig}~. Due to the 
amount of vector perturbations present, it is not
surprising that $B$-type polarization has significantly higher amplitude for
defects. Furthermore, the isocurvature  shift of the 'acoustic peaks' and
decoherence are also visible in the polarisation signal.
 
\subsection{Matter power spectra}
The first serious attempts to compute the matter distribution in scenarios 
with topological defects were done for cosmic strings~\cite{MeSc}.
But due to the problems inherent in cosmic string simulations, the accuracy 
of these results may well be less that a factor of two or three.

The situation is different for global defects. There, the first accurate
matter spectra have been computed in Ref.~\cite{PST}. This work, which 
underestimated the vector and tensor contribution to CMB anisotropies, 
still had the wrong normalization. This was corrected later 
in~\cite{ZD,PST,DKM}. For a COBE mormalized global texture model, the total 
mass  fluctuation $\sigma_R$ in a ball of radius $R=8h^{-1}$Mpc,
is about $\sigma_8 = (0.44 \pm 0.07)h$
(the error coming from the CMB normalization) in a flat universe
without cosmological constant (see Refs.~\cite{PST,DKM}.
From the observed  cluster abundance, one infers 
$\sigma_8 = (0.50 \pm 0.04) \Omega^{-0.5}$ \cite{eke} and
$\sigma_8 = 0.59^{+0.21}_{-0.16}$ \cite{lid}.  These results,
 which are obtained with the Press-Schechter formula, assume Gaussian 
 statistics. We thus have to take them with a
grain of salt, since we do not know how non-Gaussian fluctuations on
cluster scales are in the texture model. According to Ref.~\cite{free}, the 
Hubble constant lies in the interval
$h\simeq 0.73 \pm 0.06 \pm 0.08$. Hence,
in a flat CDM cosmology, taking into account the  uncertainty of
the Hubble constant, the texture scenario predicts a reasonably consistent 
value of $\sigma_8$.

However, as discussed in Refs.~\cite{ABR}, \cite{PST} and \cite{DKM},
unbiased global texture models are unable to reproduce
the power of galaxy clustering at very large scales,
 $\gsim 20 h^{-1}$ Mpc.
In order to quantify this discrepancy we compare the prediction of the linear
matter power spectrum with the decorrelated linear power spectrum of
the PSCz survey from~\cite{TegHam2000} in Fig.~\ref{matter}.
Clearly, the textures model without a cosmological constant
provides a bad fit to the large scale data.
Including a cosmological constant improves the agreement between
the shape of the theoretical spectrum and the data. However,
the required bias is too high, $b \sim 6$ and the
model fails to match the observed value of $\sigma_8$.
We further test this discrepancy by comparing the theoretical 
predictions with the
results from a wide number of infrared (Refs.~\cite{fish},\cite{tadr}) and 
optically-selected (Refs.~\cite{daco}, \cite{lin})
galaxy redshift surveys, and with the real-space power
spectrum inferred from the APM photometric sample (Ref.~\cite{baugh}). 
All the results are in rough agreement with
the PSCz comparison: For $h=0.5$ and $\Om_\La=0.8$ the shape of the 
texture power spectrum fits the data well, but the bias is very high. The 
only exception is the APM data, the shape of which cannot be fitted with 
the texture spectrum. In  Table~\ref{tabpow} we report
  for each survey the best cosmological parameter $(h, \Om_\La)$ within the
limited ones which have been analysed,
and the best value of the bias parameter obtained by $\chi^2$-minimization. 
We also indicate the value of $\chi^2$ (not divided by the number of 
data points) and the number of data points. The bias parameter
strongly depends on the data considered. This is not surprising, since
also  the catalogs are biased relative to each other.
A more detailed discussion of this analysis is found in Ref.~\cite{DKM}.

\begin{figure}
\centerline{\psfig{figure=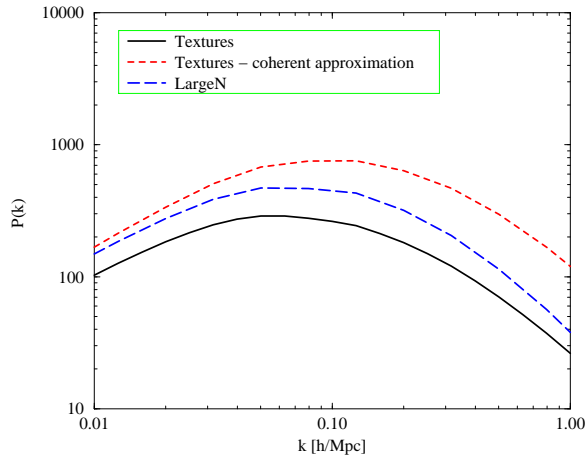,width=8.5cm}}
\caption{\label{fig22}
The dark matter power spectrum for the texture model (solid line) is
compared with the coherent approximation (short dashed) and the
large-$N$ limit (long dashed). The spectra are COBE normalized and the
cosmological parameters are $\Om_\La=0~,~h=0.5$.}
\end{figure}

\begin{figure}[ht]
\centerline{\psfig{figure=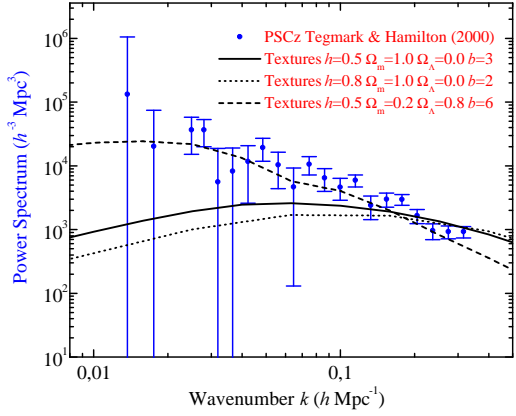,width=9.5cm, angle= -90}}
\vspace*{-2cm}
\caption{\label{matter} The
matter Power spectrum: comparison between the PSCz data and theory.}
\end{figure}

The power spectra for the large-$N$ limit and for the coherent
approximation are typically a factor 2 to 3 higher (see Fig.~\ref{fig22}), and
the biasing problem is  alleviated for these cases. For
$\Om_\La=0$ we find $\si_8=0.57h $ for the
large-$N$ limit and  $\si_8=0.94h$ for the coherent approximation.
This is no surprise since only one source function, $\Psi_s$, the
analog of the Newtonian potential, seeds dark matter fluctuations and
thus coherence always enhances the unequal time correlator.
 The dark matter Greens function
is not oscillating, so this enhancement translates directly into
the power spectrum. 

Models which are anti-coherent in the sense defined in Section~IID
reduce power on Sachs-Wolfe scales and enhance the power in the dark
matter. Anti-coherent scaling seeds are thus the most promising
candidates which may cure some of the problems of global $O(N)$ models.

The analysis described here does not take into account
the effects of non-linearities and redshift distortions.
Redshift distortions in the texture case should be less important
than in the inflationary case since the peculiar velocities are rather
low (see next paragraph).
 Non-linearities typically set in at $k \ge 0.5h$Mpc$^{-1}$ and should not
have a big effect on our main conclusions which come from much larger
scales.  Inclusion of these corrections will
result in more small-scale power and in a broadening of the spectra,
which even enhances the conflict between models and data.
Furthermore, variations of other cosmological parameters, like
the addition of massive neutrinos, hot dark matter, which are not 
considered here,
will result in a change of the spectrum on small scales but will not
resolve the  discrepancy at large scales.

\begin{table}
\begin{center}
\begin{tabular}{||c|c|c|c|c|c||}
\hline
Catalog & $h$ & $\Omega_{\Lambda}$ & Best fit bias $b$& $\chi^2$&Data points\\
\hline
 CfA2-SSRS2 101 Mpc& 0.5 & 0.8 & 9.5 &  4 & 24\\
 CfA2-SSRS2 130 Mpc& 0.5 & 0.8 & 11.1 &  4 & 19\\
 LCRS  & 0.5 & 0.4 & 3.7 &  33 & 19\\
 IRAS  & 0.5 & 0.8 & 6.3 &   9 & 11\\
 IRAS 1.2 Jy & 0.5 & 0.8 & 6.7 &   28 & 29\\
 APM & 0.5 & 0.0 & 3.3 & 1350 & 29\\
 QDOT & 0.5 & 0.8 & 7.3 & 14 & 19\\
\hline
\end{tabular}
\end{center}
\vspace{0.3cm}
\caption{\label{tabpow}
Analysis of the matter power spectrum. In the first column the
catalog is indicated. Cols.~2 and 3 specify the model parameters. In
cols.~4 and 5 we give the bias parameter inferred by $\chi^2$
minimization as well as the value of $\chi^2$. Col.~6 shows the number
of 'independent' data points assumed in the analysis.}
\end{table}

\subsection{Bulk velocities}

To confirm the missing power on 20 to 100$h^{-1}$Mpc,
we also consider the velocity power spectrum which is not plagued by the 
biasing problems of the galaxy distribution. The
assumption that galaxies are fair tracers of the velocity field seems
to us much better justified, than to assume that they are fair tracers
of the mass density. We therefore test the global texture  model against 
bulk velocity data (see also~\cite{DKM}). We
use the data by Ref.~\cite{deke} which gives the bulk flow
\be \si^2_v(R) = {H_0^2\Om_m^{1.2} \over {2 \pi^2}} \int P(k)W(kR)dk ~,\ee
in spheres of radii $R = 10$ to $60 h^{-1}$Mpc. 
These data are derived after reconstructing the $3-$dimensional
velocity field with the POTENT method (see~\cite{deke} and references 
therein).

As shown in Table~\ref{tabevl}, the COBE normalized texture 
model predicts too low velocities on large scales when compared with
POTENT results.
More recent measurements of the bulk flow lead to somewhat lower estimates
like $\sigma_{v}(R) \sim (230 \pm 90)$ at $R = 60 h^{-1}$Mpc
(see Ref.~\cite{giova}), but  a
discrepancy of about a factor of $2$ in the best case remains.
Including a cosmological constant helps at large scales, but
decreases the velocities on small scales. 
Similar results are obtained in models with cosmic strings. There, however, 
details depend on assumption about the fluid into which string loops decay.
For loops decaying into radiation, models with cosmic strings have
even less power on $50h^{-1}$Mpc.

\begin{table}
\begin{center}
\begin{tabular}{||c|c|c|c|c|c||}
\hline
R & $\sigma_v$ (R) & $\Delta_v$ & $h=0.5$& $h=1.0$& $\Omega_{\La}=0.8$\\
\hline
 10 & 494 & 170 & 145 & 205&  86 \\
 20 & 475 & 160 & 100 & 134&  78 \\
 30 & 413 & 150 &  80 &  98&  70 \\
 40 & 369 & 150 &  67 &  78&  65 \\
 50 & 325 & 140 &  57 &  65&  61 \\
 60 & 300 & 140 &  50 &  56&  57 \\
\hline
\end{tabular}
\end{center}
\vspace{0.3cm}
\caption{\label{tabevl}
Bulk velocities: Observational data from \protect\cite{deke} and theoretical
predictions. $\De_v$ estimates the observational uncertainty.
The uncertainties on the theoretical predictions are
around $\sim 30 \%$. The models $\Om_\La=0$ with $h=0.5$ and $h=1$ as well
as $\Om_\La=0.8,~h=0.5$ are given.}
\end{table}

This comparison of models with data shows that $U(1)$ cosmic strings and 
global $O(N)$ models are clearly ruled out by present large scale structure 
data. In the next chapter we shall study whether generalizations or mixed 
models may survive.
\clearpage
\section{Generalizations}
\label{Gen}

So far we have studied some specific models of topological defects and
we have found that they cannot reproduce the observed large scale
structure of the universe. Even if they basically lead to a
Harrison--Zel'dovich spectrum on large scales, the 'details' just do
not fit.  The isocurvature nature of the induced perturbations
together with the high amplitude tensor and especially vector contributions,
lead to relatively high amplitude of fluctuations on very large scales. 
Therefore, the COBE--normalized acoustic 'peaks' and the dark matter power 
spectrum typically are too low.  In addition, decoherence smears out the 
anyway too low secondary peaks in the CMB anisotropy
spectrum. 

In this chapter we first study the question whether this feature is
generic or just happens to occur in the specific models which we have
analyzed in the preceding chapters. Then we want to investigate
whether mixed models with inflationary perturbations and a defect
component might fit the data better than pure inflationary
perturbations.

\subsection{Generic properties of the unequal time correlation functions}

 We start by deriving some generic properties of power spectra
induced by topological defects. As we have seen in Chapter~2, the
resulting perturbation power spectra are fully determined by the
un-equal time correlators of the defect energy momentum tensor.

The main properties  which we shall use to
describe the defect correlators are causality, statistical homogeneity and 
isotropy and scaling. Since the nature of the defects resulting as
a 'topological relicts' from a phase transition does not enter directly
in this analysis, we speak of 'scaling causal seeds', seeds denoting any
non-uniformly distributed form of energy which provide a 
perturbation to the homogeneous background fluid.
In first order perturbation theory they
evolve according to the unperturbed (in general non-linear) equations
of motion. For simplicity, we assume the seeds to be coupled to the cosmic
fluid only via gravity. A counter example to this are the $U(1)$ cosmic
strings discussed before. As we have seen, the string CMB anisotropy power 
spectrum, especially the height of the acoustic peak, depends very
sensitively on the details of the coupling of string seeds to
matter~\cite{CHM,Nathalie}.
For uncoupled seeds the energy momentum tensor is covariantly 
conserved. To determine power spectra or other 
expectation values which are quadratic
in the cosmic perturbations, we just need to know the unequal
time correlation functions of the seed energy momentum tensor~\cite{DK,DKM},
\be
\langle \Th_{\mu\nu}(\bk,\ct) \Th^*_{\si\rho}(\bk',\ct')\rangle = 
	M^4\widehat{C}_{\mu\nu\si\rho}(\bk,\ct,\ct')\de(\bk-\bk')~,
\ee
where $M$ is a typical energy scale of the seeds ({\it e.g.} the
symmetry breaking scale for topological defects), which determines the
overall perturbation amplitude. Here we
 define Fourier transforms with the normalization
\bea 
\hat{f}({\bf k})&=&{1\over \sqrt{V}}\int d^3x f({\bf x})\exp(i{\bf kx}) ~;\\
	f({\bf x}) &=& {\sqrt{V}\over (2\pi)^3}\int d^3k \hat{f}({\bf k})
	\exp(-i{\bf kx})~.
\eea
Seeds are {\em causal}, if
\[
C_{\mu\nu\si\rho}({\bf x},\ct,\ct') \equiv {1\over M^4} \langle 
 \Th_{\mu\nu}(\by,\ct) \Th^*_{\si\rho}(\by+\bx,\ct')\rangle
\]
 vanishes for $|{\bf x}|> \ct+\ct'$; and
they are {\em scaling}, if $\widehat C$ depends on no other dimensional 
parameter than \bk, $\ct$ and $\ct'$. 
 From scaling we 
conclude that for purely dimensional reasons, we can write the
correlations functions in the form
\be
 \widehat{C}_{\mu\nu\la\rho}({\bf k},\ct,\ct') =
 {1\over\sqrt{\ct\ct'}}F_{\mu\nu\la\rho}(\sqrt{\ct\ct'}\cd{\bf k},\ct'/\ct)~,
\ee
where $F_{\mu\nu\la\rho}$ is a dimensionless  function of the four variables
 $z_i\equiv\sqrt{\ct'\ct} k_i$ and $r\equiv \ct'/\ct$, which is analytic 
in $z_i$.

We also require the seed to decay inside the horizon, which
implies
\be \lim_{{k\ct\ra \infty}}\widehat{C}_{\mu\nu\la\rho}({\bf k},\ct,\ct')= 
  \lim_{{k\ct'\ra \infty}}\widehat{C}_{\mu\nu\la\rho}({\bf k},\ct,\ct')= 0~.
\label{decay}\ee
Furthermore,
since the seeds interact with the cosmic fluid only gravitationally, $\Th$ 
satisfies covariant energy momentum conservation,
\be \Th^{\mu\nu}_{;\nu} =0 ~. \ee
With the help of these four equations, we can, for example, express the 
temporal components, $\Th_{0\mu}$ in terms of the spatial ones, $\Th_{ij}$.
The seed correlations are thus fully determined by the spatial 
correlation  functions $\widehat{C}_{ijlm}$. Statistical isotropy, scaling and
symmetry in $i,j$ and $l,m$ as well as under the transformation 
$i,j;k;\ct \ra l,m;-k;\ct'$
require the following form for the spatial correlation functions:
\bea
\lefteqn{\widehat{C}_{ijlm}({\bf k},\ct,\ct') =} \nonumber \\  
	&&{1\over \sqrt{\ct\ct'}}
	[z_iz_jz_lz_mF_1(z^2,r) + \nonumber \\  &&
  (z_iz_l\de_{jm}+z_iz_m\de_{jl}+z_jz_l\de_{im}
 +z_jz_m\de_{il})F_2(z^2,r) + \nonumber \\ &&
z_iz_j\de_{lm}F_3(z^2,r)/r +z_lz_m\de_{ij}F_3(z^2,1/r)r +
\nonumber\\ &&
+\de_{ij}\de_{lm}F_4(z^2,r) +
(\de_{il}\de_{jm}+\de_{im}\de_{jl})F_5(z^2,r)] ~,
\label{Cijlmansatz}
\eea
where the functions $F_a$ are  analytic in
$z^2\equiv \ct\ct' k^2$, and for  $ a \neq 3$ they are
invariant under the transformation $r\ra 1/r$, $F_a(z^2,r)=F_a(z^2,1/r)$.
The positivity of the power spectra 
$\widehat{C}_{ijij}({\bf k},\ct,\ct)=\lan|\Th_{ij}|^2\ran \ge 0$ leads to a 
series of positivity conditions for the functions $F_a$:
\bea
0 &\le& F_5(z^2,1) ~,  \label{pF5}\\
0 &\le& F_4(z^2,1)+2F_5(z^2,1) ~, \label{p2}\\
0 &\le& z^2F_2(z^2,1)+F_5(z^2,1) ~,\\
0 &\le& z^4F_1(z^2,1)+4z^2F_2(z^2,1)+3F_5(z^2,1)~,\\
0 &\le& z^4F_1(z^2,1)+2z^2(F_3(z^2,1)+2F_2(z^2,1)) \nonumber\\
	&& +F_4(z^2,1)+2F_5(z^2,1)~. \label{p5}\eea
\vspace{0.5cm}
Since $\widehat{C}_{ijlm}$ is the Fourier transform of a real function, 
\be
\widehat{C}_{ijlm}({\bf k},\ct,\ct')^* 
     =\widehat{C}_{ijlm}(-{\bf k},\ct,\ct')~.
\ee
Hence the 
ansatz (\ref{Cijlmansatz}) implies that the functions $F_a(z^2,r)$ are
real. Furthermore, decay inside the horizon (condition (\ref{decay})) yields
\be
 \lim_{z^2r\ra\infty}F_a(z^2,r)=\lim_{z^2/r\ra\infty}F_a(z^2,r)=0~.
\label{lim} \ee
In addition, analyticity implies that the functions $F_a$ do not
diverge in the limit $z\ra 0$, thus 
\[\lim_{z\ra 0}F_a(z^2,r)=A_a(r)   \]
with 
\[ A_a(r)=A_a(1/r)
	~\mbox{ for all } a\neq 3~. \]

As an example, we show some of these functions in 
the large $N$ limit of global scalar field seeds (see Chapter~\ref{Num}). 
In Figs.~1 and 2 we present the functions  $F_5(z^2,r)$ and $F_2(z^2,r)$.
\begin{figure}[ht]
\centerline {\psfig{figure=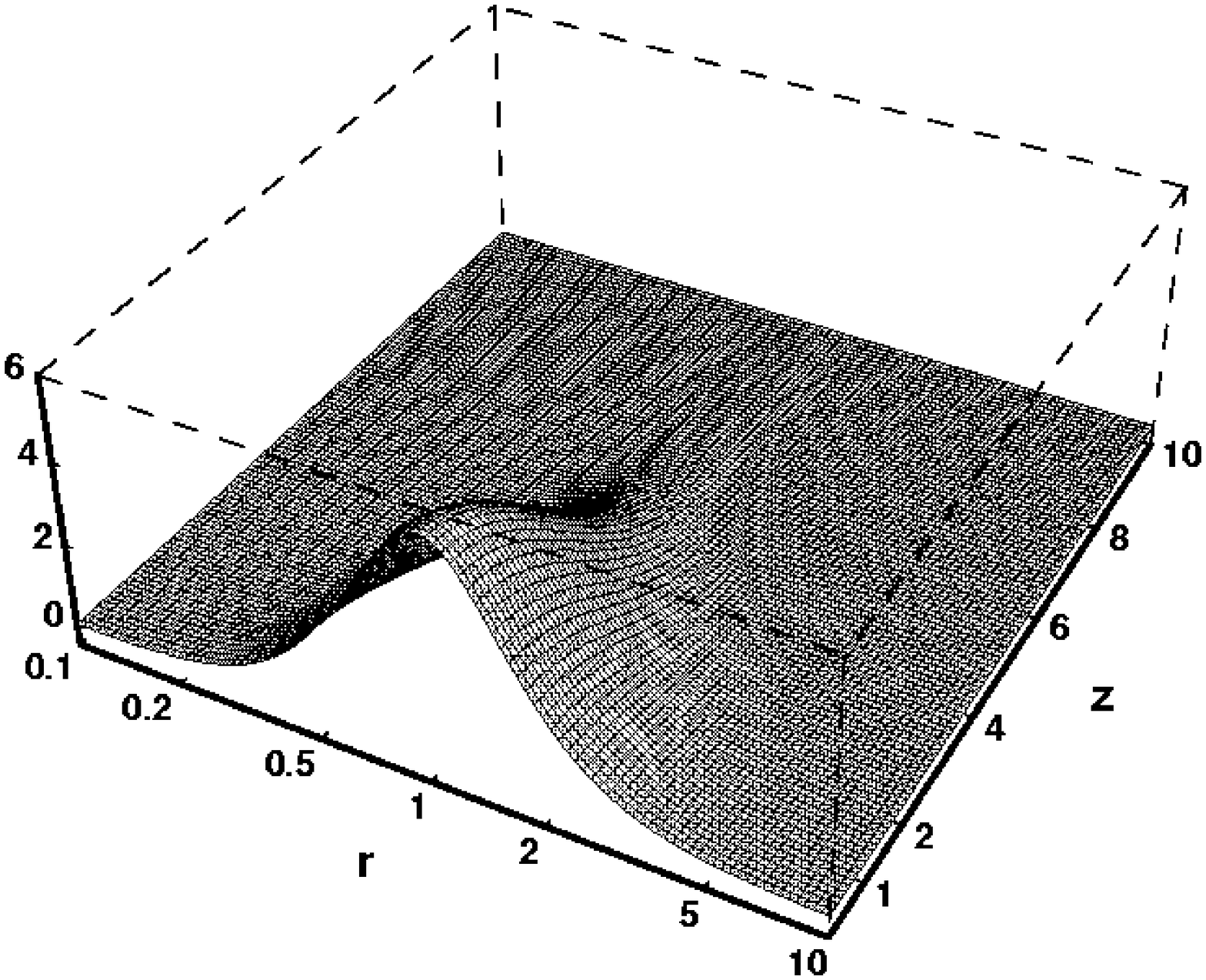,width=7cm}  
	\psfig{figure=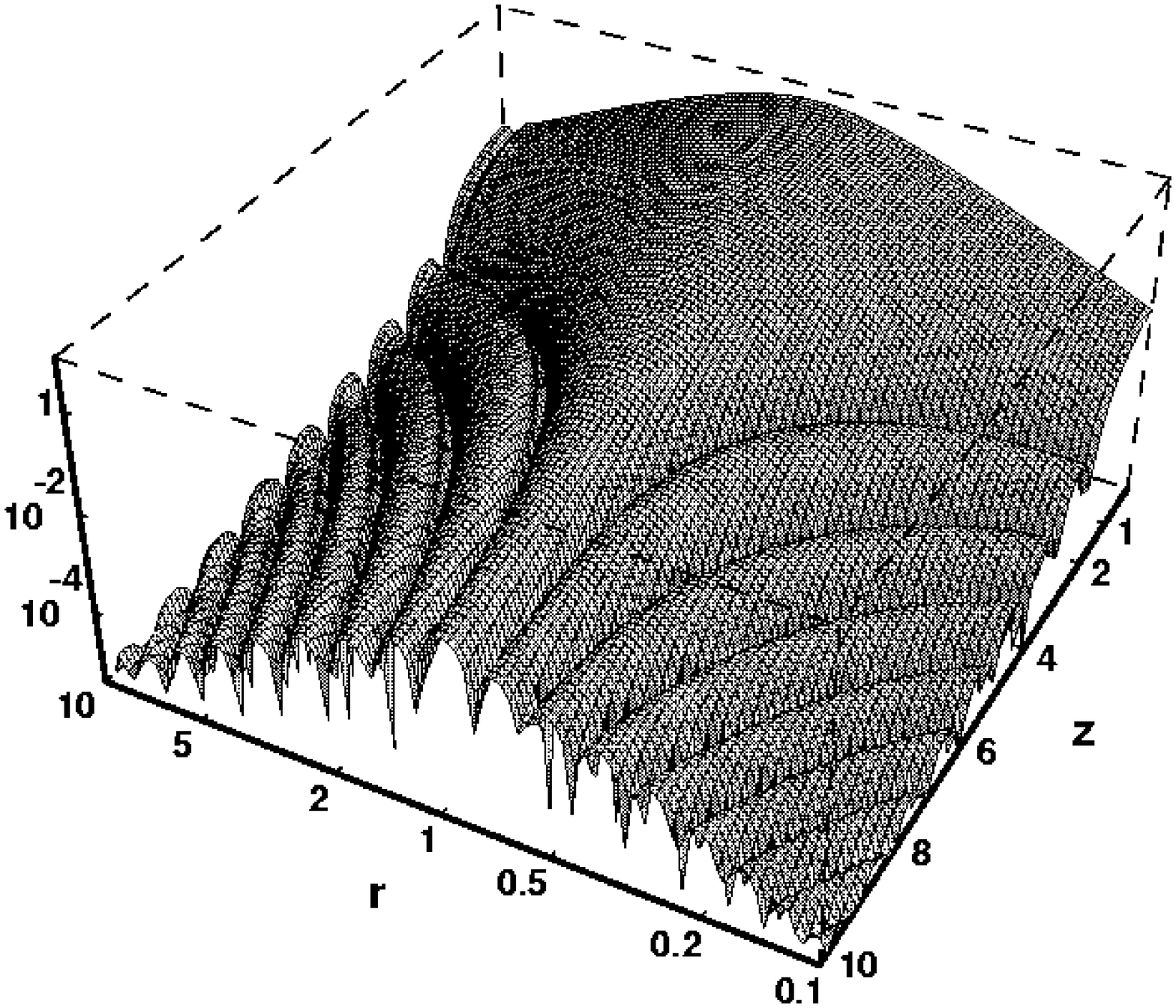,width=7cm}}
\caption{The function $F_5(z,r)$, linear (left) and $|F_5(z,r)|$
logarithmic are shown. Because of their small amplitude, the oscillations are
invisible in the linear plot.
To show the symmetry $r\ra 1/r$, the $r$-axis is chosen logarithmic in
both plots.}
\label {fig1C4}
\end{figure}
\begin{figure}[ht]
\centerline {\psfig{figure=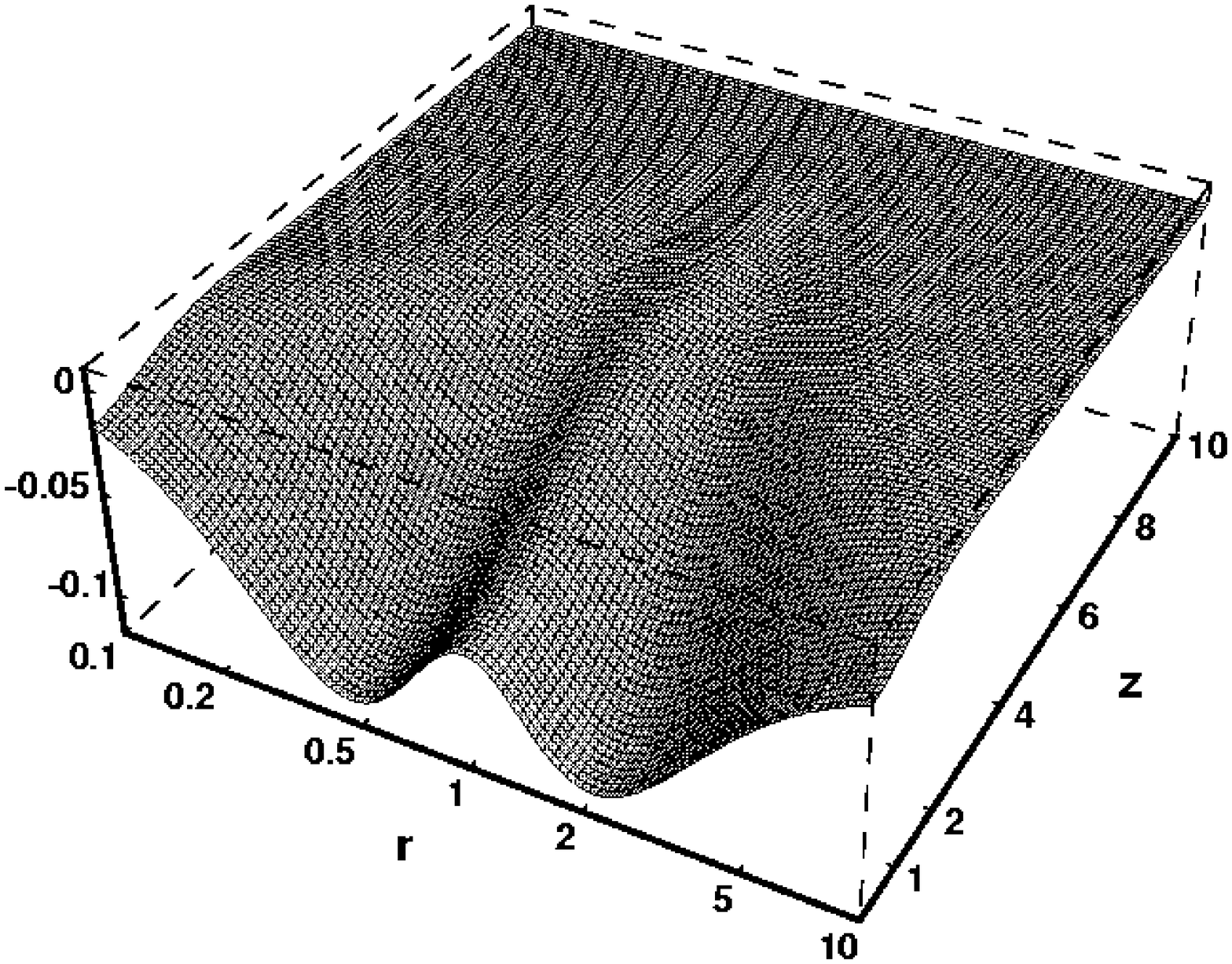,width=7cm}
   \psfig{figure=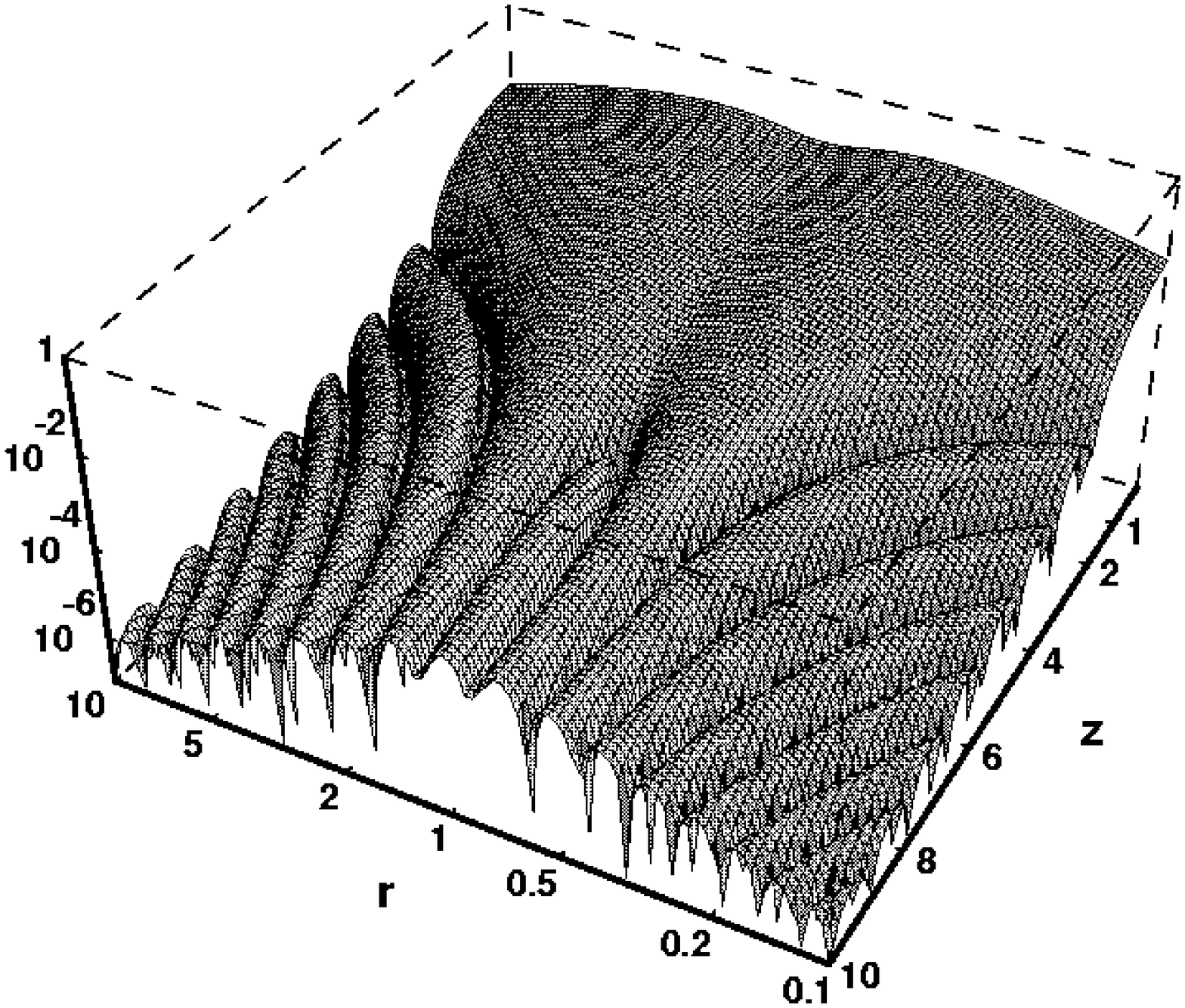,width=7cm}}
\caption{The same as Fig.~\ref{fig1C4} for the function $F_2(z,r)$.}
\label {fig2}
\end{figure}

The symmetry under the transition $r\ra 1/r$ is clearly visible. Also
the conditions that $F_a\ra 0$ if either $z\ra \infty$ or $r\ra 0$ or 
$r\ra \infty$ which follows from Eq.~(\ref{lim}) is evidently satisfied. 
For fixed $z$ the functions oscillate with a frequency which grows
with $z$. Since the amplitude decays rapidly, these oscillations are
only visible in the log-plots. The correlations always decay like
power laws, never exponentially.
 
\begin{figure}[ht]
\centerline {\psfig{figure=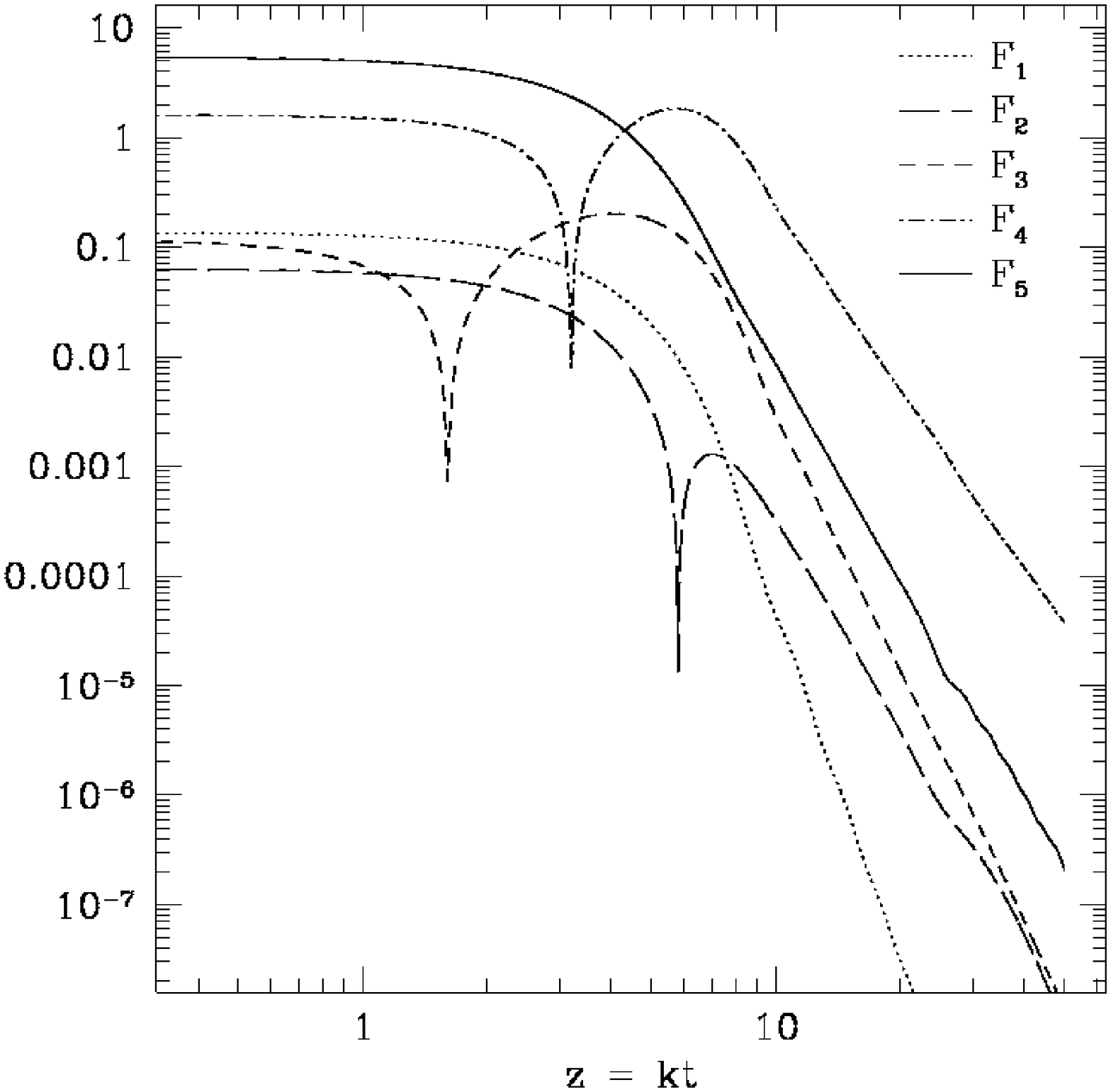,width=7cm}\quad
  \psfig{figure=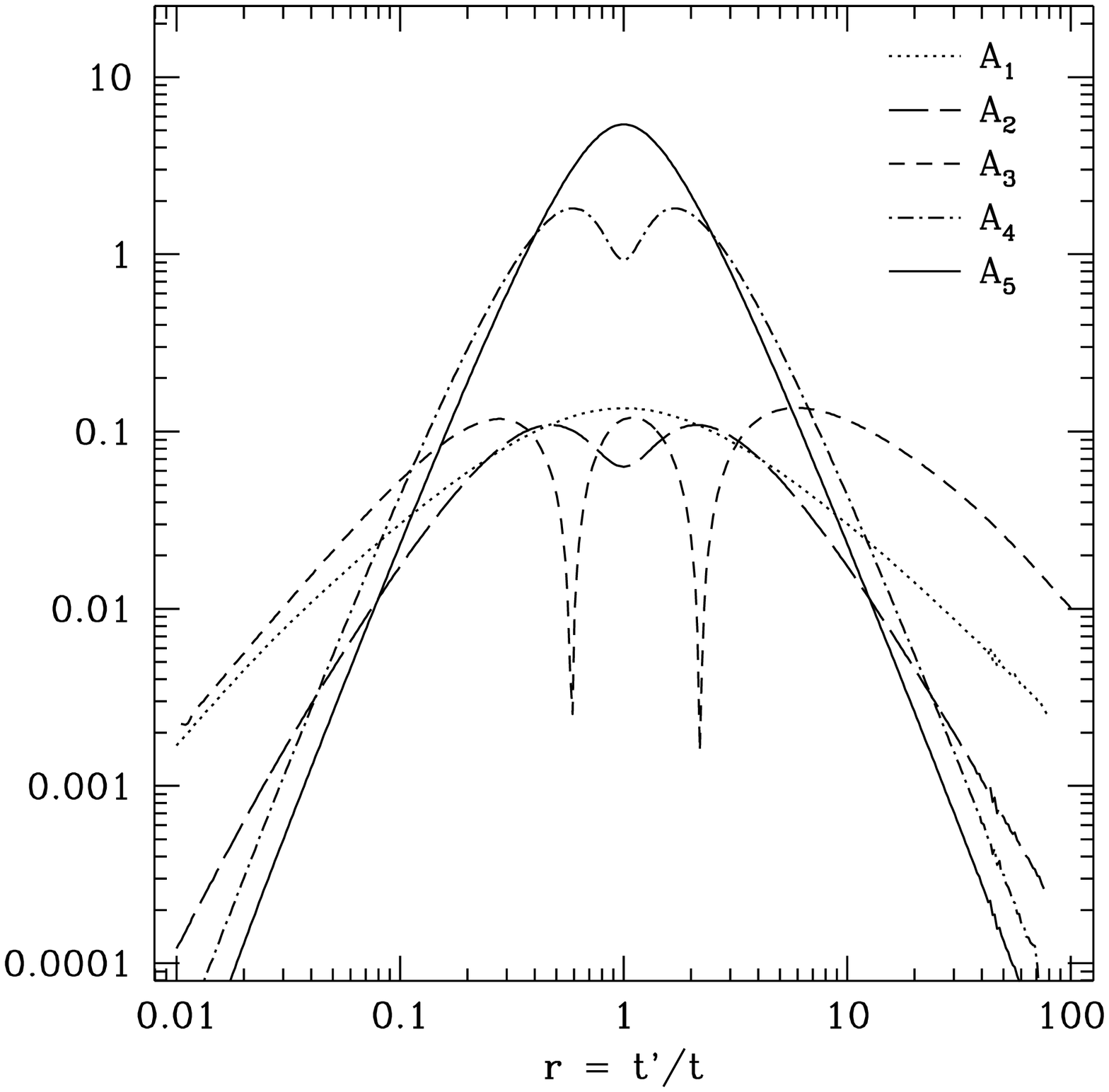,width=7.5cm}}
\caption{The functions $|F_i(z,1)|$ are shown left. The zeros are visible
as spikes in the log-plot (Further below, at $z\sim 30$, also $F_1$
passes through zero.). Right, the functions $|A_i(r)|=|F_i(0,r)|$ are shown. 
As discussed in the text, all of them except $A_3$ are symmetrical under
$r\ra 1/r$.}
\label {fig3}
\end{figure}

The equal time correlation functions, $F_1(z^2,1)$ to $F_5(z^2,1)$ 
are plotted in Fig.~\ref{fig3} left. On the right hand side we show $A_a(r)$.
All the functions, except $F_5$ which is constrained by Eq.~(\ref{pF5}) pass 
through $0$ (For $F_1$ the passage through $0$ is not visible on the plot 
since it occurs only at $z\simeq 30$). The asymptotic behavior of the  
functions can be obtained analytically.
The same is true for the functions $A_1$ to $A_5$. As argued
above, all the functions $A_i$ except $A_3$ are symmetric under $r\ra
1/r$.

\subsubsection{Scalar, vector and tensor decomposition and  CMB anisotropies}
The energy momentum tensor of seeds can be split into scalar, vector 
and tensor perturbations, since the time evolution of each of these 
components is independent. Furthermore, due to statistical isotropy,
the scalar, vector and tensor modes are uncorrelated.

We use the decomposition of $\Th_{ij}$ given in Chapter~\ref{The}, 
Eqs.~(\ref{3seed00}-\ref{Tseedjl})
The functions $F_1$ to $F_5$ determine the correlations:
To work out the correlation functions we use also Eqs.~(\ref{3fv},\ref{3fpi}) and~(\ref{2wv},\ref{3wpi}) as well as~(\ref{3tau}).
Using these identities and the ansatz (\ref{Cijlmansatz}), one easily 
verifies
\bea
\lan f_p(\ct)w_i^*(\ct')\ran &=&\lan f_\pi(\ct) w_i^*(\ct')\ran = 
	\lan f_p(\ct)\tau_{ij}^*(\ct')\ran  = 0 \\ 
 \lan f_\pi(\ct) \tau_{ij}^*(\ct')\ran &=& 
\lan w_i(\ct)\tau_{jl}(\ct')\ran = 0 \label{C0vt}\\
\lan f_p(\ct)f_p^*(\ct')\ran &=& {1\over 3\sqrt{\ct\ct'}}
	[2F_5(z^2,r)+3F_4(z^2,r)
  \nonumber \\ && + z^2(F_3(z^2,r)/r+F_3(z^2,1/r)r) + \nonumber \\
  && {4\over 3}z^2F_2(z^2,r) + {1\over 3}z^4F_1(z^2,r)]  \label{Cp}\\
\lan f_\pi(\ct)f_\pi^*(\ct')\ran &=& {1\over\sqrt{\ct\ct'}k^4}[3F_5+4z^2F_2
	+ z^4F_1]  \label{Cpi} 
\eea
\bea
\lefteqn{\lan f_p(\ct)f_\pi^*(\ct')\ran =} \nonumber\\
 &&   -\sqrt{\ct\ct'}[ {1\over 3}z^2F_1(z^2,r)
	+{4\over 3}F_2(z^2,r) 
	+rF_3(z^2,1/r)] \label{Cppi}\\
\lefteqn{\lan w_i(\ct)w_j^*(\ct')\ran =} \nonumber\\
&&  {4\over k^4\sqrt{\ct\ct'}}[F_5+z^2F_2]
	(k^2\de_{ij}-k_ik_j) \label{Cw}\\
\lefteqn{\lan\tau_{ij}(\ct)\tau^*_{lm}(\ct')\ran =} \nonumber \\
&&	{1\over\sqrt{\ct\ct'}}F_5[\de_{il}\de_{jm}
+\de_{im}\de_{jl}   -\de_{ij}\de_{lm} + 
	k^{-2}(\de_{ij}k_lk_m + \nonumber \\ &&
	\de_{lm}k_ik_j -\de_{il}k_jk_m - \de_{im}k_lk_j -\de_{jl}k_ik_m
	-\de_{jm}k_lk_i) + \nonumber \\
&&	k^{-4}k_ik_jk_lk_m] \label{Ctau}
    ~.\eea
It is interesting to note that although $\widehat{C}_{ijlm}$ is analytic, the 
correlation functions of the scalar, vector and tensor components, in 
general, are not. The reason for that is that the projection operators
onto these components are not analytic~\cite{DK}. This is important. 
It implies, {\em e.g.}, that the anisotropic stresses in general have a white 
noise and not a $k^4$ spectrum as  erroneously concluded in \cite{Ma,Sper}. 
The scalar 
anisotropic stress potential thus diverges on large scales,
$\lan |f_\pi|^2\ran \propto 1/(\ct k^4)$ for $k\ct\ll 1$. A result which is
also obtained in the large-$N$ limit and in numerical simulations
of $O(N)$ models. The power spectrum of the
scalar anisotropic stress potential $f_\pi$ is analytic if
vector and tensor perturbations are absent, $F_5=F_2=0$.
In the generic situation, $F_5(z=0,r=1)=A_5(1)\neq 0$. \footnote{Even
though the {\em potential} $f_\pi$ and thus also the Bardeen potential 
$\Psi$ (see Eq. (\ref{Bardeen}))
diverge for $k\ct\ra 0$, the physically relevant (measurable) quantities like 
$\Th_{\mu\nu}$ and $R_{\mu\nu}$ stay perfectly finite.}

Another situation where $f_\pi$ has a white noise spectrum is the case of  
perfectly coherent seeds~\cite{observ}, in other words if
\be
 \widehat{C}_{ijlm}(\bk,\ct,\ct') = A_{ij}(\bk,\ct)A_{lm}^*(\bk,\ct'). 
 \label{PerCo}
\ee
The fact that as well $\lan f_pf_p^*\ran$ as   $\lan f_pf_\pi^*\ran$ are 
white noise implies that also $\lan f_\pi f_\pi^*\ran =  \pm
|\lan f_pf_\pi^*\ran|^2/\lan f_pf_p^*\ran$ must behave like white 
noise and thus $F_5 \propto z^4$ and $F_2 \propto z^2$ on large scales.
This can also be obtained by using the analytic properties of the correlators
$\lan \Th_{0\mu}\Th_{\al\beta}^*\ran$ and energy momentum 
conservation~\cite{DLU}. 

Generically, we expect the following
relation between scalar, vector and tensor perturbations of the 
gravitational field on super-horizon 
scales, $x\equiv k\ct\ll 1$: (The equations for the scalar, vector and tensor
gravitational potentials  in terms of $f.$, {\bf w} and
$\tau_{ij}$ are given in Chapter~\ref{The}.) 
\bea
\lan |\Phi-\Psi|^2\ran &\sim& {12\ep^2\over \ct k^4}A_5(1)  \\
\lan |\Si_i|^2\ran &\sim&  {16\ep^2\ct\over k^2}A_5(1) \\
\lan |H_{ij}|^2\ran &\sim& 4\ep^2\ct^3A_5(1) ~.
\eea

If the large scale CMB anisotropies were solely induced by 
super horizon perturbations, this could be translated into a ratio between 
the scalar, vector and tensor contributions to the $C_{\ell}$'s on large 
scales, $\ell\lsim 50$. However, since the main contribution
to the CMB anisotropies is induced at horizon crossing, $x=1$ 
(see below) the above
relations cannot be translated directly and we can just learn that one 
expects, in general, contributions of the same order of magnitude from 
scalar, vector and tensor perturbations.

 We want to discuss in some detail the CMB anisotropies
induced from scalar perturbations. From
\be
\Phi +\Psi = -2\ep f_\pi~,  \label{Bardeen}
\ee
we see that even if $\Phi$ has a white noise spectrum due to 'compensation'
\cite{DS}, this generically leads to a $k^{-4}$ spectrum for $\Psi$ and 
for the combination $\Phi-\Psi$ which enters in Eq.~(\ref{dT}) below. 

This finding is in contradiction with \cite{Ma,Sper}, which predict a
white noise spectrum for $\Psi$, but it is not in conflict with the 
Harrison Zel'dovich spectrum of CMB fluctuations which has been 
obtained numerically in \cite{PST,ZD,ACSSV}. This can be seen by the 
following simple argument:
Since topological defects decay inside the horizon, the Bardeen potentials
on sub-horizon scales are dominated by the contribution from dark matter 
and thus roughly constant. The integrated Sachs Wolfe term then contributes
 only up to horizon scales. Therefore, using the fact that for defect models 
$D_g$ and $V$ are smaller than the Bardeen potentials on 
super-horizon scales, we obtain
\bea
\lefteqn{
(\De T/T)_{\ell}(k)|_{SW} \sim 
	(\Phi -\Psi)(k,x_{dec})j_{\ell}(x_0-x_{dec})}
\nonumber\\ &&
+\int_{x_{dec}}^1({\Phi}'-{\Psi}')(k,x)j_{\ell}(x_0-x)dx~, \label{dT}
\eea
where $x=k\ct$ and prime stands for derivative w.r.t. $x$. 
The lower boundary of the integrated term roughly cancels the
ordinary Sachs Wolfe  contribution and the upper boundary leads, to
\bea
k^3\lan|(\De T/T)_{\ell}(k)|^2\ran|_{SW} \sim && \nonumber \\
	\quad\quad \ep^2
	[3F_5(1)+4F_2(1)+F_3(1)]j_{\ell}^2(x_0), &&
\eea
a Harrison-Zel'dovich spectrum. The main ingredients for this result are 
the decay of the sources inside the horizon as well as scaling, the rest 
follows for purely dimensional reasons.

The parameter space of generic causal scalar seed models provided 
by the five functions $F_1$ to $F_5$ (of two
variables)  is still enormous and is rather impossible to investigate.
For a realistic
model, the parameter space is even larger due to the radiation-matter
transition which breaks scale invariance: the seed functions can be
different in the radiation and in the matter era. For global $O(N)$
defects this difference turns out not to be very important (less than
about  20\%~\cite{DKM}) it may, however, go to factors of two and more
for cosmic strings~\cite{Paul}.

\subsection{Mimicking inflation}

 Neil~Turok has constructed a model
with scaling causal seeds which perfectly reproduces the CMB
anisotropy spectrum of inflationary models~\cite{Turok}. Other
synthesized causal seed models with various heights of the acoustic
peaks are discussed in~\cite{DS,observ}. Spergel \& Zaldarriaga argued
that causal seeds can nevertheless be distinguished from inflationary
models by the induced polarization~\cite{SZ}. In their argument they 
however use that the correlator $\lan f_\pi f_\pi\ran$ be white noise.
As we have seen, this is only correct for purely scalar or perfectly coherent
seeds. Therefore, allowing for vector and tensor contributions, as well as 
for decoherence, one can in principle circumvent the Spergel \& Zaldarriaga 
argument. However, the fact that seeds notoriously lead to too low amplitude
acoustic peaks, limits very strongly the allowed vector and tensor 
contributions which enhance the CMB anisotropy spectrum in the Sachs-Wolfe, but 
not in the acoustic peak region. Furthermore, the smearing out of the
acoustic peaks induced by decoherence is not observed in the data, 
limiting the allowed amount of decoherence considerably.

We restrict the following discussion to 'perfectly coherent' models with 
purely scalar perturbations. We parameterize the seeds by
\bea
\langle\Psi_s(\bk,\ct)\Psi_s^*(\bk,\ct')\rangle &=& {\ep^2\over
\sqrt{\ct\ct'}k^4}P_1(z,r)\\ 
 \langle\Phi_s(\bk,\ct)\Phi_s^*(\bk,\ct')\rangle  &=& {\ep^2\over
\sqrt{\ct\ct'}k^4}P_2(z,r)\\ 
\langle\Psi_s(\bk,\ct)\Phi_s^*(\bk,\ct')\rangle  &=& {\ep^2\over
\sqrt{\ct\ct'}k^4}P_3(z,r)~.
\eea 
Perfect coherence then implies
\be
 P_3(z,r) = \pm \sqrt{P_1(\sqrt{z^2r},1)P_2(\sqrt{z^2/r},1)}
\label{deco}
\ee
Due to the absence of vector and tensor perturbations,
the sum $\Phi + \Psi \propto f_\pi$ is suppressed
by a factor $z^2$ on large scales, $z\ll 1$~\cite{DK}. In a first
attempt we simply set
$\Psi = -\Phi$, which implies $P_1=P_2=-P_3 \equiv P$.

We discuss two families of models (see~\cite{model}).\\
{\bf Family I}\\
To enhance the acoustic peak, we use seeds which are larger in
the radiation era than in the matter era. 
\bea
 P_r(z,1) &=& {t\over 1 +(bz)^6}  \label{Pr} \\
 P_m(z,1) &=&  {1\over 1 +(bz)^6}~,  \label{Pm}  
\eea
where here the subscripts $_r$ and $_m$ indicate the radiation and
matter era respectively. The parameters $t$ and $b$ are varied to
obtain the best fit and the amplitude $\ep$ is determined by the
overall  normalization.\\
{\bf Family II}\\
The second family of models is inspired by Ref.~\cite{Turok},
 which studies spherical exploding shells with $\rho+3p \propto \de(r-A\ct)$.
In terms of the source functions defined in Chapter~\ref{The}, this model
is characterized by
\bean
f_\rho\! + \! 3f_p &=& {1\over \al \ct^{1/2}}{\sin( Ak\ct)\over Ak\ct}\\
f_v &=& 
{E(\ct)\over k^2\ct^{3/2}}{3\over C^2}\left[\cos
(Ck\ct) - {\sin(Ck\ct)\over Ck\ct} \right]
\eean
with $\al =(\dot a/a)\ct$ and $E=(4-2/\al)/(3-12\al)$.
The functions $f_\rho$ and $f_\pi$ are then determined by energy
momentum conservation, Eqs.~(\ref{ec},\ref{3f}).
The function $E$ is chosen such that the power spectrum of $f_\pi$ is
white noise on super horizon scales, a condition which is required
for purely scalar causal seeds as we have seen in the previous section.
This leads to the Bardeen potentials
\bea
 \Phi &=& {\ep\over k^2}(f_\rho + 3{\al\over \ct}f_v)~,
 \label{phT}\\
 \Psi &=& -\Phi-2\ep f_\pi~.   \label{psT}
\eea
Here the seed functions are actually not given as random variables
but as square-roots of power spectra, and one has always to keep in
mind that we assume perfect coherence. Of course one can also regard
Eqs.~(\ref{phT},\ref{psT}) as mere definitions with
\bea
 P_1(z,1) &=& \ct k^4\Psi^2/\ep^2 ~, \\
 P_2(z,1) &=&  \ct k^4\Phi^2/\ep^2 ~, \\
 P_3(z,1) &=&   \ct k^4\Psi\Phi/\ep^2 = -\sqrt{ P_1(z,1) P_2(z,1)} ~.
\eea
With a somewhat lengthy calculation one can verify that $E$ is
chosen such that $f_\pi \propto$ const. for $z\ll 1$ and the
functions $P_i(z,1)$ are analytic in $z^2=(k\ct)^2$. This
family of models is described by the parameters $A$ and $C$, which have to
satisfy $0 < A,~ C \le 1$ for causality.
Also here  one can choose different amplitudes for the
source functions in the radiation and matter era by introduction of
the additional parameter $t\neq 1$.

Seeds generically produce isocurvature perturbations.
For a flat universe, this implies a position of the first
peak at $\ell \sim 350$, which is definitely incompatible
with the recent CMB observations (see also~\cite{juan}, \cite{enqvist}).
However, the tight constraints on the flatness of the universe 
obtained from CMB data analysis are based on the assumption of 
adiabatic primordial fluctuations. 
Using this loophole, it is possible to construct 
 closed $\Lambda$-dominated isocurvature models which have the first
acoustic peak in the observed position.

For a given seed-model, the position of the first acoustic peak is
determined primarily  by the
angle subtended by the acoustic horizon $\la_{ac}$ at decoupling time,
$\ct_{dec}$. 
The angle under which a given comoving scale $\la$ at conformal time
$\ct_{dec}$ is seen on the sky is given by
$  \theta(\la) =\la/\chi(\ct_0-\ct_{dec})$,
where
\[ \chi(y) =\left\{\begin{array}{ll}
   \sin(y) & \mbox{ if }~~ K>0 \\
   \sinh(y) & \mbox{ if }~~ K<0 \\
   y& \mbox{ if }~~ K =0~. \end{array}\right.
\]
is the comoving angular diameter distance (K denotes the curvature of 
3-space).
As the harmonic number $\ell$ is inversely proportional to the angle
$\theta$, this yields
$\ell_{\rm peak} \simeq R \ell_{\rm peak}^{\rm flat}$ ~where~ 
$R =\theta_{ac}^{\rm flat}/\theta_{ac}$ and $ \ell_{\rm peak}^{\rm flat}$ is
the peak position in a flat model with the same value of $\Om_mh^2$.
In terms of cosmological parameters one finds (see Ref.~\cite{BE} 
or~\cite{model}, a factor $1/2$ is missing in the formula of~\cite{BE}),
\[ 
R=\frac{1}{2} \sqrt{\frac{\Omega_m}{|\Omega_K|}}\chi(\ct_0-\!\ct_{\rm dec}).
\] 
An interesting point is that for
$\Omega_m \ra 0$ the quantity $R$ depends very sensitively on
$\Omega_{\Lambda}$. Thus, we can have important shifts in
the power spectrum, $R \sim 0.6$ say,  with relatively
small deviations from flatness ($\Omega_m=0.3$,
$\Omega_{\Lambda}=0.9$, $\Omega_K=-0.2$).
In Ref.~\cite{BE} the authors have shown
that the simple prescription $\ell \ra R\ell$ reproduces the CMB power
spectra for curved universes within a few percent.
On lines of constant $R$,  CMB power spectra are nearly degenerate.  
This simple prescription can be used to
rescale the flat spectrum. Furthermore, one has to ascertain that the value of
$h^2\Om_m$ used in the spectrum calculation agrees roughly with the value
preferred by the best fit value of $R$ and the super-novae
constraint~\cite{SNIa}, which can be cast in the form $\Om_m \simeq
0.75\Om_\Lambda-0.25$. $h^2\Om_m$ determines the time of equal matter
and radiation and thus influences the early integrated Sachs-Wolfe
effect, which contributes to the spectrum right in the region of the
first peak. We therefore get a better approximation if we use
the correct value for $\Om_mh^2$.

\begin{figure}[ht]
\centerline{\psfig{figure=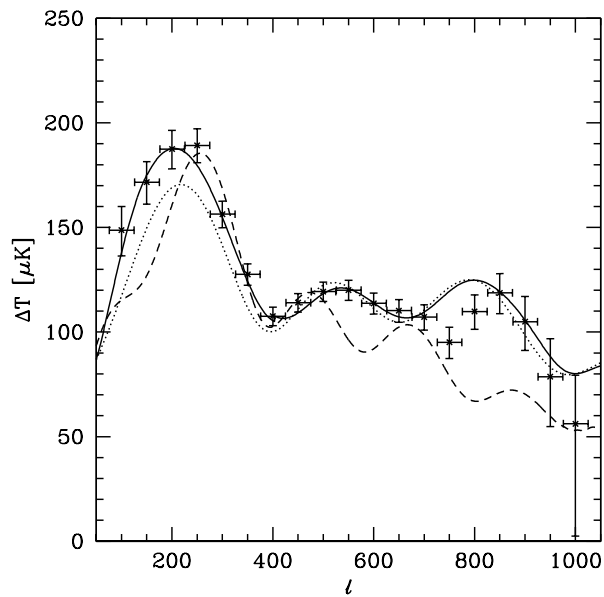,width=8.3cm}}
\caption{The CMB temperature anisotropy spectrum
$\ell(\ell+1)C_\ell^{(T)}$ for our best fit model of family I (long
dashed) and family II (solid) is compared with the new B98 
data. Family I model is closed, $\Omega \sim 1.2$ it has problems fitting 
anything. The family II model is flat and is in good
 agreement with the data ($\chi^2=4/9$), we fitted only
$\ell\leq 700$.
A standard inflationary spectrum with the same cosmological parameters
as the family II model
($h=0.65,~ h^2\Om_b=0.019,~ \Om_{cdm}=0.35,~ \Om_\La=1-\Om_m$)
is also indicated (dotted). \label{ct}}
\end{figure}

\begin{figure}[ht]
\centerline{\psfig{figure=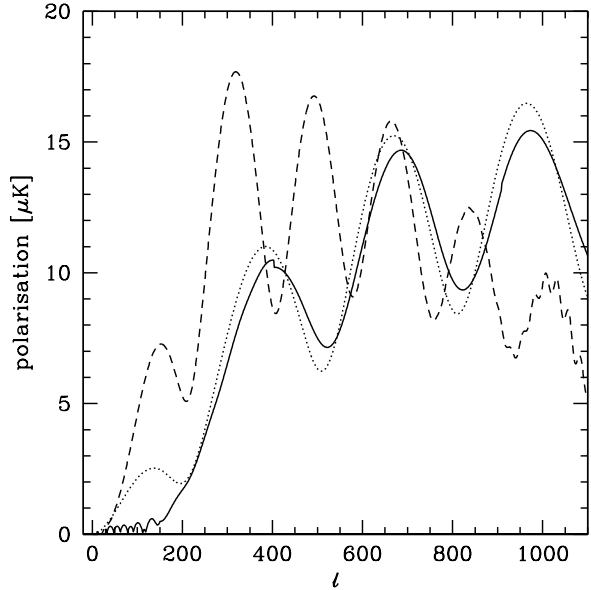,width=8.3cm}}
\caption{The CMB polarization  spectrum $C_\ell^{(E)}$s  for our best
fit model of family I  (long dashed)
and family II (solid) is compared with a standard
inflationary spectrum with the same parameters as above (dotted). 
The family I model predicts a larger polarization signal
in the band $50 \le \ell \le 500$.
On the contrary, the lack of intermediate scale polarization 
at $\ell \le 200$ in the family II model is clearly visible.
 \label{cp}}
\end{figure}

The results from an analysis of family~I are shown in  
Figs.~\ref{ct} and~\ref{cp} (long dashed lines).  
The best fit model is shown. Clearly, even though we can change the 
cosmological parameters to fit the first acoustic peak, the second and third
peaks are also shifted by this procedure and can no longer be fitted. The 
cosmological parameter $R$ is now fixed by CMB anisotropy measurements due 
to the measured inter-peak distance. This  
situation was still slightly different with the 'old' Boomerang and Maxima
data~\cite{B98,MAX} for which this model could still provide a reasonable 
fit~\cite{model}. in addition, the shifted first peak
is too narrow to fit the data well.

The 'best fit' model  corresponds to the
parameters $t=2.2,~b=1/9,~\Om_m=0.35$ and $R=0.53$. It has a value of
$\chi^2 =96$, which, for $13$ points and $4$ parameters ($t$, $b$, $R$
and the normalization), excludes it at more than $99.8\%$ c.l. if 
Gaussian statistics are assumed.
 
A much better fit can be achieved by the models of family~II.
The best fit model of this family has $\chi^2=4.0$ for $13$ points and $5$
parameters ($A,~C,~t,~R$ and the normalization) and is in good agreement with
the data (see Fig.~\ref{ct}, solid line).
The model shown corresponds to the best fit
parameters $A=0.9,~C=0.8,~t=0.9,$ and $R=1$.
In this model which is flat and causal, the first
peak in the polarization spectrum is suppressed, as
has been noted in Ref.~\cite{SZ}  (see Fig.~\ref{cp}, solid line). 

We can therefore conclude that generic causal scaling seed models for
structure formation can reproduce the recent CMB anisotropy 
data~\cite{B98,MAX}.
To achieve this agreement, we have suppressed vector and tensor
perturbations and have assumed perfectly coherent fluctuations. We
believe that it is quite improbable that topological defects 
from a GUT phase transition have such a behavior. Nevertheless, there might
be some other scale-invariant causal physical mechanism (e.g. some
spherically symmetric 'neutrino explosions', see Ref.~\cite{Turok}) leading to 
seeds of this or similar type. Clearly, we only have a satisfactory
model of structure formation if also the physical origin of the
'seeds' is clarified.

\subsection{Mixed models}

In this section we want to discuss the possibility of combining topological
defects with primordial, inflationary perturbations. 
There are at least two important reasons why such an analysis is interesting.
First of all, as we saw in the previous chapters, the predictions based
on topological defect models for the CMB power spectra are dramatically
different from what is expected in the standard inflationary model.
This means that, in principle, the two contributions can be disentangled.
Secondly, this analysis contributes in a phenomenological
but physically motivated way to the discussion, about the model dependence
of the values of cosmological parameters derived from accurate
CMB data. Inflationary models leading to a mixture of inflationary 
perturbations and defects which both contribute to the CMB anisotropies have 
been developed \eg~in Refs.~\cite{jean,DaMa,KaKa} and others.
A similar analysis, but with cosmic string and with an older
CMB dataset, can be found in \cite{contaldi2}. A mixture of global defects
and inflationary perturbations has also been considered in~\cite{BPRS}. 

We consider an
initial fluctuation spectrum with scalar initial perturbations given by their
spectral index $n_s$, in addition to topological defects which formed
at some symmetry breaking temperature $T_c$ determining the amplitude
of the seed perturbations. For definiteness we take the defects to be 
cosmic texture. It is reasonable to assume that the inflationary
fluctuations and those induced by topological defects are not
correlated and therefore the resulting perturbation spectra can be
added in quadrature. We set 
\be
 C_{\ell} = \l[{1 \over (1+r)}C_{\ell}^I + 
{{ r} \over (1+r)}C_{\ell}^D\r]A C_{10}^{COBE} \label{4cr}
\ee
for the CMB anisotropy, where $A$ is the pre factor in units of 
$C_{10}^{COBE}$, $C_{10}^I=C_{10}^D=1$ and the
parameter $r$ gives the relative amplitude of textures/inflation
at $\ell=10$.
We let vary the cosmological parameters as follows: 
$\Omega_{m}= 0.015$ to $1.0$; $\Omega_{b} = 0.015$ to $0.2$; 
$\Omega_{\Lambda}=0.0$ to $1.0$ and $h=0.25$ to $0.95$.
We restrict the analysis to flat universes.
We vary the spectral index of the primordial inflationary 
perturbations within the range $n_s=0.50$ to $1.50$.
The theoretical inflationary models are computed using the 
publicly available {\sc cmbfast} program and 
are compared with the recent BOOMERanG-98~\cite{Net}, 
DASI~\cite{DASI} and MAXIMA-1~\cite{Mnew} results.
The power spectra from these experiments were estimated in 
$19$, $12$ and $10$ bins respectively, spanning the range
$25 \le \ell \le 1000$. 
For the DASI and MAXIMA-I experiments, we use the public available
correlation matrices and window functions.
For the BOOMERanG experiment, we assign a flat window function 
in each bin $\ell(\ell+1)C_{\ell}/2\pi=C_B$, we approximate the signal 
$C_B$ to be a Gaussian variable, and we consider $\sim 10 \%$ 
correlations between neighboring bins.
The likelihood for a given cosmological model is then
 defined by 
$-2{\rm ln} L=(C_B^{th}-C_B^{ex})M_{BB'}(C_{B'}^{th}-C_{B'}^{ex})$
where  $M_{BB'}$ is the Gaussian curvature of the likelihood 
matrix at the peak. 
 We consider $10 \%$, $4 \%$  and $5 \%$ Gaussian distributed 
calibration errors for the BOOMERanG-98, DASI and 
MAXIMA-1 experiments respectively.
We also include the COBE data using Lloyd Knox's RADPack packages~\cite{Knox}.

\begin{figure}[ht]
\centerline{\epsfig{figure=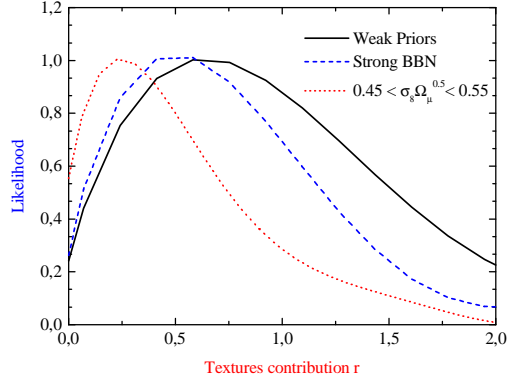,width=7.5cm}}
\caption{\label{liker} The likelihood probability distribution functions
for $r$ in the case of different external priors.}
\end{figure}

In Fig.~\ref{liker} we plot the obtained likelihood distribution
for the parameter $R$, after marginalization over the remaining
'nuisance' parameters and as function of different external priors.
The present CMB data plus an external Gaussian prior on the value of the
Hubble constant $h=0.65 \pm 0.2$ gives $r < 2.6$ at  
$95 \%$ c.l.. Hence, we cannot exclude a substantial admixture of
topological defects from CMB data alone.
Adding strong constraints on the baryon density parameter
from big bang nucleosynthesis $\Omega_bh^2=0.020 \pm 0.004$ 
gives $r < 2.1$.
Finally, combining the CMB data with LSS observation by
including a constraint $\sigma_8\Omega_m^{0.5}=0.50 \pm 0.05$
gives $r < 1.6$, always at $95 \%$ c.l..
As we can see, even assuming quite restrictive priors, the present
CMB data allow a contribution from textures as big as $r \sim 1$. 
Furthermore, models with  non-zero $r$ are in slightly
better agreement with the data, but there is no significant
evidence for non-vanishing $r$.

The main result of this analysis is that current microwave background data 
do not exclude a dominant contribution from textures on large scales, 
and marginally favor a significant fraction.

\begin{figure}[ht]
\centerline{\epsfig{figure=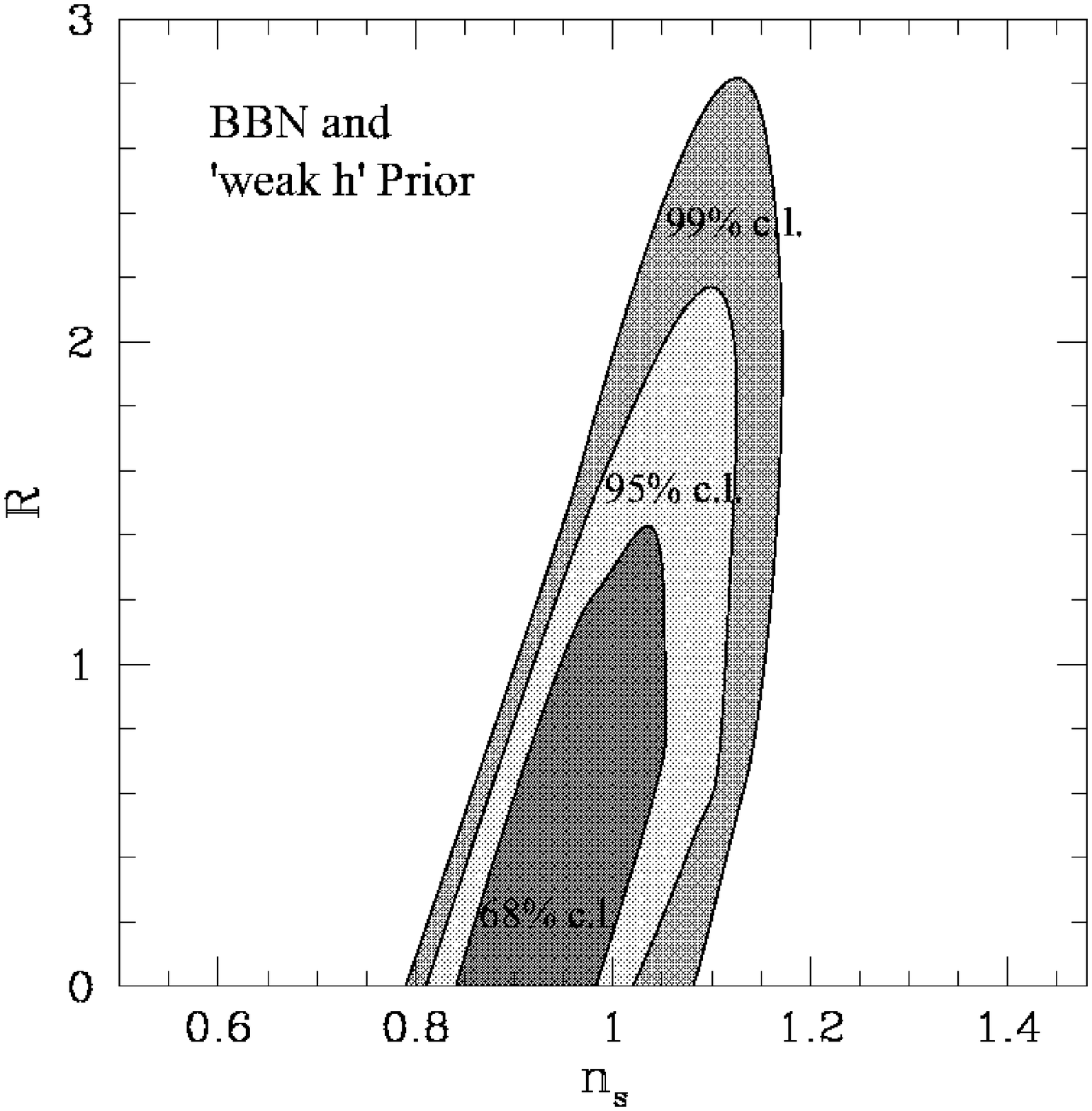,width=5.5cm} ~~
	\epsfig{figure=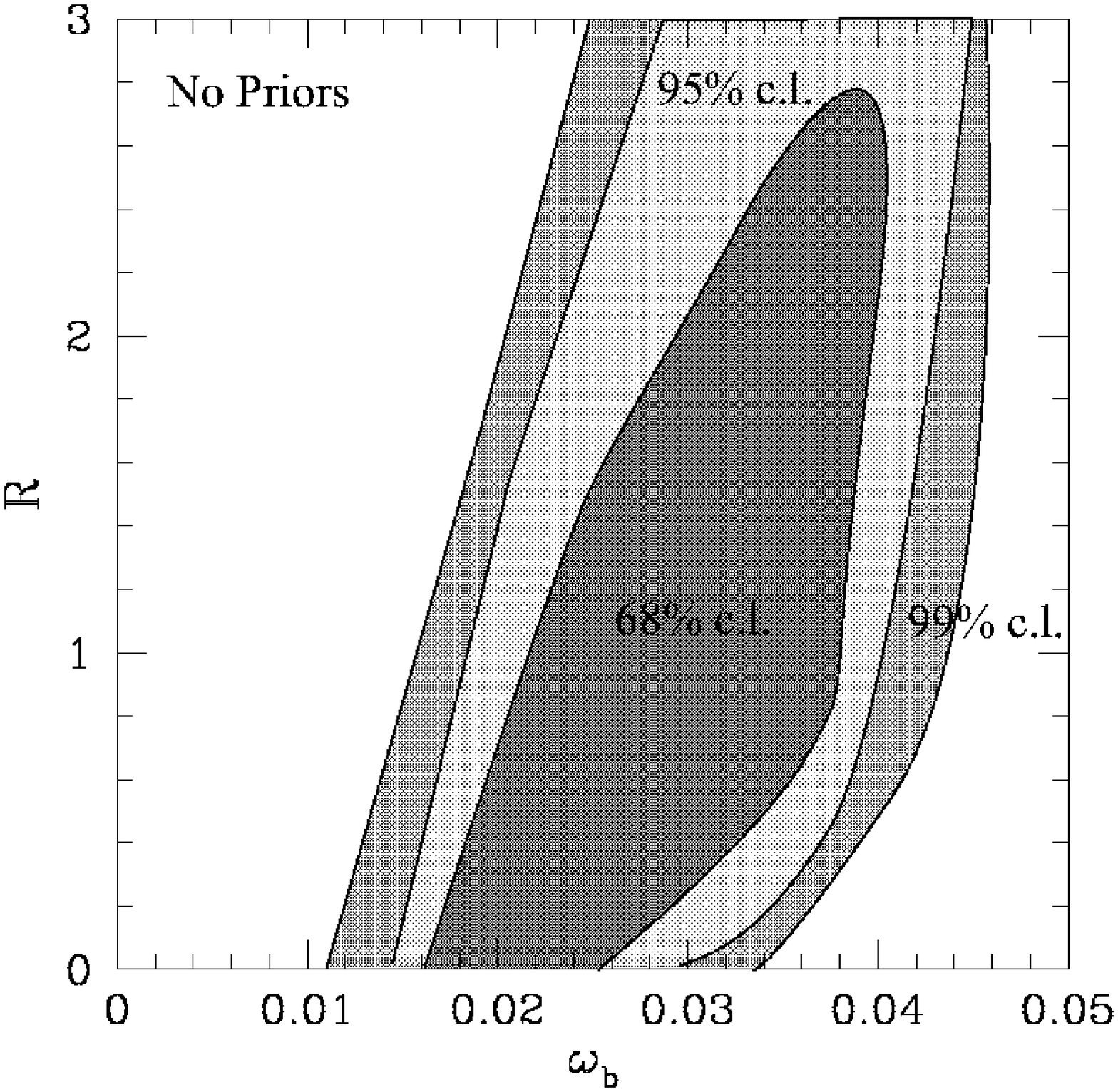,width=5.5cm}}
\caption{\label{4likrns}Left:  the likelihood contours for $r$ and $n_s$ 
are shown after marginalization over $\Om_m,~h,~\Om_b$ and $\Om_\La$ 
with weak priors on $h$ and $\Om_b$ (see text).
 Right: the likelihood contours for $r$ and 
$\om_b=\Om_bh^2$ are shown after marginalization over $n_s,~\Om_m,~h$ and 
$\Om_\La$. No priors are applied
}
\end{figure}

It is interesting to study how textures can 
affect the constraints on the remaining cosmological parameters.
Marginalizing over $\Om_m,~\Om_\La,~h$ and $\Om_b$. Assuming
the external priors $h=0.65 \pm 0.2$ and $\Omega_bh^2=0.020 \pm 0.002$,
we obtain the likelihood contours shown in Fig.~\ref{4likrns} (left panel) 
in the $r$-- $n_s$ parameter space. There is a degeneracy between the 
amplitude of the
texture component and the spectral tilt. A blue tilt ($n_s>1$) is 
found to be compatible with a larger textures contribution. 

If we instead marginalize over $\Om_m,~\Om_\La,n_S$,  without assuming 
external priors from big bang nucleosynthesis, we find the contours  shown in 
Fig.~\ref{4likrns} (right panel) in the $r$ -- $\Om_bh^2$ plane.
Without the BBN constraint, the contribution of
textures modes can be even larger, and there is clearly a 
degeneracy along the $r-\Omega_bh^2$ direction.

\begin{figure}[ht]
\centerline{\epsfig{figure=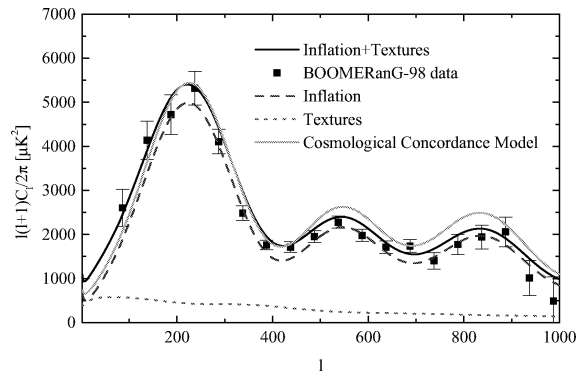,width=7.5cm}}
\caption{\label{4Clr} The best fit mixed spectrum is shown. The corresponding 
parameters are $\Om_m=0.175,~\Om_\La=0.825,~h=0.75,~\Om_b=0.045,n_s=1.02,
	~r=0.9$. Its inflationary (long dashes) and texture (short dashes) 
components as well as the spectrum from a concordance model 
(dotted, see text) are also indicated.
}
\end{figure}

The best fit spectrum with  
$\Om_m=0.175,~\Om_\La=0.825,~h=0.75,~\Om_b=0.045,n_s=1.02,~r=0.9$ together
with a 'concordance' fit, $\Om_m=0.3,~\Om_\La=0.7,~h=0.65,~\Om_b=0.04,n_s=1,
~r=0$ are shown in Fig.~\ref{4Clr}.

\clearpage
\section{Conclusion}
In this review we have discussed the role of topological defects for cosmic
structure formation. Even though the basic idea seems natural and intriguing,
we have seen that structure formation with global $O(n)$ defects and (local) 
cosmic strings is ruled out already by the CMB anisotropy data alone.
Due to the isocurvature nature of the perturbations and due to the 
importance of vector and tensor perturbations, defects do not induce the 
acoustic peaks visible in present data. Furthermore, 
non-linearities in the defect evolution tend to decohere, smear out, the 
peak  structure. Nevertheless, generic causal seeds, which are dominated by 
scalar perturbations and have nearly no decoherence, may mimic the CMB 
anisotropy spectrum of inflationary models. Such models, however, do 
certainly not represent topological defects for which non-linearity and 
therefore decoherence is one of the basic ingredients. Furthermore, such 
models could be distinguished by their polarization spectrum. Models where 
both, inflation and topological defects play a role for structure formation, 
such that the large scale CMB anisotropies from both components are 
of comparable amplitude, cannot be excluded with present data.
\vspace{2cm}\\
{\bf Acknowledgment}\\
We thank all our colleagues for the many discussions on the subject which 
contributed to this report. Especially we want to mention:
Robert 
Brandenberger, Nathalie Deruelle, Paolo deBernardis, Pedro Ferreira, Mark
Hindmarsh, Tom Kibble, Joao Magueijo, Ue-Li Pen, Patrick Peter, Mairi
Sakellariadou, Uros Seljak, David Spergel, Dani\`ele Steer, Neil Turok, 
Jean-Philippe Uzan, Tanmay Vachaspati, Alex Vilenkin and 
Nicola Vittorio.\\
This work has been supported by the European network CMBNET and by the 
Swiss NSF. RD thanks the Institute for Advanced Study for hospitality and the 
Monell Foundation for financial support.
\clearpage
{\LARGE APPENDIX}

\appendix

\section{Definitions of all gauge-invariant perturbation variables}
\label{AppVar}
In this Appendix we give precise definitions of all the
gauge-invariant perturbation variables used in the text. These
definitions, their geometrical interpretation and a short derivation
of the perturbation equations can be found in Refs.~\cite{Review,KS}.
We restrict the analysis to the spatially flat case, $K=0$.
The general case can be found in the references above.
We define the perturbed metric by
\be g = \bar{g} +a^2h ~,\ee
where $\bar{g}$ denotes the standard Friedman background, $a$ is the
scale factor and $h$ denotes the metric perturbation.

\subsection{Scalar perturbations}
Scalar perturbations of the metric are of the form
\bea h^{(S)} &=& -2A(d\ct)^2 + 2iBk_jd\ct dx^j + 2(H_L +{1\over 3}H_T)
\de_{ij}dx^idx^j \nonumber \\
 &&         -2k^{-2}H_Tk_ik_jdx^idx^j  ~ .  
\label{2h} \eea
Computing the perturbation of the Ricci curvature scalar and the shear
of the equal time slices, we obtain
\bea
   \de\bm{R} = 4 a^{-2}k^2{\cal R} &\mbox{ with }& {\cal R} =
     H_L + {1\over 3} H_T  \label{2R} \\
 \Si = a\si({k_ik^j\over k^2} - {1\over3}\de_i^j)dx^i\otimes\dd_j,
  &\mbox{ with  }& 
    \si = {1\over k}\dot{H}_T - B  ~ . \label{2sigma} 
\eea
The Bardeen potentials are the combinations
\bea
 \Phi &=& {\cal R} - (\dot{a}/a)k^{-1}\si   \label{2Phi} \\
 \Psi &=&  A  - k^{-1}[(\dot{a}/a)\si -\dot{\si}] \; . \label{2Psi}
\eea
They are invariant under infinitesimal coordinate transformations
(gauge transformations). In longitudinal gauge defined by $B=H_T=0$, the 
variable $\Phi$ represents the perturbation of the spatial metric while 
$\Psi$ is the perturbation of the $dt^2$ term, the lapse function.

To define perturbations of the most general energy momentum tensor, we
introduce the energy density $\rho$ and the energy flux $u$ as the
time-like eigenvalue and normalized eigenvector of $T^\mu_\nu$,
\[ T_{\mu}^{\;\;\nu}u^{\mu} = -\rho u^{\nu} \;\;,\;\; u^2 = -1 \; .\]
We then define the perturbations in the energy density and energy
flux field by
\be \rho = \overline{\rho}(1+\de)  \;\;, \label{2de} \ee
\be u    = u^0\dd_t +u^i\dd_i ~;     \label{2v} \ee
$u^0$ is fixed by the normalization condition, $u^0=a^{-1}(1-A)$ and 
$\overline\rho$ is the homogeneous background energy density..
In the 3--space orthogonal to $u$ we define the stress tensor by
\be \tau^{\mu\nu} \equiv P^{\mu}_{~~\al}P^{\nu}_{~~\beta}T^{\al\beta}~,\ee
where $P= u\otimes u + g$ is the projection onto the sub--space of
$T{\cal M}$ normal to $u$. It is
\[ \tau^0_0 = \tau^0_i =\tau^i_0 = 0 ~ .\]
The perturbations of pressure and anisotropic stresses can be parameterized by
\be \tau_i^{\;j} = \bar{p}[(1+\pi_L  )\de_i^{\;j} + \pi_i^{\;\;j} ]
  ~~\mbox{ , with }~ \pi^i_i = 0 \;, \label{2pi} \ee
where $\bar p$ is the background pressure.
For scalar perturbations we set
\bean   
 u^0 &=& (1-A) \;,\; {u^{(S)j}\over u^0} = -i{k^j\over k}v \\
 \mbox{ and }  && \\
\pi^{(S)i}_{~j} &=&
   (-k^{-2}k^{i}k_j+{1\over 3}\de^i_{~j})\Pi~.
\eean
Studying the behavior of these variables under gauge transformations,
one finds that the anisotropic stress potential $\Pi$ is gauge
invariant. Another gauge invariant combination from the matter variables 
alone is
\be
\Ga = \pi_L -{c_s^2\over w}\de ~. \label{GammaA}
\ee
$\Ga$ is proportional to the divergence of the entropy flux and vanishes 
for adiabatic perturbations~\cite{DSt}.
A gauge invariant velocity variable is the shear of the
   velocity field,
\be
\si^{(Sm)}_{ij} =(k^{-2}k_ik_j-{1\over 3}\de_{ij})a^3V  
	\mbox{ , ~with }~~ V =v-k^{-1}\dot{H}_T .
\ee

There are several different useful choices of gauge invariant density
perturbation variables,
\bea
  D_s &=& \de +3(1+w)(\dot{a}/a)k^{-1}\si \\
  D_g &=& \de +3(1+w){\cal R} = D_s +3(1+w)\Phi\\
  D   &=& D_s +3(1+w)(\dot{a}/a)k^{-1}V  \; .  \eea
In this work we mainly use $D_g$. Here $w=p/\rho$ denotes the enthalpy.
Clearly, these matter variables can be defined for each matter
component separately. For ideal fluids like CDM or the baryon-photon
fluid long before decoupling, anisotropic stresses vanish and
$\pi_L=(c_s^2/w)\de$, where $c_s$ is the adiabatic sound speed.

Also scalar perturbations of the photon brightness, $\iota^{(S)}$ are not gauge
invariant. To define the brightness we first consider the one particle 
photon distribution function
\[ f(x,\bp) = \bar f(\ct,p) + F(\tau,\bx,\bp) ~,\]
where $\bar f$ is just the Bose-Einstein distribution. The background Liouville
equation requires that $\bar f$ be a function of the redshift corrected 
momentum $v=ap$ only. The brightness 
perturbation is then the momentum 
integral of $F$, which depends on the photon direction, position and time
\be
\io(\bn,\bx,\ct) = {4\pi\over \bar\rho}\int dpp^3F(\bp,\bx,\ct)~.
\ee
This can be decomposed in scalar vector and tensor contributions,
\[ \io =\io^{(S)}+\io^{(V)}+\io^{(T)} ~.\]
It has been shown~\cite{Review} that the combination 
\be
\MM^{(S)} =\io^{(S)} +4{\cal R} +4ik^{-1}n^jk_j\si  \label{Mio}
\ee
is gauge invariant. This is the variable which we use here. In other
work\cite{HS} the gauge invariant variable $\Th \equiv \MM+\Phi$ has
been used.  Since $\Phi$ is independent of the photon direction $\bn$
this difference in the definition shows up only in the monopole,
$C_0$.  But clearly,
as can be seen from Eq.~(\ref{Mio}), also the dipole of $\io^{(S)}$, 
$C_1$, is gauge dependent.

The brightness perturbation of the neutrinos is defined the same way
and is not  repeated here.
  
\subsection{Vector perturbations}
Vector perturbations of the metric are of the form
\be
 h^{(V)} = 2B_jdx^jdt +   ik^{-1}(k_lH_j+k_jH_l)dx^ldx^j
  \label{2hv}~,
\ee
where $\bm B$ and $\bm H$ are transverse vector fields. The simplest
 gauge invariant variable describing the two vectorial degrees of
freedom of metric perturbations is $\bm{\Si}$,
\be  \Si_j =k^{-1}\dot{H}_j-B_j ~.  \ee
Vector anisotropic stresses are gauge invariant. They are of the form
\be  \pi_{lj}^{(V)} = ik^{-1}(k_j\Pi_l+k_l\Pi_j)   ~.\ee
The vector degrees of freedom of  the velocity field are cast in the
vorticity
\be
u_{l;j}-u_{j;l} = ia(k_j\om_l-k_l\om_j) ~\mbox{ with }~  \om_j=v_j-B_j ~.
\ee
Vector perturbations of the photon brightness are gauge-invariant.
To maintain the notation consistent we denote them by 
$\MM^{(V)}\equiv \io^{(V)}$.

\subsection{Tensor perturbations}
We define tensor perturbations of the metric by
\be
 h^{(T)} = 2H_{ij}dx^idx^j
  \label{ht}~,
\ee
where $H_{ij}$ is a  transverse traceless tensor field.

The only tensor perturbations of the energy momentum tensor are
anisotropic stresses,
\be  \pi_{ij}^{(T)} = \Pi_{ij}  ~.\ee
Again for notational consistency tensor perturbations of the photon 
brightness are denoted $\MM^{(T)}\equiv \io^{(T)}$.

Clearly, all tensor perturbations are gauge-invariant (there are no
tensor type gauge transformations).

\section{Boltzmann equation and polarization}
\label{AppBoltz}

The relativistic Boltzmann equation for the photon distribution function 
is of the form
\be
p^\mu \dd_\mu f- \Ga^i_{\al\b} p^\al p^\b \frac{\dd f}{\dd p^i}
= C[f] , \label{beq}
\ee
where $f(t,\bx,\bp)$ is the one-particle distribution function
on the mass bundle $P_m=\{(p,x)\in T\MM|g(x)(p,p)=-m^2\}$
and $C[f]$ is the collision integral which describes interactions.
The left hand side of (\ref{beq})
requires the particles to move along geodesics in the absence of
collisions. (For a thorough treatment of the kinetic approach in
general relativity see \eg~\cite{Stewart} or~\cite{Uzkin} and references
therein.)

\subsection{The collisionless case, momentum integrals}

Let us first consider the situation where collisions are negligible,
$C[f]=0$. The unperturbed Boltzmann equation then  implies that
$f$ be a function of $v=a p$ only. Setting
$f=\bar{f}(v)+F(\ct,\bx,v,\bn)$, where $\bn$ denotes the momentum
directions, leads to the perturbation equation
\be
\dd_\tau F - n^i\dd_i F =  v \frac{d\bar{f}}{dv}\l[n^i A_{,i}-
	n^i n^j\l(B_{i,j}-\dot{H}_{ij}\r) +H_L\r]. \label{bpertu}
\ee
To derive (\ref{bpertu}) we have used $p^2=0$. The Liouville
equation for particles with non-vanishing mass can be found
in Ref.~\cite{Review}.

The ansatz
\be
f(x,\bp) = \bar{f}\l(\frac{g^{(3)}(\bp,\bp)^\frac{1}{2}}{T(x,\bn)}\r)
=\bar{f}\l(\frac{v}{T(x,\bn)}\r)
\ee
with $T(x,\bn)=\overline{T}(\ct)+\De T(x,\bn)$, and where $g^{(3)}$
is the spatial part of the metric, leads to
\be
f=\bar{f}-v \frac{d\bar{f}}{dv}\frac{\De T}{\overline{T}}.
\ee

Note that in terms of  the brightness perturbation, $\imath$ defined in 
Appendix\ref{AppVar},
\be
\imath=\frac{4\pi}{\bar{\rho}a^4}\int_0^\infty F v^3 dv
~~~~\mbox{ we have }~~
\frac{\De T}{T} = \frac{1}{4} \imath~.
\ee

Comparing this with equation (\ref{bpertu}), we obtain
\bea
\lefteqn{\dd_\ct \l(\frac{\De T}{T}\r) + n^i \dd_i \l(\frac{\De T}{T}\r)
    =} \nonumber\\
 &&-\l[n^i A_{|i}-\l(B_{i|j}-\dot{H}_{ij}\r)n^i n^j+H_L\r]. \label{tpertu}
\eea

The fact that gravitational perturbations of Liouville's equation can
be cast entirely into {\em temperature perturbations} of the original 
distribution is not astonishing. This is just an expression of gravity 
being ``achromatic'', \ie~independent of the photon energy.

We now decompose (\ref{tpertu}) into scalar, vector and tensor
components. Since components with spin higher than 2 are
not sourced by the right hand side of equation (\ref{tpertu}) and
since they are suppressed at early times, when collisions
are important, we neglect them.


For the {\em scalar} contribution to $\De T/T$ we obtain from (\ref{tpertu})
(in Fourier space)
\bea
\lefteqn{\dd_\ct \l(\frac{\De T}{T}\r)^{(S)}+i n^\ell k_\ell \l(\frac{\De T}{T}\r)^{(S)}     =} \nonumber \\
&&-\l[n^i A_i+n^i n^j k_i k_j \l(B-\dot{H}\r)+H_L+\frac{1}{3} k^2\dot{H}\r].
\label{scaldt}
\eea

This equation is not manifestly gauge-invariant. However, defining
\be
\frac{1}{4} \MM^{(S)}=\l(\frac{\De T}{T}\r)^{(S)}+H_L+\frac{1}{3}k^2 H
+i n^\ell k_\ell\l(\dot{H}-B\r) ,
\ee
it reduces to
\be
\dd_\ct\MM^{(S)}+ \mu k\MM^{(S)}
=4 i \mu k(\Phi\!-\!\Psi) , \label{scalbright}
\ee
where $\Phi$ and $\Psi$ are the Bardeen potentials and we have set
$\mu=n^jk_j/k$. Since the
right hand side of (\ref{scalbright}) is gauge invariant,
the left hand side must be so as well and we conclude that
$\MM^{(S)}$ is a gauge-invariant variable (a direct proof of this,
analyzing the gauge transformation properties of the distribution
function, can be found in~\cite{Review}).

${1\over 4}\MM^{(S)}$ coincides with the scalar temperature fluctuations
up a to a gauge dependent monopole and dipole contribution.

The vector and tensor contributions to $\De T/T$ are gauge invariant
by themselves and we denote them by ${1\over 4}\MM^{(V)}$ and 
${1\over 4}\MM^{(T)}$.
In the absence of collisions, they satisfy the equations
\bea
\dot{\MM}^{(V)} +i\mu k\MM^{(V)}
&=& -4 i\mu n^m k \Si_m^{(V)}\\
\dot{\MM}^{(T)} +i\mu k\MM^{(T)}
&=& 4 n^\ell n^m \dot{H}_{m\ell} .
\eea

The components of the energy momentum tensor are obtained by
integrating the second moments of the distribution function 
over the mass shell,
\be
T^{\mu\nu} = \int_{P_m(x)} p^\mu p^\nu f(p,x) \mu(x,p)~,
\ee
with the invariant measure $\mu(x,p)={\sqrt{|\det(g)}|\over p^0}d^3p$.
One finds 
\bea
D_g^{(\ga)} &=& \frac{1}{4\pi}\int\MM^{(S)} d\Om 
	=\si_0^{(S)}=\MM_0^{(0)}~, \label{int1}\\
V^{(\ga)} &=& \frac{3 i}{16\pi}\int \mu \MM^{(S)} d\Om \label{Vrad}
	=\frac{3}{4}\si_1^{(S)}=\frac{1}{4} \MM_1^{(0)} ~,\\
\Pi^{(\ga)} &=& \frac{-9}{8\pi}\int \l(\mu^2-\frac{1}{3}\r) \MM^{(S)} d\Om 
=3 \si_2^{(S)}=\frac{3}{5} \MM_2^{(0)}~, \label{int3}\\
\Ga^{(\ga)} &=& 0 ~,\\
\bm V^{(V)} &=& \frac{1}{4\pi} \int \bn \MM^{(V)} d\Om   \\
	&=& \frac{1}{3}\l(\si_{1,0}^{(V)}+\si_{1,2}^{(V)},
	\si_{2,0}^{(V)}+\si_{2,2}^{(V)},0\r) \nonumber \\
	&=& \frac{-1}{3\sqrt{2}} \l(\MM_1^{(+1)}+\MM_1^{(-1)},
	i(\MM_1^{(+1)}-\MM_1^{(-1)}),0\r)~, \nonumber \\
\bm\Pi^{(V)} &=& \frac{3}{2\pi} \int \mu \bn \MM^{(V)} d\Om \\
	&=& \frac{-6i}{5} \l(\si_{1,1}^{(V)}+\si_{1,3}^{(V)},
	\si_{2,1}^{(V)}+\si_{2,3}^{(V)},0 \r)  \nonumber \\
	&=& \frac{\sqrt{6}}{5} \l(i(\MM_2^{(+1)}+\MM_2^{(-1)}),
	\MM_2^{(+1)}-\MM_2^{(-1)},0\r) ~, \nonumber \\
\l(\Pi_{ij}^{(T)}\r) &=& \frac{3}{4\pi}\l( \int n_i n_j \MM^{(T)} d\Om\r) 
	= \l( 
	\begin{array}{ccc} -\pi^{(T)}_+     & \pi^{(T)}_\times & 0 \\
			   \pi^{(T)}_\times & \pi^{(T)}_+      & 0 \\
                   0                & 0                & 0 \end{array} \r) ~,
\eea
\bea
	\mbox{with }~~~ \pi_i^{(T)} &=& 
		\frac{2}{35}\l(7 \si_{0,i}^{(T)}+10 \si_{2,i}^{(T)}
		+3 \si_{4,i}^{(T)}\r) \nonumber \\
	\mbox{and }~~~ \pi_+^{(T)} &=& \frac{-\sqrt{3}}{5 \sqrt{2}}
		\l(\MM_2^{(+2)}+\MM_2^{(-2)}\r)~, \nonumber \\
	\pi_\times^{(T)} &=& \frac{-\sqrt{3}i}{5 \sqrt{2}}
		\l(\MM_2^{(+2)}-\MM_2^{(-2)}\r) \nonumber
	\eea
	 \nonumber
Here, we have also expressed the result in terms of 
the multipole moments defined by the expansions (\ref{devel})
and (\ref{newexp}) for a wave vector $\bk$ pointing in $\bz$-direction..

The expressions for the neutrino fluid perturbations in terms of $\NN$ are 
identical. 

\subsection{Collisions, polarization}

Let us now turn to the collision term. Just before the process of 
recombination during which the fluid description of radiation breaks down,
the temperature
is $\sim 0.4 \mr{~eV}$. The electrons and nuclei are non-relativistic
and the dominant collision process is non-relativistic Thomson
scattering.

Thomson scattering depends also on the polarization
of the incoming radiation field. It is therefore necessary to treat
polarization. This is usually done by introducing the
{\em Stokes parameters}~\cite{jack,koso,melvit,chandra}:

For a harmonic electro-magnetic wave with associated electric field
\be
{\bf E} (\bx,t)=\l({\bm\ep}_1 E_1+{\bm\ep}_2 E_2\r)
e^{ip \bm{n\cd x} - i\om t} ~,
\ee
where $\bn$, $\bm{ \ep}_1$ and $\bm{\ep}_2$ form an orthonormal
basis and the complex field amplitudes are parameterized
as $E_j=a_j e^{i\de_j}$, the Stokes parameters are given by
\bea
I &=& a_1^2 + a_2^2\\
Q &=& a_1^2 - a_2^2\\
U &=& 2 a_1 a_2 \cos(\de_2-\de_1)\\
V &=& 2 a_1 a_2 \sin(\de_2-\de_1) .
\eea
$I$ is the intensity of the wave (whose perturbation $\MM$ has interested 
us so far), while $Q$ is a measure of the strength of linear polarization 
(the ratio of the principal axis of the polarization ellipse). $U$ and $V$ 
give phase information (the orientation of the ellipse). $V$ also gives the 
amplitude of circular polarization. For non-relativistic Thomson scattering 
$V$ is completely decoupled and (since it vanishes at early times) is 
therefore never generated.

As $Q$ and $U$ vanish in the background, perturbations cannot couple
to them (since such terms are 2nd order), and the equation
corresponding to (\ref{scaldt}) for $Q$ and $U$ become (neglecting scattering)
\be
\dd_\ct(Q;U)+i n^\ell k_\ell(Q;U)  =0 . \label{scalstokes}
\ee

The differential cross section of Thomson scattering for a photon
with incident polarization ${\bm \ep}_{(i)}$ scattering into the
 outgoing polarization ${\bm \ep}_{(s)}\equiv {\bm \ep}'$ is~\cite{jack}
\be
\frac{d\si}{d\Om} = \frac{3}{8\pi}\si_T 
\l|{\bm\ep}_{(s)}^* {\bm\ep}_{(i)}\r|^2 .
\ee

\begin{figure}[ht]
\centerline{\epsfig{file=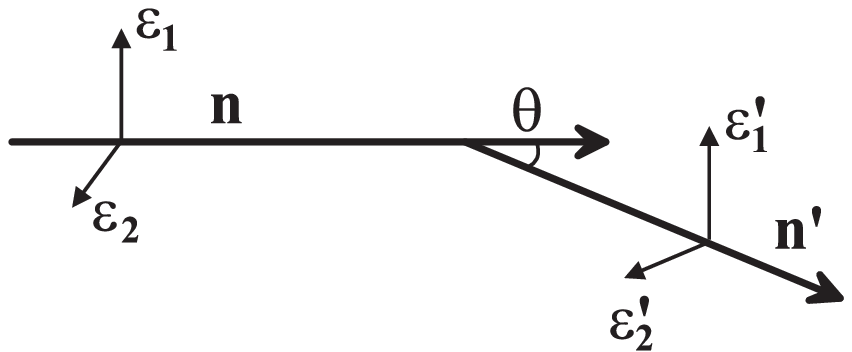,width=6cm}}
\caption{\label{scat1}
Definition of the angles and vectors for Thomson scattering
in the  $(\bn,{\bm \ep}_2)$ plane.}
\end{figure}
It is often convenient to introduce the two `partial' intensities
$I_1\equiv a_1^2 = (I+Q)/2$ and $I_2 \equiv a_2^2 = (I-Q)/2$. A
wave scattered in the $(\bn,{\bm \ep}_2)$ plane (see figure 
\ref{scat1}) by an angle $\tha$ has the intensities
\bea
I_1^{(s)} &=& \frac{3\si_T}{8\pi} I_1^{(i)} \nonumber \\
I_2^{(s)} &=& \frac{3\si_T}{8\pi} I_2^{(i)} \cos^2\tha \label{thomscat1a} ,
\eea
or, expressed in terms of the Stokes parameters,
\be
\l( \begin{array}{c} I^{(s)} \\ Q^{(s)} \end{array} \r) = \frac{3\si_T}{16\pi}
\l( \begin{array}{cc} 1+\cos^2\tha & \sin^2\tha \\
	\sin^2\tha & 1+\cos^2\tha \end{array} \r)
\l( \begin{array}{c} I^{(i)} \\ Q^{(i)} \end{array} \r) . \label{thomscat1b}
\ee

A rotation in the $({\bm\ep}_1,{\bm\ep}_2)$ plane doesn't change
the intensity of the wave, but it changes $Q$ and $U$ to
\bea
Q' &=& Q \cos(2\phi) + U \sin(2\phi) \label{thomrota} \\
U' &=& -U \sin(2\phi) + Q \cos(2\phi)\label{thomrotb} ~.
\eea

\begin{figure}[ht]
\centerline{\epsfig{file=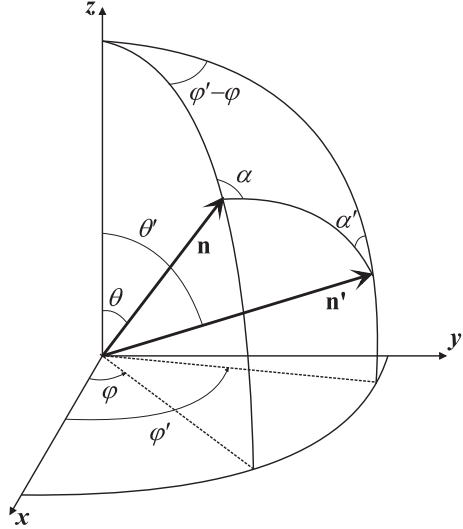,width=6cm}}
\caption{\label{scat2}
Definition of the angles and vectors for Thomson scattering
in the general case. The polarization vectors are oriented
like in figure \ref{scat1}.}
\end{figure}
To determine the cross section that a given 'initial' wave \\
$(I^{(i)},Q^{(i)},U^{(i)})$
propagating in direction $\bn$ be scattered into a wave $(I^{(s)},Q^{(s)},U^{(s)})$
with direction $\bn'$, we need to go through the following steps
(we will use the plane  $({\bm y}, {\bm z})$ as reference
plane, see figure (\ref{scat2}) for definitions of the angles
and vectors):
\begin{enumerate}
\item Rotate around $\bn$ such that the plane $(\bn,\bn')$ turns into 
the plane $(\bn {\bm z})$. One needs to apply
the rotation (\ref{thomrota},\ref{thomrotb}) for $\phi=\al$ to the
Stokes parameters.
\item Rotate the new plane $(\bn,\bn')$ around  ${\bm z}$
into the reference plane  $({\bm y}, {\bm z})$. This operation does not 
influence the incoming Stokes parameters..
\item Now we are in the known case of (\ref{thomscat1a}) 
and (\ref{thomscat1b}). Hence we can apply the
scattering matrix.
\item We then rotate the scattering plane back around ${\bm z}$
into the old $({\bm z},\bn')$ plane. This does  not
change the scattered Stokes parameters.
\item Finally we rotate around $\bn'$ by the angle $\al'$
to reach the original state. To do this, we have to apply
the rotation matrix (\ref{thomrota},\ref{thomrotb}) again, but now for
$\phi=\al'$.
\end{enumerate}

Following the steps outlined above, we recover the scattering matrix in the 
basis $(I_1,I_2,U)$ given in equations (\ref{chandra_eq}) - (\ref{ch_eq_2}) 
(see also~\cite{chandra}). $V$ is completely decoupled from the other
parameters and follows an evolution which is independent of the
rest. Hence by starting with $V(t \ll t_{dec}) = 0$ it will stay
zero and can be neglected. The scattering matrix $P$, which determines
the (non vanishing) scattered Stokes parameters from the initial ones,
\be
 \left( \begin{array}{c} I_1^{(s)} \\ I_2^{(s)} \\ U^{(s)} \end{array} \right)
 = {\si_T\over 4\pi} P
 \left( \begin{array}{c} I_1^{(i)} \\ I_2^{(i)} \\ U^{(i)} \end{array} \right)
\ee
is then given by
\be
  P = \l[P^{(0)} + \sqrt{1-\mu^2}\sqrt{1-\mu^{\prime2}} P^{(1)}
  +P^{(2)}\r] , \label{chandra_eq}
\ee
where
\be
P^{(0)} = \frac{3}{4} \l( \begin{array}{ccc}
 2 (1-\mu^2) (1-\mu^{\prime2})+\mu^2\mu^{\prime2} & \mu^2 & 0  \\
 \mu^{\prime2}                                    & 1     & 0  \\
 0                                                & 0     & 0 
\end{array} \r) ,
\ee
\be
P^{(1)}=\frac{3}{4} \l( \begin{array}{ccc}
 4 \mu \mu' \cos(\phi'-\phi) & 0 & 2\mu\sin(\phi'-\phi)  \\
 0                           & 0 & 0                    \\
 -4\mu'\sin(\phi'-\phi)      & 0 & 2\cos(\phi'-\phi)    
\end{array} \r) ,
\ee
\be
P^{(2)} = \frac{3}{4} {\scriptsize\l( \begin{array}{ccc}
 \mu^2\mu^{\prime2}\cos[2(\phi'-\phi)] & -\mu^2\cos[2(\phi'-\phi)] & \mu^2\mu'\sin[2(\phi'-\phi)]  \\
 -\mu^{\prime2}\cos[2(\phi'-\phi)]     & \cos[2(\phi'-\phi)]       & -\mu'\sin[2(\phi'-\phi)]    \\
 -2\mu\mu^{\prime2}\sin[2(\phi'-\phi)] & 2\mu\sin[2(\phi'-\phi)]   & 2\mu\mu'\cos[2(\phi'-\phi)]
\end{array} \r)}~ .   \label{ch_eq_2}
\ee

As we are in an isotropic situation, we will perform all the
calculations in a special coordinate system with $\bk \parallel \hat{\bz}$
and $\bn, \bn'$ as in Fig.~\ref{scat2}.
Clearly the results are independent of this coordinate choice.

The matrix $R$ connecting $(I_1,I_2,U)$ with $(I,Q,U)$ is
given by
\be \label{matrixR}
\l( \begin{array}{c}
	I_1 \\
	I_2 \\
	U \end{array} \r)
= \l( \begin{array}{c}
	1/2 (I+Q) \\
	1/2 (I-Q) \\
	U \end{array} \r)
= \frac{1}{2} \l( \begin{array}{rrr}
	1 &  1 & 0 \\
       	1 & -1 & 0 \\
	0 &  0 & 2 \end{array} \r) 
\l( \begin{array}{c}
	I \\
	Q \\
	U \end{array} \r)
\equiv R \l( \begin{array}{c}
	I \\
	Q \\
	U \end{array} \r).
\ee
	
To calculate the collision term including polarization , we 
change into the $(I_1,I_2)$ basis.
For each of the two intensities $\la\in\{1,2\}$ we then have
the collision term given by
\be
C[f^{(\la)}]=\frac{df_+^{(\la)}}{d\ct}-\frac{df_-^{(\la)}}{d\ct} ,
\ee
where $f_+^{(\la)}$ and $f_-^{(\la)}$ denote the distribution of photons 
in the polarization state $\la$ scattered into
respectively out of the beam due to Thomson scattering.

In the matter  (baryon/electron) rest frame,
which we indicate by  a prime, we know that
\[ {df_+^{(\la)\prime}\over dt'}(p,\bn)= {\si_Tn_e\over 4\pi}\int
     P^\la_{~\de}({\bn,\bn}') f^{(\de)\prime}(p',\bn')d\Om'  \; , \]
where $n_e$ denotes the electron number density
and $P^\la_\de$ is the $2\times2$ upper left corner of the normalized Thomson 
scattering matrix (\ref{chandra_eq}).
In the baryon rest frame which moves with four velocity $u$, the photon
energy is given by
\[ p' = p_\mu u^\mu \; . \]
We denote the photon energy with respect to a tetrad adapted to
the slicing of space-time into $\{ \ct=$ constant$\}$  hyper--surfaces by $p$  :
\[ p =  p_\mu n^\mu \; ,~~~\mbox{ with }~~
  n = a^{-1}[(1-A)\dd_\ct +B^i\dd_i] ~,~ \]
The lapse function and  the shift vector of the slicing are given by
  $\al= a(1+A)$ and  $\bm{\beta}=
	-B^{i}\dd_i$ .  In first order,
\[ p_0 = ap(1+A) - ap n_iB^i~~,\]
and to zeroth order $ p_i = ap n_i$.
Furthermore, the baryon four velocity has the form
$ u^0 = a^{-1}(1-A)~~,~~~ u^i = u^0v^i $ up to first order.
This yields
\[ p' = p_\mu u^\mu = p(1+ n_i(v^i-B^i)) \; . \]
Using this identity and performing the integration over photon energies,
we obtain
\bean 
\rho_\ga{d\io_+^{(\la)}( \bn)\over d\ct'} &=& a\rho_\ga\si_Tn_e\l[1+
4 n_i(v^i- B^i) + \r.  \\  && \l.
 {1\over 4\pi}\int\io^{(\de)}( \bn')P^\la_\de( \bn, \bn')d\Om'\r] ~ .
\eean

Photons which are scattered leave the beam with the probability
given by the Thomson cross section
 (see \eg~\cite{LP})
\[ 
{df^{(\la)}_-\over dt'} = \si_Tn_ef^{(\la)\prime}(p',\bn) ~, 
\]
so that we finally have
\bea
C^{(\la)\prime} &=& \frac{4\pi}{\rho_\ga a^4} 
	\int dp\l({df^{(\la)}_+\over dt'} -{df^{(\la)}_-\over dt'}\r)p^3 
	=\frac{1}{2}\si_Tn_e\big[4 n_i(v^i-B^i)  -\io^{(\la)}
 \nonumber \\  
&& ~~~ 
 + {1\over 4\pi} \int
  \io^{(\de)}( \bn') P^\la_\de({\bn,\bn}') d\Om'\big] ~ .
\eea

By setting $C^{(\MM)} = C^{(1)}+C^{(2)}$ and $C^{(Q)} = C^{(1)}-C^{(2)}$
we transform the collision integral back to the normal stokes
parameters. The term for $U$ has the same form as the one for
$Q$, just with the corresponding matrix elements in the integral.
The Boltzmann equation then finally becomes (setting 
$\EE\equiv (\MM,Q,U)$ and for the flat case, $\ka = 0$):

\bea
\lefteqn{\dot{\MM} + i \mu k \MM = 4 i \mu k (\Phi-\Psi+n^m \Si^{(V)}_m) 
  + 4 n^\ell n^m \dot{H}_{m \ell}} \nonumber \\
 &&+ a \si_T n_e \l[-\MM -4 i \mu V_b+4 n^\ell \om_{b,\ell} +
	\int d\Om' P^\al_1 \EE'_\al \r] \label{BMAp} \\
\lefteqn{\dot{Q} + i \mu k Q 
= a \si_T n_e \l[-Q + \int d\Om' P^\al_2 \EE'_\al \r]} \label{BVAp} \\
\lefteqn{\dot{U} + i \mu k U 
= a \si_T n_e \l[-U + \int d\Om' P^\al_3 \EE'_\al \r]}, \label{BUAp}
\eea
where we have to use the scattering matrix transformed into the 
$(\MM,Q,U)$ basis,
\be
 P  = P_S +P_V +P_T 
\ee
with
\bea
P^{(S)} &=& R^{-1}P^{(0)} R \\
  &=& \frac{3}{8} \l( \begin{array}{cc}
	   3-\mu^2-\mu^{\prime 2} + 3 \mu^2 \mu^{\prime2} & 
	(1-3\mu^2)(1-\mu^{\prime2})  \\
	 (1-\mu^2)(1-3\mu^{\prime2}&3 (1-\mu^2)(1-\mu^{\prime2})
	\end{array} \r) \\
P_V &=& \sqrt{1-\mu^2} \sqrt{1-\mu^{\prime2}} R^{-1} P^{(1)} R \\
  &=& \frac{3}{2} \sqrt{1-\mu^2} \sqrt{1-\mu^{\prime2}}
	\l( \begin{array}{ccc}
	\mu \mu' C & \mu \mu' C & - \mu S \\
	\mu \mu' C & \mu \mu' C & - \mu S \\
	\mu' S    & \mu' S    & C \end{array} \r)  \\
P_T &=& R^{-1} P^{(2)} R \\
  &=& \frac{3}{8}\scriptsize{ \l( \begin{array}{ccc}
	(1-\mu^2)(1-\mu^{\pr2})C_T   & -(1-\mu^2)(1+\mu^{\pr2})C_T 
	& 2 (1-\mu^2)\mu' S_T  \\
	- (1+\mu^2)(1-\mu^{\pr2})C_T & (1+\mu^2)(1+\mu^{\pr2})C_T  
	& -2(1+\mu^2)\mu' S_T \\
	-2\mu(1-\mu^{\pr2})S_T & 2\mu(1+\mu^{\pr2})S_T  & 4\mu\mu' C_T
	\end{array} \r)}
\eea
with $C = \cos(\phi-\phi')$, $S = \sin(\phi-\phi')$ and   \\
 $C_T = \cos(2(\phi-\phi'))$, $S_T = \sin(2(\phi-\phi'))$.
The parts $P_S,~P_V,~P_T$ of $P$ describe the scattering of the scalar,
vector and tensor contribution to $\EE$ respectively.

The functions $\MM$, $Q$ and $U$ depend on the wave vector $\bk$, the
photon direction $\bn$ and conformal time $\ct$.
We choose for each $\bk$-mode a reference system with 
$z$-axis parallel to $\bk$. 
For scalar perturbations we achieve in this way 
azimuthal symmetry --- the right-hand side of the Boltzmann equation and
therefore also the brightness $\MM^{(S)}$ depend only
on $\mu =  (\hat{\bk}\cd \bn)$ and can be expanded
in Legendre polynomials. The right-hand side of
the Boltzmann equation also determines the azimuthal
dependence of vector and tensor perturbations.
One can continue with this approach, but the resulting equations for $Q$ 
and $U$ and especially also their power spectra depend explicitly on the 
coordinate system. Therefore, we prefer an approach which is inherently 
covariant under rotation.

\subsection{Electric and magnetic polarization}
Since the Stokes parameters explicitly depend on the 
coordinate system, and Eqs.~(\ref{BVAp}) and (\ref{BUAp}) transform in a 
rather complicated way under rotations of the coordinate system.
A  method to characterize CMB polarization due to non-relativistic 
Thomson scattering, which is more convenient than 
the Stokes parameters since its transformation properties are very simple,
has been developed a couple years ago~\cite{SZ1,zalsel,Kam1,Kam2,HSZ}. A 
detailed derivation of this method goes beyond the scope of this report. Here 
we just repeat the definitions and the main results. We set
\be
	\bm{\cal T} = (\MM, Q+ iU,Q-iU) \label{TTAp}
\ee

In terms of this three component vector the collision integral above can 
we written (in vector form) as
\bea
 \bm{C}[\bm{\cal T}] &= & a\si_Tn_e[-\bm{\cal T} +
	\l({1\over 4\pi}\int\d\Om'\MM' 
 +(\bn\cd{\bf v}_b),0,0\r)  \nonumber \\
 &&  +{1\over 10}\sum_{m=-2}^2\int\d\Om' 
	P^{(m)}(\bn,\bn')\bm{\cal T}' \label{colA} 
\eea
From Eqs.~(\ref{chandra_eq}) to (\ref{matrixR}) one can determine the 
scattering matrix for the vector $\bm{\cal T}$
\be
  P^{(m)} = \left( \begin{array}{ccc}
  Y_2^{m'}Y_2^m & -\sqrt{{3\over 2}} {_2Y_2^{m'}}Y_2^m & 
	-\sqrt{{3\over 2}}_{-2}Y_2^{m'}Y_2^m \\
  -\sqrt{6}Y_2^{m'}{_2Y_2^m} & 3 _2Y_2^{m'}{_2Y_2^m} & 
	3 _{-2}Y_2^{m'}{_2Y_2^m} \\
 -\sqrt{6} Y_2^{m'}{_{-2}Y_2^m} & 3 _2Y_2^{m'}{_{-2}Y_2^m} & 
	3 _{-2}Y_2^{m'}{_{-2}Y_2^m} 
\end{array}\right) \label{TscatA}
\ee
where $_sY_l^{m'} = {_sY_l^{m*}(\bn')}$ and $_sY_l^{m} = {_sY_l^{m*}}(\bn)$ 
are the spin-weighted spherical harmonics~\cite{NP,Kam2,HW97}. 

We now decompose the Fourier components of the 
temperature anisotropy $\MM$ and the polarization variables $Q$ and $U$ as

\bea
 \MM &=& \sum_{\ell}\sum_{m=-2}^2\MM_{\ell}^{(m)}
	{ _0 G^m_{\ell}}, \label{newexp}\\
 Q \pm iU  &=& \sum_{\ell}\sum_{m=-2}^2
	(E_{\ell}^{(m)} \pm iB_{\ell}^{(m)}){ _2G^m_{\ell}}(\bn).
\eea
Here $m=0$ is the scalar mode, $m=\pm1$ are the vector and $m=\pm2$ are
the tensor modes. The functions $_sG^m_{\ell}$ are closely related to the spin 
weighted harmonics $_sY^m_{\ell}$:
\[
 _sG^m_{\ell}(\bn) = (-i)^{\ell}\sqrt{{4\pi\over 2\ell +1}} {_sY^m_{\ell}}(\bn)
\]

The evolution equations in term of these variables can be given in the 
following compact form~\cite{HSZ}
\bea
\lefteqn{ \dot\MM_{\ell}^{(m)} -k\l[{_0\ka^m_{\ell}\over 2\ell-1}
	\MM_{\ell-1}^{(m)}
-{_0\ka^m_{\ell+1}\over 2\ell+3}\MM_{\ell+1}^{(m)}\r] = }\nonumber \\
  &&	-n_e\si_Ta\MM_{\ell}^{(m)}  +S_{\ell}^{(m)} ~~~ (\ell\ge m)\\
\lefteqn{ \dot E_{\ell}^{(m)} -k\l[{_2\ka^m_{\ell}\over 2\ell-1}
E_{\ell-1}^{(m)}  -{2m\over \ell(\ell+1)}B_{\ell}^{(m)}
-{_2\ka^m_{\ell+1}\over 2\ell+3}E_{\ell+1}^{(m)}\r] = } \nonumber \\ && 
  -n_e\si_Ta[E_{\ell}^{(m)} + \sqrt{6}C^{(m)}\de_{\ell,2}] ~~~ (\ell\ge 2)\\
\lefteqn{ \dot B_{\ell}^{(m)} - k\l[{_2\ka^m_{\ell}\over 2\ell-1}
  B_{\ell-1}^{(m)}   +{2m\over \ell(\ell+1)}E_{\ell}^{(m)}
-{_2\ka^m_{\ell+1}\over 2\ell+3}B_{\ell+1}^{(m)}\r] = } \nonumber \\ &&  
  -n_e\si_TaB_{\ell}^{(m)} ~~~ (\ell\ge 2)~.
\eea
where we have set
\be \begin{array}{cc}
 S_0^{(0)} =n_e\si_Ta\MM^{(0)}_0, & S^{(0)}_1 =n_e\si_Ta4V_b +4k(\Psi-\Phi), \\
 S^{(0)}_2 =n_e\si_TaC^{(0)}, & S^{(1)}_1 =n_e\si_Ta4\sqrt{2}\om_b,  \\
 S^{(1)}_2 =n_e\si_TaC^{(1)} +4\sqrt{2/3}k\Si, 
	&   S^{(2)}_2 =n_e\si_TaC^{(2)} -8\sqrt{2/3}\dot H 
\end{array}   \nonumber
\ee
and $C^{(m)}= {1\over 10}[\MM^{(m)}_2 -\sqrt{6}E^{(m)}_2] $.
The coupling coefficients are
\[ 
	_s\ka^m_{\ell} =\sqrt{{(\ell^2-m^2)(\ell^2-s^2)\over \ell^2}}.
\]
Note that for scalar perturbations, $m=0$, $B$-polarization is not sourced 
and we have $B_{\ell}^{(0)}\equiv 0$.

Finally, we want to connect the intensities $\MM^{(m)}_{\ell}$ with the 
more familiar expansion of the 
scalar $(S)$, vector $(V)$ and tensor $(T)$ contributions to the 
brightness function in terms of Legendre polynomials. Usually one sets
\[ \MM = \MM^{(S)} +  \MM^{(V)} +  \MM^{(T)} ~.\]
Here $\MM^{(S)}$ only depends on $\mu=(\bn\cdot\bk)/k$ and the 
$\bn$-dependence of $\MM^{(V)}$ and $\MM^{(T)}$ can be written as  
\bea 
\MM^{(V)}(\mu,\phi) &=& \sqrt{1\!-\!\mu^2}\l[ \MM_1^{(V)}(\mu) \cos\phi 
	\!+\! \MM_2^{(V)}(\mu) \sin\phi\r] \label{v1}\\
\MM^{(T)}(\mu,\phi) &=& (1-\mu^2) \l[ \MM_+^{(T)} \cos(2\phi) + 
	\MM_\times^{(T)} \sin(2\phi)\r], \label{t1}
\eea
where $\phi$ is the azimuthal angle in the plane normal to $\bk$.
 Each of the functions $\MM_{\bullet}^{(S,V,T)}(\mu)$ is now expanded 
in Legendre polynomials
\be
\MM_{\bullet}^{(S,V,T)} =\sum_{\ell}(-i)^{\ell}(2\ell+1)
	\si_{\bullet,\ell}^{(S,V,T)}P_{\ell}(\mu)~.  \label{devel}
\ee
The coefficients $\si_{\bullet,\ell}^{(S,V,T)}$ are then related to 
$\MM_{\ell}^{(m)}$ via  the identities
\bea
\MM_{\ell}^{(0)} &=&  (2\ell+1) \si_{\ell}^{(S)} \\
\MM_{\ell}^{(\pm 1)} &=& \sqrt{\ell(\ell+1)}
   \l[\l(\si^{(V)}_{2,\ell-1}+\si^{(V)}_{2,\ell+1}\r) \r. \nonumber \\
  && ~~~ \l. \pm i \l(\si^{(V)}_{1,\ell-1}+\si^{(V)}_{1,\ell+1}\r)\r]\\
\MM_{\ell}^{(\pm 2)} &=& -\sqrt{(\ell+2)!\over (\ell-2)!} 
	\l[ {1\over 2\ell+3}
 \si_{\uparrow\downarrow,\ell+2}^{(T)} + {2(2\ell+1)\over (2\ell-1)
  (2\ell+3)}\si_{\uparrow\downarrow,\ell}^{(T)}\r. \nonumber \\
 && ~~~ \l. +
{1\over 2\ell-1} \si_{\uparrow\downarrow,\ell-2}^{(T)}\r]\\
 && \mbox{ with } \si_{\uparrow\downarrow,\ell}^{(T)}= 
\si_{+,\ell}^{(T)}\mp i\si_{\times,\ell}^{(T)} \nonumber
\eea
We do not repeat this correspondence for the Stokes parameters $Q$ and $U$ 
since it is rather complicated and not very useful as it depends on the 
coordinate system chosen. 

\subsection{Power spectra}
Since they are functions on a sphere, the observed CMB anisotropies and
polarization are conveniently expanded in spherical
harmonics: $\delta T (\bn,\ct_0)/T_0 = \sum_{\ell m} a_{\ell m} Y_{\ell}^m
 (\bn)$.
The coefficients $a_{\ell m}$ are random variables with zero mean and
rotationally invariant
variances, $C_\ell \equiv \langle \mid a_{\ell m} \mid ^2 \rangle$.
The  correlation function of the anisotropy pattern then has the
standard expression:
\be
 \left\langle {\delta T\over T_0}(\bn_1){\delta T\over
T_0}(\bn_2)\right\rangle = {1\over 4\pi} \sum_\ell (2\ell+1) {C_\ell} P_\ell
(\cos\theta)
\ee
 where $\cos\theta = \bn_1 \cdot \bn_2$ and $\lan\cdots\ran$ denotes 
 ensemble average. We use the Fourier transform normalization
\be
  \hat{f}(\bk) = {1\over V}\int f({\bf x})\exp(i\bk\cd{\bf x})d^3x~,
\ee
with some normalization volume $V$. Using statistical homogeneity 
we have
\bea  \lefteqn{
\left\langle {\delta T\over T_0}(\bn_1)
 {\delta T\over T_0}(\bn_2)\right\rangle  = 
{1\over V}\int d^3x \left\langle {\delta T\over T_0}({\bf x},\bn_1)
	{\delta T\over T_0}({\bf x}, \bn_2)\right\rangle} \nonumber \\
&=& {1\over (2\pi)^3}\int d^3k \left\langle  {\delta T\over T_0}(\bk,\bn_1)
	{\delta T\over T_0}(\bk, \bn_2\right\rangle)~.
\eea
Inserting our ansatz (\ref{devel}) for  ${\delta T\over T_0} ={1\over 4}\MM$,
and using the addition theorem for spherical harmonics, \\
$P_\ell(\bn_1\cdot\bn_2) = \frac{4\pi}{2\ell+1} 
	\sum_m Y_{\ell m}^*(\bn_1) Y_{\ell m}(\bn_2)$,
we find
\bea   \lefteqn{
\left\langle {\delta T\over T_0}(\bn_1){\delta T\over T_0}(\bn_2)\right\rangle
 = {1\over 8\pi}\sum_{\ell,\ell',m,m'}(-1)^{(\ell-\ell')}Y_{\ell m}(n_1)
    Y^*_{\ell' m'}(n_2)} \nonumber\\
~~~~~~  & ~~ &\times\int k^2dkd\Om_{\hat{\bk}}Y_{\ell m}^*(\hat{\bk})
	Y_{\ell' m'}(\hat{\bk})
  \langle\si_\ell\si^*_{\ell'}\rangle(k) \nonumber \\
~~~~~~   &=&{1\over 32\pi^2}\sum_\ell(2\ell+1)P_\ell(\bn_1\cd\bn_2) \int
	k^2dk \langle\si_\ell\si^*_\ell\rangle(k)~, 
\label{ClSa}\eea
from which we conclude

\be
C_\ell^{\MM\MM,(S)} =  {1 \over {8 \pi}}\int
 k^2dk\langle|\si^{(S)}_\ell(t_0,k)|^2\rangle ~,
\label{ClSapp}\ee
where the superscript $^{(S)}$ indicates that Eq.~(\ref{ClSapp}) gives
the contribution from {\em scalar} perturbations and $^{\MM\MM}$ means
that it is the contribution to the intensity perturbation.

The $QQ$, $UU$, $\MM Q$, $\MM U$ and $QU$ correlators depend with the Stokes 
parameters on the particular coordinate system chosen.
It is much more convenient to express the polarization power spectra in 
terms of the variables $E$ and $B$ which are independent of the coordinate 
system. Furthermore, since $B$ is parity odd, its correlators with $\MM$ 
and $E$ vanish. One finds the simple general expression~\cite{HSZ}
\be
 (2\ell+1)^2C_\ell^{XY(m)} = {n_m\over 8\pi}\int k^2dkX_{\ell}^{(m)}
	Y_{\ell}^{(m)*}~,  \label{PSpec}
\ee
where $n_m=1$ for $m=0$ and $n_m=2$ for $m=1,2$, accounting for the number 
of modes. Here $X$ and $Y$ run over $\MM$, $E$ and $B$.
\clearpage

\end{document}